\documentclass[12pt,letterpaper]{article}

\usepackage{jheppub_mod}
\usepackage{slashed}
\usepackage{bigints}
\usepackage{cancel}
\usepackage{stackrel}
\usepackage{tikz}
\usepackage{rotate}
\usepackage{rotating}

\def\Z{\mathbb{Z}}
\def\R{\mathbb{R}}

\def\ov{\overline}

\def\Tr{\text{Tr}}
\def\IM{\text{Im}\,}
\def\RE{\text{Re}\,}
\def\ov{\overline}
\def\1{{\bf 1}}
\def\2{{\bf 2}}
\def\3{{\bf 3}}
\def\4{{\bf 4}}
\def\6{{\bf 6}}

\def\targ#1#2{\genfrac{[}{]}{0pt}{}{#1}{#2}}
\def\targ2#1#2{\genfrac{}{}{0pt}{}{#1}{#2}}

\definecolor{mygr}{rgb}{0,0.6,0}
\definecolor{mygrey}{rgb}{0,0.1,0.2}
\definecolor{myblue}{rgb}{0,0.5,0.9}
\definecolor{myblue2}{rgb}{0,0.5,0.5}
\definecolor{myblue3}{rgb}{0,0.7,0.9}
\definecolor{myblue4}{rgb}{0,0.6,0.6}
\definecolor{myorange}{rgb}{1,0.5,0}
\definecolor{mypurple}{rgb}{0.6,0,1}
\definecolor{mygolden}{rgb}{1,0.8,0.2}
\definecolor{mycyan}{rgb}{0,1,1}
\definecolor{mymagenta}{rgb}{1,0,1}
\definecolor{mykiwi}{rgb}{0.8,1,0.5}
\definecolor{mybrown}{cmyk}{0.14, 0.42, 0.56, 0.2}
\definecolor{myturq}{cmyk}{0.99, 0, 0.2, 0.4}
\definecolor{myaubergine2}{cmyk}{0.4, 0.5, 0, 0.1}
\definecolor{myaubergine}{cmyk}{0.6,0.85,0,0}
\definecolor{CycleGreen}{cmyk}{0.52,0,1,0}
\definecolor{CycleBrown}{cmyk}{0, 0.4, 0.9, 0.2}

\makeatletter
\newsavebox\myboxA
\newsavebox\myboxB
\newlength\mylenA

\newcommand*\xoverline[2][0.75]{%
\sbox{\myboxA}{$\m@th#2$}%
\setbox\myboxB\null
\ht\myboxB=\ht\myboxA%
\dp\myboxB=\dp\myboxA%
\wd\myboxB=#1\wd\myboxA
\sbox\myboxB{$\m@th\overline{\copy\myboxB}$}
\setlength\mylenA{\the\wd\myboxA}
\addtolength\mylenA{-\the\wd\myboxB}%
\ifdim\wd\myboxB<\wd\myboxA%
   \rlap{\hskip 0.5\mylenA\usebox\myboxB}{\usebox\myboxA}%
\else
    \hskip -0.5\mylenA\rlap{\usebox\myboxA}{\hskip 0.5\mylenA\usebox\myboxB}%
\fi}
\makeatother

\title{Phases of Inflation}

\date{\today}

\author[a]{Gary Shiu,}
\author[b]{Wieland Staessens}

\affiliation[a]{Department of Physics, University of Wisconsin-Madison, Madison, Wisconsin, USA}
\affiliation[b]{Instituto de F\'isica Te\'orica UAM-CSIC, Cantoblanco, 28049 Madrid, Spain}

\emailAdd{shiu@physics.wisc.edu}
\emailAdd{wieland.staessens@csic.es}

\abstract{Motivated by the 4d effective field theories for closed string axions in Type II string compactifications with D-branes, we consider chiral gauge theories coupled to multiple axions. We discuss how well-known non-perturbative dynamical phenomena, such as gauge instantons, fermion confinement and Nambu-Jona-Lasinio interactions, give rise to non-trivial vacuum configurations in the IR. The fluctuations about the IR vacuum are captured by some remaining closed string axions and infladrons (scalar chiral condensate excitations), which acquire dynamical masses. By employing the full power of the effective field theory, we investigate the applicability of these IR theories as inflationary models (natural, monodromy, Starobinsky) and connect different types of inflationary scenarios to different phases of the non-Abelian gauge theory or the Nambu-Jona-Lasinio four-fermion couplings. The back-reaction of the infladrons flattens the axion potential in natural-like inflationary models, such that the tension with current CMB data with respect to the spectral index and the tensor-to-scalar ratio can be partially alleviated.      
 }

\preprint{MAD-TH-17-09\\ IFT-UAM/CSIC-18-70}

\begin{document}

\maketitle

\section{Introduction}
In the last four decades, observational cosmology has produced a staggering amount of experimental data and transformed our understanding of the universe's evolution at its core. Besides establishing the $\Lambda$-CDM model as the main contender to describe the properties of the cosmos in a handful of (independent) parameters, telescope experiments such as COBE, WMAP and Planck have also offered a glimpse into the earliest epochs of our expanding universe through the cosmic microwave background (CMB). More precisely, the observed homogeneous and isotropic nature of the CMB poses a series of new conundrums, such as the horizon problem and the flatness problem, which can be resolved by assuming a period of cosmic inflation~\cite{Starobinsky:1979ty,Starobinsky:1980te,Guth:1980zm,Linde:1981mu,Brout:1977ix} prior to the radiation- and matter-dominated era of our universe. As an extra treat, the paradigm of cosmic inflation also predicts the observed anisotropies in the CMB resulting from primordial curvature perturbations. At present, the fit of the $\Lambda$-CDM to the nearly scale-invariant and nearly Gaussian CMB power spectrum indicates~\cite{Ade:2015lrj} more than ever that inflation driven by a slowly rolling, single scalar field represents the most plausible scenario. 

The smoking gun that would settle once and for all the occurrence of an early inflationary epoch in our universe is the potential observation of B-mode polarization in the CMB due to primordial gravitational waves~\cite{Lyth:1996im}. Moreover, the detection of B-mode polarization would equally imply that the inflaton belongs to a so-called large field inflationary model, in which the inflaton transverses over super-Planckian distances in field space during inflation. Large field inflationary models are inherently sensitive to the UV-completion with gravity and only models embedded in a full-fledged quantum gravity have a chance to stand scientific scrutiny. In this sense, a promising arena for theoretical cosmology is presented by four-dimensional string compactifications, in which scalar fields abundantly arise as the geometric moduli of some internal space equipped with fancy mathematical structures. Using geometric moduli as inflaton candidates happens to be a well-explored avenue and has delivered some mixed results~\cite{Baumann:2014nda}. It still appears rather challenging to concoct sufficiently flat scalar potentials for a single geometric modulus whose mass has to be parametrically lighter than all the other geometric moduli.  

Alternatively, stringy axions~\cite{Witten:1984dg,Choi:1985je,Barr:1985hk,Svrcek:2006yi} represent inflaton candidates that are particularly well-motivated from string compactifications, as they enjoy perturbative shift symmetries that can help to constrain the form of and corrections to the inflaton potential. The last years have witnessed a tremendous activity in string cosmology around the construction (and destruction\footnote{In the last couple of years, attention has shifted more towards the study of quantum gravity constraints that potentially imperil the viability of large field inflation with axions.}) of inflation models with axions, which can be brought back to a set of prototype models: natural inflation~\cite{Freese:1990rb}, axion monodromy inflation~\cite{Silverstein:2008sg,McAllister:2008hb}, aligned natural inflation~\cite{Kim:2004rp}, N-flation~\cite{Dimopoulos:2005ac} and kinetic alignment~\cite{Bachlechner:2014hsa}. It is quite surprising that a limited set of building blocks in Type II string theory, namely fluxes or D-branes, already gives rise to such a variety of models. However, a complete and global Type II string compactification has to combine all building blocks in such a way that geometric moduli are stabilized and D-branes support the chiral gauge theories and particle spectrum of the Standard Model; for a review of Type II string model building see e.g.~\cite{Ibanez:2012zz,Blumenhagen:2005mu,Blumenhagen:2006ci} and for recent attempts in Type IIA see e.g.~\cite{Honecker:2012qr,Ecker:2014hma,Ecker:2015vea,Berasaluce-Gonzalez:2016kqb,Berasaluce-Gonzalez:2017bib}. Compactifications including D-branes with chiral gauge theories are characterized by additional four-dimensional mixing effects among multiple axions, such that their rich structure allows for parametrically large effective axion decay constants~\cite{Shiu:2015uva,Shiu:2015xda}, an indispensable first step towards natural inflation. Yet, to arrive at genuine inflation models in this setting, one also has to explain how (unstabilized) axions and possibly other scalar excitations acquire their mass. 

This last reflection is precisely the motivation behind this present work. To explain the generation of mass we employ the chiral nature of the gauge theories supported by the D-branes and various chiral symmetry breaking mechanisms arising from non-perturbative effects. The typical example in particle physics are gauge instanton solutions for non-Abelian gauge theories, which break the axial $U(1)$ symmetry explicitly and lift the mass of the $\eta'$-boson. And even though the mysterious nature of (fermion) confinement has still not been clarified, there is strong evidence that the formation of fermionic condensates plays an important role in the generation of hadron masses. Moreover, four-fermion interactions appear quite naturally in our setting upon integrating out a massive chiral $U(1)$ gauge boson, such that the Nambu-Jona-Lasinio (N-JL) mechanism offers an alternative way to generate non-trivial IR vacua in which scalar (composite) fields acquire a dynamical mass. Hence, the richness of our UV setting translates into a variety of mechanisms to dynamically generate scalar potentials for the infrared degrees of freedom, which in turn yields a diversity in inflationary models depending on the phase of the gauge theory or the four-fermion interactions. Generically, the inflation model arising in the IR will consist of multiple scalar fields, but for specific choices of the parameters a mass hierarchy between the various scalar fields emerges such that the set-up can be reduced to a single field inflationary model. Each single field inflationary model can be recognized in terms of a specific slow-roll prototype: natural, axion monodromy or Starobinsky inflation. The non-perturbative dynamics makes it rather difficult to compute the inflation potential from first principle, but with the help of effective field theory (EFT) the scalar IR degrees of freedom can be identified, which consist of fundamental stringy axions and composite bound states. Upon interpreting the effective scalar potentials as inflationary potentials and computing the cosmological observables tied to cosmic inflation, such as the spectral index and the tensor-to-scalar ratio, we can identify the regions of the parameter space for which inflation occurs over a timespan of 50 to 60 e-folds.

To present our ideas as lucidly as possible, we organize the paper in the following way: in section~\ref{S:MixAxSMG} we motivate the four-dimensional effective field theory for closed string axions arising upon dimensional reduction of Type II superstring theory on Calabi-Yau orientifolds. We also review how the various four-dimensional mixing effects can give rise to effective decay constants within a broader window than usually assumed in the literature, due to regions in the moduli space where the effective decay constant is enhanced or suppressed. As a third element, we identify the global symmetries of the chiral gauge theories which leave the action invariant and discuss how they are broken by the IR vacuum configuration in case of strong dynamics. Explicit breaking comes in the form of gauge instantons and the $\theta$-vacuum, while dyanmical breaking occurs in the confinement phase of the non-Abelian gauge theory or in the Nambu-Goldstone phase for the Nambu-Jona-Lasinio four-fermion interaction. In section~\ref{S:EFTInfAx} we work out the low-energy effective field theory for a one-generational chiral gauge theory at strong coupling, with the vacuum excitations consisting of one closed string axion and two scalar composite states formed from bilinear fermionic bound states. In case the closed string axion comes with an effective decay constant that is parametrically larger than the strong coupling scale, a natural-like inflationary model can be extracted and the machinery of effective field theories provides the necessary tools to assure that perturbative quantum corrections do not drastically change the vacuum configuration. This section concludes with a study of the back-reaction effects due to two massive bound states on the inflationary trajectory of the closed string axion. It is shown that this back-reaction flattens the periodic axion potential around its maxima and moves the spectral index and tensor-to-scalar ratio closer towards the 95\% confidence region in the $(n_s, r)$-plane. Section~\ref{S:PhaseInf} discusses other single field inflationary models that can be obtained from the same UV set-up, such as axion monodromy-like models and Starobinsky-like models. The latter type arises in the Nambu-Goldstone phase for strongly coupled four-fermion interactions. And last but not least, section~\ref{S:con} combines the conclusions of our work.

\section{Mixing Axions and Dynamical Mass Generation}\label{S:MixAxSMG}
\subsection{Stringy Axions in Effective Field Theories}\label{Ss:StrAxII}
The low-energy effective field theory of a superstring compactification to four dimensions is enriched with fields and interactions beyond those of particle physics or the inflationary paradigm. A typical example is the presence of four dimensional ${\cal N}=1$ supersymmetry in case the compactification manifold is a Calabi-Yau manifold for heterotic or Type I superstring theory and a Calabi-Yau orientifold for Type II superstring theory. Another well-known example is the appearance of axions (or CP-odd scalars with shift symmetry) following the Kaluza-Klein reduction of differential $p$-forms in the massless closed string spectrum. These massless gauge potentials are for instance abundantly present in the Ramond-Ramond (RR) sectors of the type II superstring theory, and complement the Neveu-Schwarz (NS) 2-form inherent to all superstring theories but Type I. Let us, for lucidity, elaborate briefly the case for Type IIA superstring theory compactified on a Calabi-Yau orientifold\footnote{For the compactification to give rise to geometric moduli and axions in four dimensions, the internal Calabi-Yau orientifold has to allow for non-vanishing Hodge numbers~$h_-^{11}$ and~$h^{21}$. The Hodge number~$h_-^{11}$ encodes the possible K\"ahler deformations of the internal space and counts the number of non-trivial harmonic (1,1)-forms supported by the Calabi-Yau orientifold, or equivalently the number of orientifold-odd two-cycles. The Hodge number $h^{21}$ captures the possible complex structure deformations of the Calabi-Yau orientifold and the number of non-trivial harmonic $(2,1)$-forms supported on the manifold.} and focus on four-dimensional axions $b^A$ arising from the NS 2-form $B_2$ and the axions~$c^i$ arising from the RR 3-form $C_3$, though we emphasize that similar considerations can be made for other string theory compactifications~\cite{Svrcek:2006yi}. Upon dimensionally reducing the ten-dimensional Type IIA supergravity action along the six-dimensional orientifold, a four-dimensional effective action arises for the axions $b^a$ and $c^i$, which will determine their interactions at low energy compared to the Kaluza-Klein and string scale. Universal to each compactification are the kinetic terms for the axions,
\begin{equation}\label{Eq:KinTermsC3Axions}
{\cal S}^{\rm kin}_{\rm axions} = \int - \frac{1}{2} \sum_{A,B=1}^{h_-^{11}} {\cal G}_{AB} (t) db^A \wedge \star_4 db^B - \frac{1}{2} \sum_{i,j=0}^{h^{21}} {\cal G}_{ij} (u) dc^i \wedge \star_4 dc^j.  
\end{equation}
which arise from the dimensional reduction of the kinetic terms for the NS 2-form and RR 3-form respectively. The metric ${\cal G}_{AB}(t)$ depends on the K\"ahler moduli $t^A$, the saxionic partners of $b^A$ in the ${\cal N}=1$ multiplet. The metric ${\cal G}_{ij}(u)$ on the axion moduli space depends explicitly on the complex structure moduli $u^i$, which form the saxionic partners of $c^i$ in the ${\cal N}=1$ chiral multiplet for a supersymmetric compactification on the Calabi-Yau orientifold.\footnote{The factorability of Calabi-Yau moduli space into a K\"ahler and complex structure moduli space is valid for string compactifications without D6-branes or where D6-branes wrap rigid three-cycles without deformation moduli. In all other cases, the moduli space no longer factorizes and the analysis becomes more intricate.} In practice, we will assume that the K\"ahler moduli and complex structure moduli acquired a non-zero {\it vev} below the Kaluza-Klein scale $M_{KK}$ by virtue of moduli stabilization mechanisms~\cite{DeWolfe:2005uu,Camara:2005dc}, such that we can treat the metric entries as continuous parameters in the theory for the remainder of the paper.\footnote{This does not a priori mean that the axionic partners are stabilized as well. In type IIA string theory, for instance, only one linear combination of complex structure axions is stabilized by turning on a perturbative NS-flux $H_3$, while all orthogonal directions remain flat directions.}  

Hence, an ${\cal N}=1$ Calabi-Yau orientifold compactification of Type IIA string theory naturally comes with a plethora of axions $b^A$ and $c^i$, whose four-dimensional (global) shift symmetry is inherited from the gauge symmetry of the higher-dimensional $p$-forms. Consequently, any axion-coupling in the four-dimensional superpotential or scalar potential is highly constrained by the perturbative shift symmetry, which is not expected to be broken randomly. Yet, a standard lore dictates that global symmetries in a full theory of quantum gravity ought to be broken due to gravitational effects~\cite{Abbott:1989jw,Coleman:1989zu,Kallosh:1995hi,Banks:2010zn}. Otherwise these four-dimensional effective field theories risk being incompatible with quantum gravity and are thereby destined to end up in the swampland~\cite{Vafa:2005ui,Ooguri:2006in}. Fortunately, higher-dimensional couplings for the gauge potentials $B_2$ and $C_3$, due to effects as perturbative fluxes along the internal dimensions or D-branes wrapping along specific internal cycles, are expected to induce~\cite{Montero:2017yja} additional interactions for their respective four-dimensional axions, such that their global symmetries in four dimensions are (spontaneously) broken or gauged.   

The {\it breaking} of the shift symmetry is triggered by coupling the axion to a four-form~$F_4$ through a polynomial:
\begin{equation} \label{Eq:KS-GenStruct}
{\cal L}_{KS} = ( e_0 + q_A b^A + \frac{1}{2}  m_{AB} b^A b^B + \ldots ) F_4,  
\end{equation}
which represents a (natural) generalization of the formulation for gauging axion symmetries (in particular, for QCD) \cite{Dvali:2005an,Dvali:2005zk}, which has  subsequently been adopted for inflation in \cite{Kaloper:2008fb}. The coefficients in the polynomial correspond to flux quanta resulting from internal flux threading non-trivial cycles along the compact directions. The invariance of the lagrangian under the axion shift symmetry $b^A \rightarrow b^A +\lambda^A$ is preserved, provided that the flux quanta shift appropriately, namely $e_0 \rightarrow e_0 - q_A \lambda^A - \frac{1}{2} m_{AB} \lambda^A \lambda^B$ and $q_A \rightarrow q_A - m_{AB} \lambda^B$ for the polynomial above. Upon dualization of the four-form in favour of the flux quanta, one obtains a scalar potential for the axions with the polynomial structure in $b^A$ as in (\ref{Eq:KS-GenStruct}). The generalized axion-4-form coupling can be extracted both for $B_2$-axions and $C_3$-axions in the presence of closed string fluxes~\cite{Bielleman:2015ina}, and continues to hold even for open string axions (associated to D-brane displacement moduli)~\cite{Escobar:2015fda,Escobar:2015ckf,Carta:2016ynn}, as can be checked explicitly through the dimensional reduction of the ten-dimensional Chern-Simons actions.  This intimate connection between flux vacua in Type IIA compactifications and the axion-4-form coupling allows for an elegant realization of axion monodromy.

The four-form $F_4$ does not have to be fundamental in order for the axion-4-form coupling (\ref{Eq:KS-GenStruct}) to emerge, as it can also correspond to the topological density $\Tr(G\wedge G)$ of a non-Abelian gauge group, with $G = d{\cal A} + {\cal A} \wedge {\cal A}$ the non-Abelian field strength. Historically, such a coupling was identified as the anomalous coupling between an axion and the non-Abelian gauge instantons:
\begin{equation}\label{Eq:AnomCoupling}
{\cal L}_{\rm anom} = \frac{1}{8 \pi^2} \sum_i n_i c^i \, \Tr(G\wedge G).
\end{equation}
In type IIA string theory this coupling can be explicitly obtained for the $C_3$-axions by virtue of the dimensional reduction of the D6-brane Chern-Simons action. In this expression, the integers $n^i \in \Z$ represent the topological wrapping numbers of the three-cycle wrapped by the D6-brane stack expanded in a basis of three-cycles, meaning that $n^i\neq 0$ if the D6-brane stack wraps around a particular basis three-cycle. Physically, gauge instantons represent finite energy solutions to the Euclidean action for the non-Abelian gauge theory, which connect two distinct (winding number) vacua through quantum mechanical tunneling. Instantons minimize the energy of the Euclidean action, as they satisfy the (anti)self-duality conditions $G_{\mu \nu} = \pm \varepsilon_{\mu\nu\rho\sigma} G_{\rho \sigma}$ in Euclidean spacetime. The perturbative shift symmetry of the collective axionic direction $\sum_i n_i c^i$ induces simultaneously a shift of the continuous $\theta$-parameter associated to the $\theta$-vacuum of the non-Abelian gauge theory in analogy to the discussion in section~\ref{Ss:ChiralSymBreaking}, but its continuous component is broken explicitly once a particular $\theta$-vacuum has been selected. To maintain the invariance of the path integral in the presence of a gauge instanton background, the axionic direction $\sum_i n_i c^i$ is subject to the {\it collective} discrete periodicity,
\begin{equation}
\sum_i n_i c^i \simeq \sum_i n_i c^i + 2 \pi.
\end{equation}
Apart from gauge instantons, string theory also provides additional non-perturbative effects which do not have a field theoretic counter-part, namely D-brane instantons. Geometrically, a D-brane instanton in Type IIA corresponds to a Euclidean D2-brane (E2-brane) wrapping a three-cycle along the internal Calabi-Yau orientifold and its coupling to the axion $c^i$ is captured by a non-perturbative contribution to the superpotential:
\begin{equation}\label{Eq:InstSuperPot}
{\cal W}_{np} \ni A (e^{-T^A} , \Phi)  \, e^{-S_{E2} + i \, c_i}
\end{equation}
with the amplitude potentially depending on the K\"ahler moduli $T^A$ and on open string excitations collectively denoted by $\Phi$. The E2-brane instanton action $S_{E2}$ scales with the volume of the wrapped three-cycle $\Gamma_i$:
\begin{equation}
S_{\rm E2} = \frac{1}{g_s} \frac{1}{\ell_s^3} {\rm Vol} (\Gamma_i) .
\end{equation}
The non-perturbative nature of the E2-brane instantons is encoded in the dependence on the string coupling $g_s$, and effectively breaks the continuous shift symmetry of the axion $c^i$ to a discrete one:
\begin{equation}\label{Eq:ED2period}
c^i \rightarrow c^i + 2 \pi.
\end{equation}  
Hence, the coupling of closed string axions to E2-brane instantons constrains the topology of the $c^i$-axion moduli space turning it into a $h^{21}+1$ dimensional torus.

Instead of breaking the axion shift symmetry, one could promote it to the global part of a {\it gauge} symmetry by introducing St\"uckelberg couplings to an Abelian one-form. In case the Type IIA orientifold compactification includes D6-branes, a closed string axion $c^i$ can acquire a St\"uckelberg coupling whenever a single D6-brane wraps an (orientifold-odd) three-cycle $\Delta^i$. In that case the kinetic term (\ref{Eq:KinTermsC3Axions}) for axion $c^i$ has to be amended through the replacement 
\begin{equation}
dc^i \rightarrow dc^i + k^i A,
\end{equation} 
where $A$ represents the four-dimensional $U(1)$ gauge potential supported by the D6-brane with $k^i$ the associated St\"uckelberg charge.\footnote{Technically, the St\"uckelberg couplings can be obtained through dualisation of the Hodge dual two-forms appearing in the reduction of the RR 5-form $C_5$, as reviewed in more detail in~\cite{Shiu:2015xda}.} These St\"uckelberg couplings are an immediate consequence of the Green-Schwarz mechanism required for anomaly-cancellation in case the local $U(1)$ symmetry is anomalous, and follow from the D6-brane Chern-Simons action upon dimensional reduction. Under influence of the St\"uckel-berg coupling the $U(1)$ gauge field acquires a longitudinal component, which makes this gauging scenario an excellent starting point to investigate inflation by virtue of massive vector bosons~\cite{Golovnev:2008cf} in Type IIA compactifications.

Instead, we keep our focus on the closed string axions and dedicate our attention to compactifications in which breaking and gauging of the axion shift symmetries simultaneously occurs. To this end, we consider Type IIA compactifications on Calabi-Yau orientifolds with D6-branes and zoom in on the generic four-dimensional effective action for closed string axions:\footnote{A simpler form of this lagrangian was recently considered purely from field theoretic considerations~\cite{Dvali:2017mpy} to argue for the spontaneous breaking of chiral flavor symmetry in chiral gauge theories.} 
\begin{align}
{\cal S}_{\rm CSA} &= \int - \frac{1}{2} \sum_{i,j} {\cal G}_{ij} (dc^i -k^i_a A^{(a)}) \wedge \star_4 (dc^j -k^j_b A^{(b)}) - \frac{1}{8\pi^2} \left(\sum_{\alpha, i} n_i^\alpha c^i \right) \Tr(G_\alpha\wedge G_\alpha) \nonumber  \\ 
& \qquad - \frac{1}{g_{(a)}^2} F^{(a)} \wedge \star_4 F^{(a)} - \frac{1}{g_{(\alpha)}^2} \Tr(G_\alpha \wedge \star_4 G_\alpha ) + {\cal L}_{\rm fermions} \label{Eq:GenericAction}\\
&\qquad - \frac{1}{2} \sum_{A,B} {\cal G}_{AB} db^A \wedge \star_4 db^B - V_{\rm flux}(b^A, c^i) \star_4 \1 \notag
\end{align}
with $i,j \in \{1,\ldots, h^{21}+1 \}$, $A,B \in \{1, \ldots, h^{11}_- \}$, $a,b \in \{1, \ldots, N_1 \}$ running over the number $N_1$ of $U(1)$ gauge groups with gauge potential $A^{(a)}$ and field strength $F^{(a)}$ (in the open string sector) and $\alpha \in\{1,\ldots N_G\}$ running over the number $N_G$ of non-Abelian gauge groups with field strength $G_\alpha$ supported by stacks of coincident D6-branes. In order for this action to be fully consistent, we also require the presence of chiral fermions whose contribution to the gauge anomalies cancel the non-invariance of the anomalous couplings between the charged axions $\sum_i n_i^\alpha c^i$ and topological density $\Tr(G_\alpha \wedge G_\alpha)$ under the $U(1)$ transformations. We will discuss the importance and implications of these chiral fermions in full detail in section~\ref{Ss:ChiralSymBreaking}. The last line also takes into consideration the perturbative scalar potential $V_{\rm flux}$ capturing the (generalized) axion-4-form couplings for the closed string axions induced by internal NS- and RR-fluxes.

The set-up in (\ref{Eq:GenericAction}) is not just restricted to Type IIA compactifications with D6-branes, but can also be obtained from Type IIB orientifold compactifications with D7-branes, where closed string axions arise from the RR-forms $C_2$ and $C_4$. A~more complete (and rigorous) analysis of Type II superstring compactifications with D-branes can be found in the literature~\cite{Ibanez:2012zz,Blumenhagen:2006ci,Jockers:2004yj,Haack:2006cy,Kerstan:2011dy,Grimm:2011dx,Camara:2011jg}. For our purposes, it suffices to argue how the effective action (\ref{Eq:GenericAction}) finds its microscopic justification from string theory compactifications and consider that effective action as the starting point to investigate the low energy behavior for closed string axions.

\subsection{Effective Axion Decay Constants}\label{Ss:EffADC}
Their inherent shift symmetry and natural abundance in string compactifications have made axions very beloved candidates for large field inflationary models in string cosmology. A naive string model builder would attempt to embed natural inflation~\cite{Freese:1990rb} in a controlled string environment and realise rather quickly that this is not attainable when considering a single axion. As the axion decay constant for a closed string axion is set by the {\it vev} of their saxion partner (K\"ahler or complex structure modulus), trans-Planckian axion periodicities -- needed to address the $\eta$-problem -- are unequivocally excluded when maintaining perturbative control of the compactification~\cite{Banks:2003sx}. Clearly, in order to surpass this obstacle and realise controllable inflationary models with stringy axions, we need to dig deeper into the toolbox of field theory and string compactifications. It turns out that slow-roll single field inflationary models can still be constructed with a single axion by turning on other dynamical effects than instanton contributions to break the shift symmetry of the axion, as discussed around equation (\ref{Eq:KS-GenStruct}). Stringy elements such as D-branes and internal fluxes allow to generate an inflationary potential with monodromy, hence the name {\it axion monodromy}~ \cite{Silverstein:2008sg,McAllister:2008hb,Berg:2009tg,Marchesano:2014mla,Blumenhagen:2014gta,Hebecker:2014eua,McAllister:2014mpa}, while keeping the axion decay constant sub-Planckian. Despite its elegant simplicity, axion monodromy models still require a lot of work to arrive at a fully controlled string computation with moduli stabilization and face various open questions regarding moduli back-reaction and other quantum gravity constraints~\cite{Blumenhagen:2014nba,Baume:2016psm,Blumenhagen:2017cxt,Landete:2017amp}.  

Extending the field content to two axions presents the next conspicuous avenue towards large field inflation, for which the axions are subject to two non-perturbative effects such that an almost flat direction is created along one particular direction in the axion field space. This almost flat axionic direction arises by tuning a handful of parameters in the scalar potential to obtain an effective trans-Planckian decay constant, which captures the basic idea behind {\it aligned natural inflation}~\cite{Kim:2004rp} and its variants~\cite{Choi:2014rja,Higaki:2014pja,Tye:2014tja,Ben-Dayan:2014zsa,Burgess:2014oma}. This minimal extension is unfortunately also plagued with various drawbacks: on the one hand, embeddings of this alignment mechanism in Type IIB string theory set-ups~\cite{Long:2014dta,Ben-Dayan:2014lca} point generically to non-Abelian gauge groups with unnaturally large rank, and on the other hand the inflationary trajectory is expected to be lifted due to uncontrolled gravitational instantons~\cite{Montero:2015ofa} (or more generally, expectations from the weak gravity conjecture \cite{ArkaniHamed:2006dz, Brown:2015iha,Brown:2015lia,Hebecker:2015rya, Rudelius:2015xta,Heidenreich:2015wga}). Considering $N_{\rm ax}\gg2$ axions allows to take into account {\it kinetic mixing} between axions due to a non-diagonal metric on the axion moduli space, whose topology is determined by at least $N_{\rm ax}$ instanton corrections in the scalar potential. A mismatch between the eigenbasis for the kinetic terms and the eigenbasis for the potential terms eventually requires to identify an invariant diameter for the $N_{\rm ax}$-dimensional hypercube, whose length serves as a proxy for large field excursions. This diameter can be parametrically enhanced due to its scaling with the number $N_{\rm ax}$ of axions~\cite{Bachlechner:2014gfa,Bachlechner:2015qja,Junghans:2015hba}:  
\begin{equation}\label{Eq:ParaEnhNaxions}
f_{\rm diag} \sim N^{p}_{\rm ax} f_{\rm max}, \qquad \qquad p \in \left\{\frac{1}{2}, 1, \frac{3}{2} \right\}
\end{equation}  
with $f_{\rm max}$ the largest eigenvalue of the metric on the axion moduli space. Despite this enhancement, it still remains difficult~\cite{Long:2016jvd} to achieve trans-Planckian displacements along the diameter of the axion moduli space. 

\begin{figure}[h]
\begin{center}
\begin{tabular}{c@{\hspace{0.8in}}c}
\includegraphics[width=5cm, height=5cm]{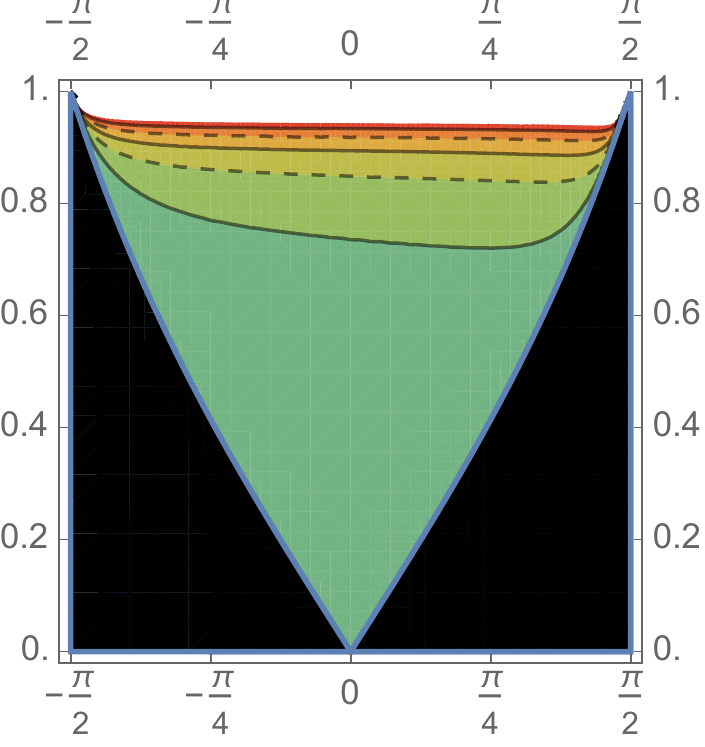} \begin{picture}(0,0) \put(-77,-6){$\chi$}  \put(-180,72){$\sqrt{\frac{{\cal G}_{22}}{{\cal G}_{11}}}$} \end{picture}  & \includegraphics[width=5cm, height=5cm]{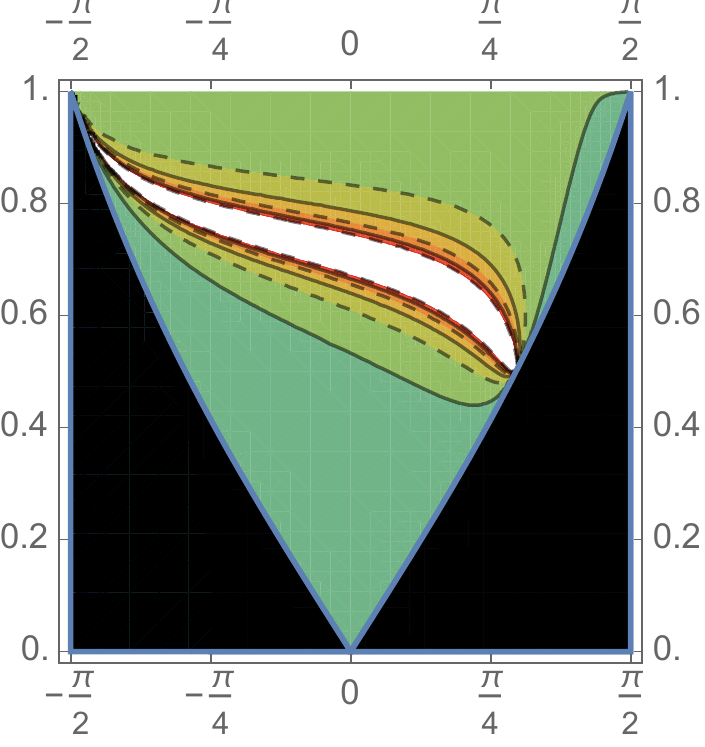} \begin{picture}(0,0) \put(-77,-6){$\chi$}  \put(-180,72){$\sqrt{\frac{{\cal G}_{22}}{{\cal G}_{11}}}$}  \end{picture} 
\end{tabular}
\caption{Contour plot for effective axion decay constant $f_\xi$ in (\ref{Eq:NewADCFullMixing}) as a function of the moduli-dependent quantities $\chi$ and $\sqrt{{\cal G}_{22}/{\cal G}_{11} }$ with discrete parameters $ k^1 = k^2 =  n_1 =  n_2$ (left) and $ k^1 =   k^2 = - 2 n_1 = - n_2$ (right). Black regions correspond to unphysical values for $f_{\xi}$ (due to a negative value for $\lambda_-$), while the physical values follow the color-coding from small (green) to large (red).     \label{Fig:ContEx1}}
\end{center}
\end{figure}
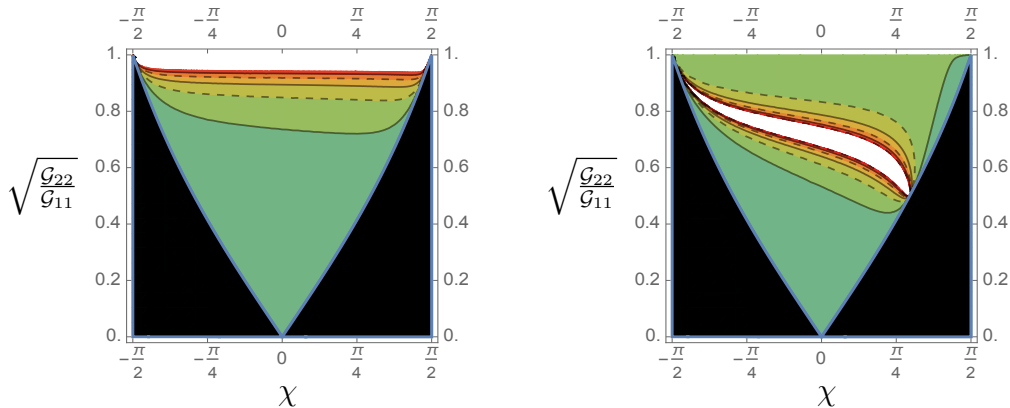

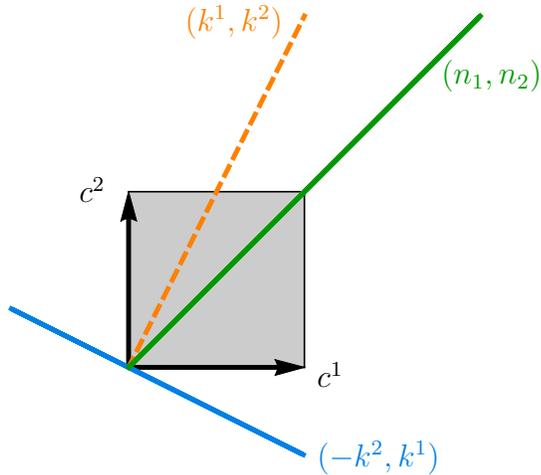
\begin{figure}[h]
\begin{center}
\includegraphics[scale=0.6]{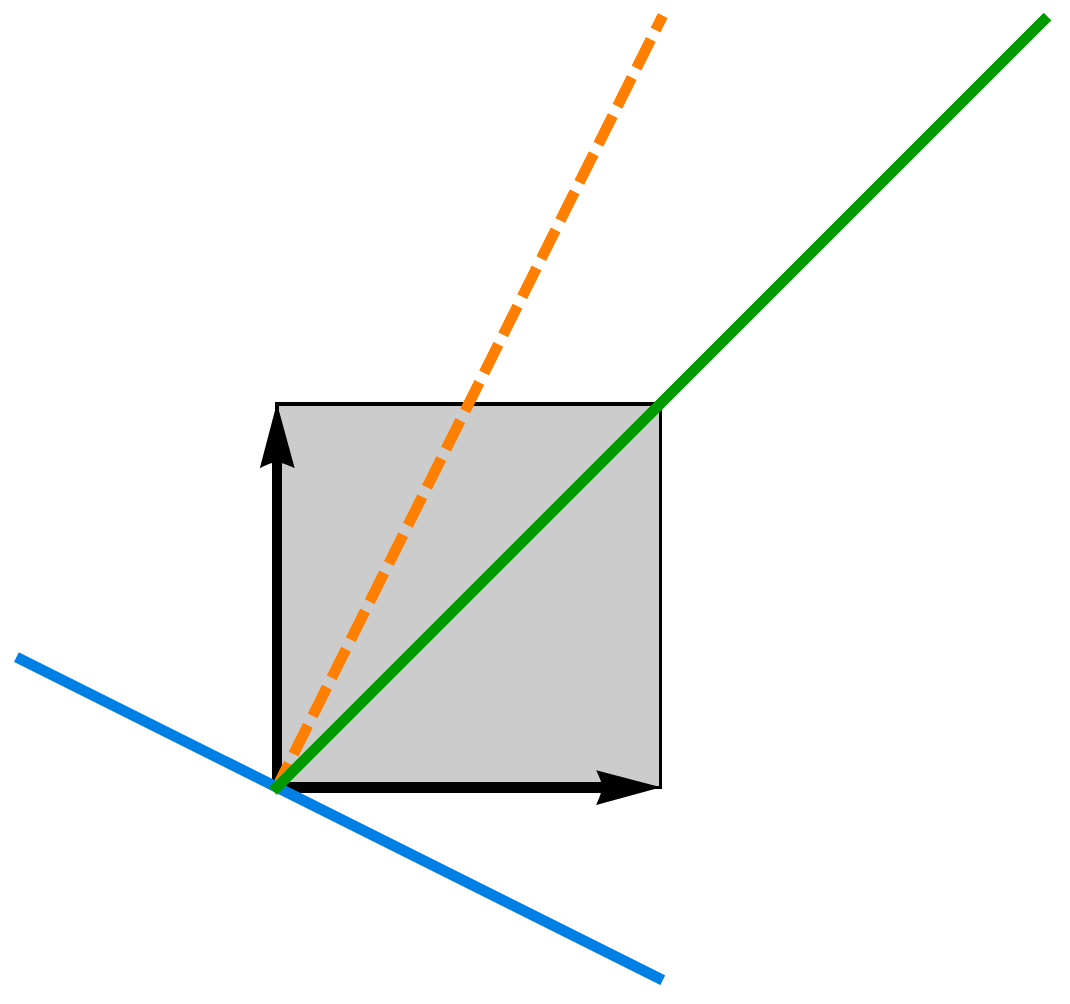} \begin{picture}(0,0) \put(-70,30){$c^1$}  \put(-160,100){$c^2$} \put(-120,165){\color{myorange}$(k^1,k^2)$} \put(-23,145){\color{mygr}$(n_1,n_2)$} \put(-70,0){\color{myblue}$(-k^2,k^1)$} \end{picture}
\caption{Axion moduli space for two axions $(c^1,c^2)$ with the orbit under the $U(1)$-St\"uckelberg gauging indicated through the orange, dashed line. The blue line represents the orthogonal direction to the St\"uckelberg axion, while the green line indicates the axionic direction coupled to the non-Abelian gauge group.  \label{Fig:AxionModuliSpace}}
\end{center}
\end{figure}

Going back to action (\ref{Eq:GenericAction}) we realise that the dimensional reduction of Type~II superstring theory offers yet another element not taken into account in the scenarios discussed above, namely the possibility of gauging the axion shift symmetry for the RR-axions $c^i$ and its effect on the effective axion decay constant. This was exactly the question posed in~\cite{Shiu:2015uva,Shiu:2015xda} and can already be addressed by restricting (\ref{Eq:GenericAction}) to a minimal set-up consisting of $N_{\rm ax} = 2$ axions, one Abelian $U(1)$ gauge group and one non-Abelian $SU(N_c)$ gauge group:    
\begin{eqnarray}\label{Eq:GeneralLagrangianN2}
{\cal S}_{\rm CSA}^{\rm N=2} &= & \bigintsss \left[- \frac{1}{2}\, \sum_{i,j=1}^2 {\cal G} _{ij} (d c^{i}-k^{i}A) \wedge \star_4 (d c^{j}-k^{j}A) -\frac{1}{g_{1}^{2}} F \wedge \star_4 F  -\frac{1}{g_{2}^{2}} \Tr(G\wedge \star_4 G) \right. \notag \\
&& \left. \qquad  +\frac{1}{8\pi ^2}\left( \sum_{i=1}^2 n_i c^i \right) \text{Tr} ( G \wedge G ) + {\cal L}_{\rm fermions} \right]. 
\end{eqnarray}
Upon diagonalizing the metric ${\cal G}_{ij}$ and identifying the axionic direction $\zeta$ absorbed by the $U(1)$ gauge boson in the St\"uckelberg mechanism, we can deduce the effective axion decay constant $f_\xi$ for the orthogonal axionic direction $\xi$:\footnote{In \cite{Hebecker:2015rya} an alternative method was proposed using perturbative fluxes in the Type IIB superpotential to enforce a flat axionic direction.}
\begin{equation}\label{Eq:NewADCFullMixing}
f_{\xi} = \frac{\sqrt{\lambda_+ \lambda_-}  \sqrt{ \lambda _- \,(k^-)^2 + \lambda _+\,(k^+)^2}}{ \cos \frac{\chi}{2} \left(  \lambda_+  k^+  n_2  + \lambda_- k^- n_1 \right) + \sin \frac{\chi}{2}  \left( \lambda_- k^- n_2 - \lambda_+  k^+  n_1  \right)},
\end{equation} 
with $\lambda_{\pm} \equiv \frac{1}{2}\left[ ({\cal G}_{11} + {\cal G}_{22}) \pm \sqrt{4 {\cal G}_{12}^2 + ({\cal G}_{11}-{\cal G}_{22})^2} \right]$ the eigenvalues of metric ${\cal G}_{ij}$ and $\chi$ the rotation angle allowing for its diagonalization:  
\begin{equation}
\sin \chi=\frac{2{\cal G}_{12}}{\sqrt{4 {\cal G}_{12}^2 + ({\cal G}_{11}- {\cal G}_{22})^2}}, \qquad \text{with } \,0\leq \chi <2\pi.
\end{equation}
Under the $SO(2)$ diagonalisation matrix for the metric ${\cal G}_{ij}$ also the St\"uckelberg charges rotate into the charges $k^\pm$: 
\begin{equation}\label{Eq:ChargesNewBasis}
\left(\begin{array}{c}  k^- \\ k^+ \end{array} \right) = \left(  \begin{array}{lr} \sin \frac{\chi}{2} & - \cos \frac{\chi}{2} \\  \cos \frac{\chi}{2} & \sin \frac{\chi}{2}   \end{array} \right)   \left(\begin{array}{c}  k^1 \\ k^2 \end{array} \right).
\end{equation}
From an effective field theory viewpoint, the axion decay constant (\ref{Eq:NewADCFullMixing}) depends on a number of discrete parameters $(k^1,k^2,n_1,n_2)$ and a set of stabilised moduli fields hidden within the three independent metric components ${\cal G}_{ij}$. To depict its dependence on the continuous moduli, the axion decay constant can be represented by two-dimensional countourplots~\cite{Shiu:2015xda} as in figure~\ref{Fig:ContEx1}, which are spanned by the moduli ratio $\sqrt{{\cal G}_{22}/{\cal G}_{11}}$ and the rotation angle $\chi$. The values of the axion decay constant are then measured in units of the remaining continuous parameter $\sqrt{{\cal G}_{11}}$ in such contourplots, which can contain regions where the axion decay constant parametrically enhances as its denominator tends to blow up, without having to fine-tune any of the discrete parameters. 
In explicit Type II model building scenarios such an enhancement was noticed in areas of the closed string moduli space which exhibit a higher form of isotropy~\cite{Shiu:2015xda} among the complex structure (IIA) or K\"ahler (IIB) moduli. We cannot stress enough that this type of enhancement is different from the parametric enhancement discussed in equation (\ref{Eq:ParaEnhNaxions}). In the first place, the dimension of the moduli space hypercube is reduced due to the $U(1)$ gauge boson eating away an axionic direction in case the axions carry St\"uckelberg charges, such that any large $N_{\rm ax}$ enhanced displacement would have to take place along axionic directions perpendicular to this St\"uckelberg direction. A second difference is that this parametric enhancement already occurs for two axions and does not require $N_{\rm ax}\gg2$. Coming back to the examples in figure \ref{Fig:ContEx1}, we observe that the white strips contain regions where the axion decay constant $f_\xi$ enhances by a factor ${\cal O}(10^2-10^3)$, while the dark green zones encompass areas with a suppressed axion decay constant by a factor ${\cal O}(10^{-2}-10^{-3})$. If the eigenvalues of the moduli metric ${\cal G}_{ij}$ are of the order ${\cal O}(10^{16})$~GeV, the plots in figure~\ref{Fig:ContEx1} suggest a much wider axion window for the closed string axion decay constant than previously assumed in the literature~\cite{Banks:2003sx,Svrcek:2006yi}:
\begin{equation}\label{Eq:CSAxionWindow}
10^{13} \text{ GeV} \lesssim f_\xi \lesssim 10^{19} \text{ GeV},
\end{equation}      
due to the misalignment between the axion basis in the kinetic and potential terms. In order to obtain a better geometric picture of this enhancement process, it is useful to take a look at the axion moduli space for the set-up with two axions in (\ref{Eq:GeneralLagrangianN2}), as depicted in figure \ref{Fig:AxionModuliSpace}. 
The two-torus topology of the axion moduli space follows from the periodicity~(\ref{Eq:ED2period}) imposed by the Euclidean D2-brane instantons coupling to the closed string axions $c^i$, as explained in the previous section. More explicitly, D2-brane instantons define a lattice $\Gamma: c^i \simeq c^i + 2\pi$ on the (perturbative) ambient space $\R^2$ such that the physical axion moduli space arises only  upon identification under the lattice and thus corresponds to a two-torus $T^2\cong \R^2/\Gamma$. For simplicity, we start from a square two-torus, where we subsequently gauge one axionic direction by virtue of the St\"uckelberg couplings $(k^1,k^2)$. Geometrically, this St\"uckelberg gauging identifies all points along an $S^1$ on the two-torus (represented by the dashed, orange line) and the resulting axion moduli space corresponds to the quotient space $T^2/U(1)$ consisting of the $U(1)$-invariant gauge orbits, each represented by a point along the orthogonal direction (indicated through the blue line). Thus, topologically the St\"uckelberg gauging reduces the two-torus to a one-sphere~$S^1$. However, this orthogonal direction generically does not coincide with the axionic direction coupling anomalously to the non-Abelian gauge group in~(\ref{Eq:GeneralLagrangianN2}) and represented in figure~\ref{Fig:AxionModuliSpace} through the green line. This mismatch between the axionic eigenvector in the kinetic terms on the one hand and the axionic direction appearing in the non-perturbative coupling on the other hand is the underlying {\it raison d'\^etre} for the effective decay constant $f_\xi$ in (\ref{Eq:NewADCFullMixing}), with a non-trivial moduli dependence in numerator and denominator. The tunability of the effective axion decay constant is then intimately linked to a gauge slicing of the two-torus, where the representatives of the gauge orbits do not lie along the axionic direction perpendicular to the gauging direction (i.e.~the blue line). Indeed, when the green line aligns with the blue line, the effective decay constant (\ref{Eq:NewADCFullMixing}) reduces to the expression:
\begin{equation}\label{Eq:SpCaseFeffPerp} 
f_\xi \stackrel{(n_1,n_2) = (-k^2, k^1)}{\longrightarrow} \frac{\sqrt{\lambda_+ \lambda_-}}{\sqrt{\lambda_+ (k^+)^2 + \lambda_- (k^-)^2}},
\end{equation}     
which can obviously not enhance parametrically and remains sub-Planckian for sub-Planckian eigenvalues $\lambda_\pm$. Once the angle between the green line and blue line is non-vanishing, the effective axion decay constant $f_\xi$ takes the more generic expression~(\ref{Eq:NewADCFullMixing}), whose full moduli-dependence allows for a considerable widening of the axion window as pointed out in equation (\ref{Eq:CSAxionWindow}).

\subsection{Chiral Symmetry and its Inevitable Breaking}\label{Ss:ChiralSymBreaking}
Quantum consistency for the set-up in (\ref{Eq:GeneralLagrangianN2}) requires the presence of a set of chiral fermions $\psi_L^i$ and $\psi_R^i$ with $i\in \{1,\ldots, n_F\}$ transforming in the bifundamental representation under the non-Abelian $SU(N_c)$ gauge group and the local $U(1)$ gauge group: 
\begin{equation}\label{Eq:ChiralSpectrumFermions}
\begin{array}{c|@{\hspace{0.1in}}c@{\hspace{0.2in}}c}
& SU(N_c) & U(1)\\
\hline
\psi_L^i & R^i_1 & q^i_L\\
\psi_R^i & R^i_2 & q^i_R\\
\end{array}
\end{equation}
With this spectrum of massless chiral fermions, we can specify the fermionic part of the lagrangian by introducing the vier-bein $e_\mu{}^a$, see e.g.~\cite{Birrell:1982ix}:
\begin{equation}
{\cal L}_{\rm fermions}= \det(e_\mu{}^a) \sum_{i=1}^{n_f} \left(   \ov \psi_L^i\, i \sigma^\mu \stackrel{\leftrightarrow}{D}_\mu \psi_L^i + \ov \psi_R^i\, i \ov \sigma^\mu \stackrel{\leftrightarrow}{D}_\mu \psi_R^i \right),
\end{equation}
with the covariant derivatives acting as follows on the chiral fermions:
\begin{equation}\label{Eq:FermKinTerms}
\begin{array}{rcl}
D_\mu \psi_L^i &=& (\nabla_\mu  + i  q_L^i A_\mu +  {\cal A}_\mu^a T^a _{R_1^i} )\psi_L^i = \tilde D_\mu \psi_L^i + i  q_L^i A_\mu \psi_L^i , \\
D_\mu \psi_R^i &=& (\nabla_\mu  + i  q_R^i A_\mu +  {\cal A}_\mu^a T^a_{R_2^i} )\psi_R^i =\tilde D_\mu \psi_R^i  + i  q_R^i A_\mu \psi_R^i .
\end{array}
\end{equation}
The derivative operator $\nabla_\mu$ represents the covariant derivative for curved spacetimes acting on the chiral fermions and contracts with the spacetime dependent $(\sigma^\mu,\ov\sigma^\mu)$ matrices (corresponding to Dirac's gamma-matrices in the Weyl representation).   
The non-invariance of the fermionic measure in the path integral under a chiral $U(1)$ transformation then ensures the cancelation of the mixed $U(1) - SU(N_c)^2$ anomaly. Following the arguments in section 2.2.1 of~\cite{Shiu:2015xda} the axion $\zeta$ eaten by the $U(1)$ gauge boson can be eliminated from the effective lagrangian by going to the unitary gauge. When the massive $U(1)$ gauge boson is integrated out by virtue of its equations of motion, the following effective lagrangian remains below the St\"uckelberg mass scale $M_{\rm St}$:
\begin{eqnarray}
{\cal S}^{\rm N=2, unitary}_{\rm axion} &=& \bigintssss \left[ - \frac{f_{\xi}^2}{2} d\xi  \wedge \star_4  d\xi  + \frac{1}{8\pi^2} \xi \Tr(G\wedge G) 
+\frac{1}{2M_{\rm St}^2} {\cal J}_{\psi} \wedge \star_4  {\cal J}_{\psi} +  {\cal L}_{\rm fermions}^{\rm kin}   \right]. \nonumber\\\label{Eq:LagrFullU1UnitaryWithoutA}
\end{eqnarray}
Hence, we are left with a single axion coupled anomalously to a strongly coupled non-Abelian $SU(N_c)$ gauge theory and a collection of chiral fermions charged under the $SU(N_c)$ gauge group. The $U(1)$ symmetry continues to act as a global (anomalous) chiral $U(1)$ symmetry on this model, while a second remnant of the local $U(1)$ symmetry are the ($M_{\rm St}$-suppressed) four-point fermion couplings ${\cal J}_\psi \wedge \star_4 {\cal J}_\psi$, with the conserved current ${\cal J}_\psi$ given (in local flat coordinates) by :
\begin{equation}\label{Eq:CurrentExpression}
{\cal J}_\psi^\mu = \sum_{i} \left[ (q_L^i) \ov \psi_{L}^i \gamma^\mu   \psi_{L}^i  + (q_R^i) \ov \psi_{R}^i  \gamma^\mu  \psi_{R}^i \right].
\end{equation}
Using standard Fierz-identities we can rewrite the four-fermion interactions ${\cal L}_{4\psi}$ in~(\ref{Eq:LagrFullU1UnitaryWithoutA}) as follows:
\begin{eqnarray}
{\cal L}_{4\psi} &=&  \frac{1}{ M_{\rm St}^2} \sum_{i,j=1}^{n_F} \left[ q_L^j q_R^i \, \ov \psi_R^i \psi_L^j  \, \ov \psi_L^j  \psi_R^i+ q_L^i q_R^j \,  \ov \psi_L^i  \psi_R^j  \, \ov \psi_R^j \psi_L^i    \right]  \label{Eq:4fermCouplings} \\
&&+ 
\frac{1}{2 M_{\rm St}^2} \sum_{i,j=1, i\neq j}^{n_F} \left[ q_L^i q_L^j (\ov \psi_{L}^i \gamma^\mu   \psi_{L}^i) ( \ov \psi_{L}^j \gamma_\mu   \psi_{L}^j)+ q_R^i q_R^j (\ov \psi_{R}^i \gamma^\mu   \psi_{R}^i) ( \ov \psi_{R}^j \gamma_\mu   \psi_{R}^j)   \right] . \notag 
\end{eqnarray}
If we assume from now on that the left-and right-chiral fermions transform both in the fundamental representation, i.e.~$R_1^i = R_2^i$, and that the $U(1)$ charges are generation-independent, i.e.~$q_L^i = q_L$ and $q_R^i = q_R$ for all $i\in \{1, \ldots, n_F\}$, we can translate the four-fermion interactions into more recognizable Nambu-Jona-Lasinio (N-JL) type interactions~\cite{Nambu:1961tp,Nambu:1961fr}:
\begin{eqnarray}
{\cal L}_{4\psi} &=&    \frac{q_L q_R}{ 2 M_{\rm St}^2} \sum_{i,j=1}^{n_F} \left[  \left(\ov \psi^i \psi^j  \, \ov \psi^j  \psi^i - \ov \psi^i \gamma^5\psi^j  \, \ov \psi^j \gamma^5  \psi^i   \right)  \right] \label{Eq:NJL4fermCouplings} \\
&& +
\frac{1}{2 M_{\rm St}^2} \sum_{i,j=1, i\neq j}^{n_F} \left[ q_L^2  (\ov \psi_{L}^i \gamma^\mu   \psi_{L}^i) ( \ov \psi_{L}^j \gamma_\mu   \psi_{L}^j)+ q_R^2  (\ov \psi_{R}^i \gamma^\mu   \psi_{R}^i) ( \ov \psi_{R}^j \gamma_\mu   \psi_{R}^j)    \right] . \notag
\end{eqnarray}
By writing out the four-fermion interactions explicitly, one can easily observe that the classical lagrangian (\ref{Eq:LagrFullU1UnitaryWithoutA}) remains invariant under the global remnant of the chiral $U(1)$ symmetry, which combines a vector and axial $U(1)$ symmetry:
\begin{equation}\label{Eq:ChiralU1Fermions}
\left\{\begin{array}{l}
\psi_L^i \rightarrow e^{i \alpha\, q_L} \psi_L^i,  \\ 
\\
\psi_R^i \rightarrow e^{i \alpha\, q_R} \psi_R^i,
\end{array} \right. \qquad \text{ or } \qquad
\left\{\begin{array}{l}
\psi^i \rightarrow e^{i  q_+ \alpha} e^{i  q_-  \alpha\gamma^5} \psi^i, \\ 
\\
\ov \psi^i \rightarrow \ov \psi^i e^{-i  q_+ \alpha} e^{i  q_-  \alpha\gamma^5}.
\end{array}\right.
\end{equation}
To write down how the $V-A$ type $U(1)$ symmetry acts on Dirac-fermions, we introduced the charges $q_\pm = \frac{q_R \pm q_L}{2}$ and as along as $q_-\neq 0$, the chiral $U(1)$ symmetry will always contain an axial $U(1)$ part. For generation-independent $U(1)$ charges and representations under the $SU(N_c)$ gauge group there is an accidental global chiral $SU(n_F)_L\times SU(n_F)_R$ symmetry:
\begin{equation}
\left( \begin{array}{c} \psi^1_L \\ \vdots \\ \psi^{n_F}_L  \end{array} \right) \rightarrow U_L \left( \begin{array}{c} \psi^1_L \\ \vdots \\ \psi^{n_F}_L  \end{array} \right), \qquad  \left( \begin{array}{c} \psi^1_R \\ \vdots \\ \psi^{n_F}_R  \end{array} \right) \rightarrow U_R \left( \begin{array}{c} \psi^1_R \\ \vdots \\ \psi^{n_F}_R  \end{array} \right),
\end{equation}
acting on the chiral fermions in the fundamental representation with $U_L \in SU(n_F)_L$ and $U_R \in SU(n_F)_R$.

If we want to know the fate of these chiral symmetries in the low-energy limit of the theory, it is paramount to understand the vacuum configuration in the infrared of this set-up. In addition, determining the vacuum also allows to identify the mass-generating effects and the associated mass spectrum in the infrared. In first instance, the vacuum configuration for the non-Abelian gauge theory is studied in Euclidean spacetime and corresponds to a gauge configuration with a vanishing field strength $G_{\mu\nu} = 0$ at (Euclidean) infinity~\cite{Jackiw:1976pf,Callan:1976je,Callan:1977gz}. For a non-Abelian gauge theory, the gauge-invariant vacua with vanishing field strength correspond to pure gauge potentials $A_\mu = g^{-1} \partial_\mu g$, which define maps from a three-sphere at infinity into the non-Abelian gauge group $SU(N_c)$. By classifying these maps according to the homotopy group $\pi_3(SU(N_c)) = \Z$ distinct vacuum configurations can be labeled by the integer-valued winding number or Pontryagin class:        
\begin{equation}
n = \frac{1}{6 \pi^2} \int d^3 x \varepsilon_{ijk}  \Tr(g^{-1} \partial_i g\, g^{-1} \partial_j g\, g^{-1} \partial_k g ),
\end{equation}
expressed here in the temporal gauge $A_0 =0$ for the Euclidean theory. None of the winding vacua $|n\rangle$ is, however, able to play the role of the ground state, as they are not truly gauge-invariant. More precisely, by acting with a large gauge transformation $V_\lambda = e^{i \lambda^a T^a}$, for which $\lambda^a \slashed{\rightarrow} 0$ at Euclidean infinity $|x|\rightarrow \infty$, on a winding vacuum $|n\rangle$ we end up with a different winding vacuum $|m\rangle$ with $m\neq n$. Moreover, in the presence of chiral fermions the winding vacua do not respect the cluster decomposition theorem. To obtain a gauge-invariant ground state in line with the cluster decomposition theorem, we should define a state that forms an eigenstate of the unitary operator $V_\lambda$, which implies that its eigenvalue is of the form $e^{i \theta}$. The $\theta$-vacuum can be constructed from the winding vacua:
\begin{equation}
|\theta\rangle = \sum_{n=-\infty}^{+\infty} e^{in\theta} |n\rangle,
\end{equation} 
and all physical quantities ought to be evaluated in the $\theta$-vacuum. Nonethless, the precise value for parameter $\theta$ cannot be coined by any further physical reasoning. The physical importance of the $\theta$-vacuum cannot be overstated, as it encodes the vacuum energy associated to the instanton solutions, which are responsible for the quantum mechanical tunneling between two distinct winding vacua. When evaluating the theory in the $\theta$-vacuum, the gauge instanton solutions then require the addition of an effective $\theta$-term to the original, classical lagrangian: 
\begin{equation}\label{Eq:thetaterm}    
{\cal L}_{\theta} = \frac{\theta}{8 \pi^2} \Tr(G_{\mu \nu} G_{\rho \sigma} \varepsilon^{\mu\nu\rho\sigma}).
\end{equation}
The topogical nature of this term can be brought to light by writing the topological density $\Tr(G_{\mu \nu} G_{\rho \sigma} \varepsilon^{\mu\nu\rho\sigma})$ in terms of a total derivative of the current $K^\mu \equiv \frac{1}{4\pi^2}   \varepsilon^{\mu\nu\rho\sigma} \Tr(A_\nu \partial_\rho A_\sigma + \frac{2}{3} A_\nu A_\rho A_\sigma)$, such that its importance in the lagrangian is tied to the boundary conditions of the instanton solutions.

In the presence of massless chiral fermions charged under the non-Abelian gauge groups some additional considerations have to be made to identify the proper structure of the vacuum. In first instance, one has to point out that the axial part of global $U(1)$ symmetry is no longer conserved in the $\theta$-vacuum given the instanton solutions and that the violation of the conserved current equation is proportional to the topological density of the non-Abelian gauge theory:
\begin{equation}\label{Eq:U(1)ConsEq}
\langle \partial_\mu J^\mu_{U(1)}  \rangle =  \frac{n_F q_- g^2}{16\pi^2}  \langle\Tr(G_{\mu \nu} \tilde G^{\mu \nu})\rangle 
\end{equation} 
with the $U(1)$ current given (in local coordinates) by:
\begin{equation}
J^\mu_{U(1)} = \sum_{i=1}^{n_F} \left( q_+ \ov\psi^i \gamma^\mu \psi^i + q_-\ov\psi^i \gamma^\mu \gamma^5\psi^i \right). 
\end{equation}
A more physical interpretation~\cite{tHooft:1986ooh,Vainshtein:1981wh} of this equation imposes itself by integrating it over a spacetime volume containing gauge instantons and anti-instantons: each instanton takes a left-handed Weyl fermion and converts it into a right-handed Weyl-fermion, whereas an anti-instanton acts in the reverse manner. As a result the charge associated to the $U(1)$ current changes by two orders $\Delta Q_{U(1)} = 2q_-$, compatible with the non-conservation of the associated current and a breaking of the $U(1)$ symmetry to a discrete $\Z_{|q_-|n_F}$ symmetry. This symmetry breaking can equally be deduced by investigating~\cite{Fujikawa:1979ay} the behavior of the path integral under the $U(1)$ symmetry, in which the $\theta$-parameter of the $SU(N_c)$ vacuum is forced to shift,
\begin{equation}
\theta \longrightarrow \theta - 2 q_- n_F \alpha, \qquad \alpha = \frac{k \pi}{q_-n_F} \text{ with } k\in \Z,
\end{equation} 
such that the path integral remains invariant under the chiral Abelian symmetry. 

Instantons are not the only non-perturbative effects that can trigger a dynamical breaking of the global $U(1)$ symmetry. In case the interactions of the theory allow for a vacuum configuration in which bound states of fermionic bilinears acquire a non-vanishing vacuum expectation value $\langle \ov \psi^i_L \psi^j_R \rangle \neq 0$, the corresponding fermionic condensate breaks the remaining discrete $\Z_{|q_-|n_F}$ symmetry further down to a $\Z_2$ symmetry:
\begin{equation}
0 \neq \langle \ov \psi_L^i \psi_R^j \rangle \longrightarrow e^{2i q_- \alpha} \langle \ov \psi^i_L \psi^j_R \rangle, \qquad \alpha = 0 \, \text{ or } \, \alpha =n_F \frac{ \pi}{q_- n_F}.
\end{equation}
The model at hand exhibits two distinct dynamical mechanisms to induce the formation of (gauge-invariant) bound states consisting of fermion-antifermion pairs, whose non-vanishing vacuum expectation values will also trigger a spontaneous breaking of the accidental non-Abelian $SU(n_f)_L \times SU(n_F)_R$ chiral symmetries besides the chiral $U(1)$ symmetry:       
\begin{itemize}
\item[(1)] \underline{Nambu-Jona-Lasinio mechanism\label{P:NJLMech}}~\cite{Nambu:1961tp,Nambu:1961fr}: self-interactions between the fermions due to the four-fermion interactions (\ref{Eq:NJL4fermCouplings}) can induce the formation of fermion bound states $\langle \ov \psi^i \psi^i \rangle\neq 0$, which in turn yields  dynamical masses for the fermions given by:
\begin{equation}\label{Eq:FermMassNJL}
m_i = - \frac{q_L q_R}{M_{\rm St}^2} \langle \ov \psi^i \psi^i \rangle.
\end{equation}
This dynamical process is self-consistent provided there exists a non-trivial solution to the mass-gap equation which relates the dynamical masses of the fermions to the one-loop self-energy of the fermions. Such a non-trivial solution exists if the coupling of the four-fermion interactions exceeds a critical value, which for this set-up can be brought back to a lower bound on the $U(1)$ gauge coupling at the St\"uckelberg scale:
\begin{equation}\label{Eq:NJLU1GaugeCond}
g_1^2\big|_{M_{\rm St}}   > \frac{4 \pi^2}{n_F N_c |q_L q_R|}.
\end{equation}
In this vacuum configuration we expect $n_F^2-1$ massless Goldstone bosons associated to the broken directions of the global $SU(n_F)_L \times SU(n_F)_R$ symmetry. The Goldstone boson corresponding to the broken chiral $U(1)$ symmetry will not be (entirely) massless given the explicit breaking of the $U(1)$-current (\ref{Eq:U(1)ConsEq}) in the $\theta$-vacuum.   
\item[(2)] \underline{$SU(N_c)$ Confinement}: in case the $SU(N_c)$ gauge theory finds itself in the strongly coupled regime, an attractive, long-range strong force is expected to act between a fermion and an antifermion, such that the states in the spectrum consist of interacting bound states of fermion-antifermion pairs. Unfortunately, the process of fermion confinement and chiral symmetry breaking in strongly coupled gauge theories with massless fermions is not yet fully understood, and the mass-gap equation cannot be solved as for the N-JL models. Yet, under mild assumptions regarding the confining force one can argue~\cite{Casher:1979vw} that chiral symmetry has to be spontaneously broken and at the same time deduce some restrictions on the scales and field content in order to guarantee the confinement phase sets in. The first condition follows from requiring that the non-Abelian gauge theory is strongly coupled in the IR and weakly coupled in the UV, which boils down to requiring that the one-loop beta-function is negative, or said otherwise in terms of the field content:
\begin{equation}\label{Eq:OneLoopCondition}
11 N_c > 2 n_F.
\end{equation}
Secondly, we want to avoid an IR free fixed point in this gauge theory with massless fermions, which can occur if the two-loop contribution to the beta-function cancels the one-loop contribution. Avoiding such a Banks-Zaks fixed point~\cite{Banks:1981nn} imposes a second constraint on the field content (consisting of massless fermions in the fundamental or anti-fundamental representation):
\begin{equation}\label{Eq:Banks-ZaksCondition}
n_F < \frac{34 N_c^3}{13N_c^2 - 3}.
\end{equation} 
Note that when the latter condition (\ref{Eq:Banks-ZaksCondition}) is satisfied for the field content, the constraint (\ref{Eq:OneLoopCondition}) derived from the one-loop beta function coefficient will be automatically satisfied as well. Under these assumptions the non-Abelian gauge theory will become strongly coupled at the energy scale $\Lambda_s$:
\begin{equation}
\Lambda_s = M_{\rm string} e^{- \frac{8 \pi^2}{\beta_0 g_2^2}},
\end{equation}
with the gauge coupling $g_2^2$ evaluated at the string mass scale $M_{\rm string}$ and $\beta_0 = \frac{11}{3} N_c - \frac{2}{3} n_F$ the one-loop beta function coefficient. In the strongly coupled regime, fermion confinement then takes place by which the fermions combine into gauge-invariant bound states of fermion-antifermion. As pointed out above, in the confinement phase chiral symmetry will be broken spontaneously due to a non-trivial vacuum configuration for the these bound states $\langle \ov\psi_L \psi_R\rangle \neq 0$. An intrinsic energy scale can be associated with the fermion condensate, which is estimated to be of the order of the strong coupling scale, i.e.~$\langle \ov \psi^i_L \psi^j_R \rangle \sim - \Lambda_s^3 \delta^{ij}$. In the process of chiral symmetry-breaking dynamical masses $M_{\rm dyn}$ for the chiral fermions are expected to arise from the strongly coupled non-Abelian interactions of the order~${\cal O}(\Lambda_s)$. 
\end{itemize}
It is obvious from the previous considerations that the theory in (\ref{Eq:LagrFullU1UnitaryWithoutA}) can exhibit an intricate vacuum structure depending on the particular phase of the gauge theories in the infrared, such that the global chiral symmetries are broken and dynamical fermion masses are generated. In the remainder of the paper, we will investigate the implications of the $SU(N_c)$ confinement phase on the mass spectrum of bound fermion-antifermion states and its appropriateness to discuss models of cosmological inflation. The repercussions of the N-JL type symmetry breaking, with applications to cosmological inflation, will be discussed in section~\ref{S:PhaseInf}. 

\begin{figure}[h]
\begin{center}
\includegraphics[scale=0.6]{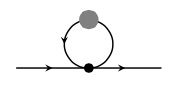} \begin{picture}(0,0) \put(-70,0){$ \frac{q_L q_R}{M_{st}^2}$} \put(-20,10){$\psi^i$}  \put(-120,10){$\ov\psi^i$} \end{picture}
\caption{One-loop fermion self-energy induced by the perturbative four-fermion interactions of equation~(\ref{Eq:NJL4fermCouplings}), with the grey blob representing the dynamical fermion mass $M_{\rm dyn}$. \label{Fig:4FermSelfInter}}
\end{center}
\end{figure}
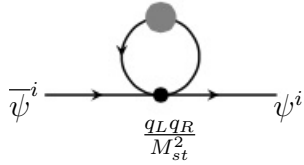

One could still ask which role the perturbative four-fermion interactions play in the $SU(N_c)$ confinement phase. To address this question we consider the two-point function of one of the fermions $\psi^i$:
\begin{equation}\label{Eq:FermCondensate}
\langle \ov \psi^i(x)  \psi^i(y)\rangle = -  \langle\Tr S_F(0)\rangle,
\end{equation}    
which corresponds to evaluating the trace of the fermion propagator $S_F(x-y)$ in the vacuum taken at the point $x-y = 0$ to assure Lorentz-invariance. It is impossible to determine the fermion propagator at strong non-Abelian coupling, but we can infer its functional structure based on symmetry arguments such as Poincar\'e and parity invariance in momentum space:
\begin{equation}
S_F^{-1}(p) = A(p^2)\left( i\slashed{p} + M(p^2) \right),
\end{equation}   
where we introduced the dynamical function $A(p^2)$ and mass function $M(p^2)$ to encode all the dynamics involving the fermions. The fermion self-interactions $\Sigma_{4\psi}$ induced by the perturbative four-fermion interactions~(\ref{Eq:NJL4fermCouplings}) contribute undoubtedly to the mass function and the one-loop contribution can be estimated to take the form:
\begin{equation}\label{Eq:OneLoop4Fermion}
\Sigma_{4\psi} (p) = \frac{2 q_L q_R}{M_{st}^2}\int \frac{d^4 p}{(2\pi)^4} \Tr\left(\frac{i}{\slashed{p} - M_{\rm dyn}}\right),
\end{equation}
where $M_{\rm dyn}$ represents the fermion mass generated dynamically when the chiral symmetry is spontaneously broken in the $SU(N_c)$ confinement phase. A graphical depiction of the one-loop diagram responsible for this contribution is given in figure~\ref{Fig:4FermSelfInter}.
The fermion self-interactions will subsequently seed into the fermion condensate (\ref{Eq:FermCondensate}), which acts as an order parameter for chiral symmetry-breaking. As such, we observe that the perturbative four-fermion interactions are expected to contribute additionally to the fermion masses in the  $SU(N_c)$ confinement phase:
\begin{equation}\label{Eq:FermMassConfinement}
m_{ij} = - \frac{q_L q_R}{M_{\rm St}^2} \langle \ov \psi^i_L \psi^j_R \rangle \delta_{ij},
\end{equation} 
through diagrams such as in figure~\ref{Fig:4FermSelfInter}. The effective lagrangian describing the fermionic fluctuations over the symmetry-breaking vacuum can be schematically written as:
\begin{eqnarray}\label{Eq:EFTFermionConfinement}
{\cal L}_{\rm fermions} = {\cal L}_{\rm fermions}^{\rm kin}  + {\cal L}_{4\psi} + {\cal L}_{\cancel{U(1)}},
\end{eqnarray}
where ${\cal L}_{\rm fermions}^{\rm kin}$ corresponds to the kinetic terms in (\ref{Eq:FermKinTerms}) for the fermions (excluding the coupling to the $U(1)$ gauge theory), ${\cal L}_{4\psi}$ gives the four-fermion interactions as in (\ref{Eq:NJL4fermCouplings}), and ${\cal L}_{\cancel{U(1)}}$ captures the $U(1)$-breaking fermion interactions induced by the intricate vacuum structure. The Lagrangian component ${\cal L}_{\cancel{U(1)}}$ contains two separate parts and reads explicitly,
\begin{eqnarray}
{\cal L}_{\cancel{U(1)}} &=& - \sum_{i,j=1}^{n_F}  \left(  m_{ji} \ov \psi_R^i \psi_L^j  + m_{ij} \ov \psi_R^j  \psi_L^i  + m_{ij}^\dagger \ov  \psi_L^j \psi_R^i +  m_{ji}^\dagger  \ov \psi_L^i  \psi_R^j \right) \notag\\
&& \qquad+ \frac{e^{-\frac{8\pi^2}{g_2^2}}}{g_2^{4N_c}} \left[ {\cal C}_{n_f} e^{i\theta} \det\left( \ov \psi^i_R  \psi^j_L \right) + h.c. \right]. \label{Eq:DynMassTermFermions}
\end{eqnarray} 
The dominant contribution to the fermion masses in the first line are dynamically generated in the confinement phase of the $SU(N_c)$ gauge theory, but they also acquire contributions from the fermion self-energy due to the four-fermion interactions according to~(\ref{Eq:FermMassConfinement}). The appearance of fermion masses indicates a breaking of the chiral $U(1)$ and $SU(n_F)_L\times SU(n_F)_R$ symmetry. Whether or not the Goldstone bosons associated to the broken symmetry generators are entirely massless depends on the occurrence of fermion masses whose microscopic origin can be decoupled from the $SU(N_c)$ confinement process. One could imagine adding perturbative fermion masses resulting from Yukawa interactions as in QCD or consider more involved strongly coupled gauge dynamics leading up to fermions masses similarly to the mechanisms used to generate quark and lepton masses in the extended techni-colour scenario~\cite{Farhi:1980xs}, which would break the chiral symmetries explicitly and lift the masses of the Nambu-Goldstone bosons. The second line captures the interactions between the chiral fermions and the gauge instantons by virtue of the so-called 't Hooft determinant~\cite{tHooft:1976rip,tHooft:1976snw}. This effective interaction follows from computing the vacuum-to-vacuum transition amplitude for the non-Abelian gauge theory with chiral quarks in the $\theta$-vacuum. More explicitly, it is the integration over the fermionic zero modes and gauge boson modes that yields this particular non-vanishing contribution to the effective lagrangian, with the determinant taken over the number of generations $n_F$. The exponential pre-factor $e^{-\frac{8\pi^2}{g_2^2}}$ -- typical for gauge instantons -- signals the non-perturbative nature of the interaction, with $g_2^2$ corresponding to  the one-loop renormalized gauge coupling of the non-Abelian gauge group evaluated at the strong coupling scale $\Lambda_s$. The pre-factor $C_{n_f}$ is a dimensionful parameter whose mass dimension is set by the number of fermion generations. Its precise expression follows from integration over the collective coordinates of the instantons, as will be reviewed in more detail in section~\ref{Sss:InterInfladrons}. The parameter $\theta$ represents the $\theta$-vacuum contribution of the non-Abelian gauge group. An underlying assumption leading up to the second line of (\ref{Eq:DynMassTermFermions}) is the dilute instanton gas approximation: the size of the instantons is sufficiently small, such that they can be treated as non-overlapping, non-interacting particles similar to a gas of identical particles. As the determinant is taken over the number of fermion generations, the 't Hooft operator only break the chiral $U(1)$ symmetry, and not its non-Abelian counterpart.

\section{EFT with Infladrons and Axions}\label{S:EFTInfAx}
\subsection{Strong Dynamics and Effective Field Theory}\label{Ss:StDynEFT}
\subsubsection{A one-generational Toy Model}\label{Sss:FermInfladrons}
In the previous section, we have argued that the $U(1)$ chiral symmetry is broken in the confinement phase of the non-Abelian gauge theory due to the instanton background and that accidental non-Abelian chiral symmetries are broken by the formation of (gauge-invariant) bound states consisting of a fermion and an antifermion. In order to be more explicit, we consider a one-generational model and work out the details for the effective description, starting from the following UV-action:
\begin{eqnarray}
{\cal S}_{\rm EFT}^{n_f =1} &=& \bigintssss \left[ - \frac{f_{\xi}^2}{2} d\xi  \wedge \star_4  d\xi  -\frac{1}{g_{2}^{2}} \Tr(G\wedge \star_4 G)   + \frac{1}{8\pi^2} \xi\, \Tr(G\wedge G)   \right. \notag \\
&& \qquad  + \det(e_\mu{}^a) \left( \ov \psi_L\, i \sigma^\mu \stackrel{\leftrightarrow}{\tilde D}_\mu \psi_L + \ov \psi_R\, i \ov \sigma^\mu \stackrel{\leftrightarrow}{\tilde D}_\mu \psi_R \right) \notag  \\
&& \qquad \left. +  \frac{q_L q_R}{ M_{\rm St}^2}  \left( \ov \psi_L \psi_R  \, \ov \psi_R  \psi_L + \ov \psi_R \psi_L  \, \ov \psi_L  \psi_R  \right) 
\right]. \label{Eq:EFTNf1}
\end{eqnarray}
The covariant derivative  $\tilde D_\mu$ acting on the fermions was introduced in equation (\ref{Eq:FermKinTerms}). Note that the second line of four-fermion interactions in (\ref{Eq:4fermCouplings}) is absent for $n_F=1$, such that the four-fermion interactions reduce to pure N-JL interactions: 
\begin{equation}\label{Eq:NJLInter1Gen}
{\cal L}_{4\psi} = \frac{q_L q_R}{ M_{\rm St}^2}  \left( \ov \psi_L \psi_R  \, \ov \psi_R  \psi_L + \ov \psi_R \psi_L  \, \ov \psi_L  \psi_R  \right) =  \frac{q_L q_R}{2 M_{\rm St}^2}  \left( \ov \psi \psi  \, \ov \psi  \psi + \ov \psi i\gamma^5\psi  \, \ov \psi i\gamma^5  \psi \right)   .  
\end{equation}
While the classical lagrangian (\ref{Eq:EFTNf1}) is manifestly invariant under the chiral symmetry acting on the fermions as in (\ref{Eq:ChiralU1Fermions}), the anomalous nature of the $U(1)$ symmetry implies that the associated $U(1)$ current is explicitly broken by the non-trivial vacuum structure of the non-Abelian gauge theory as pointed around equation (\ref{Eq:U(1)ConsEq}). The physical significance of this non-conservation manifests itself when computing the vacuum-to-vacuum amplitude in the presence of background gauge instantons. The dominant contribution to this transition amplitude comes~\cite{tHooft:1976snw,Bernard:1979qt,Vainshtein:1981wh,Shifman:2012zz} from the one-instanton transition amplitude $Z_1$ from winding vacuum $|0\rangle$ to winding vacuum $|1\rangle$:
\begin{eqnarray}\label{Eq:InstPreFactor} 
Z_1 &=&   \int d^4x \int_{\rho_0}^{\infty} d\rho\, \rho^{3n_f - 5}  \frac{C_1}{(N_c-1)! (N_c-2)!} \left(\frac{8\pi^2}{g_2^2(\rho)}\right)^{2N_c}  e^{- \frac{8\pi^2}{g_2^2(\rho_0) } + \ln(\frac{\rho}{\rho_0}) (\frac{11}{3} N_c - \frac{2}{3} n_F)  + C_2  } \notag \\
&& \qquad \qquad \qquad \qquad \qquad \times \left[ \det(\ov\psi_L \psi_R) + h.c. \right],
\end{eqnarray}
where the integration is taken over the collective zero-modes, being the instanton size~$\rho$ and the position $x^\mu$. To regularize the integral, we introduced the UV-regulator  $\rho_0  $, at which scale the $SU(N_c)$ gauge coupling is evaluated. The constants $C_1$ and $C_2$ arise due to the regularization procedure used in computing the one-instanton pre-factor and therefore depend on the subtraction scheme. The exponential factor results from evaluating the non-Abelian Yang-Mills action for the one-instanton solution and the classical action obtains a one-loop correction from the massless fields upon dimensional renormalization. The two-fermion vertex $\det(\ov\psi_R \psi_L)$ arises by integrating over the fermionic measure from the path integral in the instanton background. Given that the fermions are massless at the classical level, one might expect that their functional determinant vanishes due the existence of a fermion zero mode in the one-instanton background. However, the four-fermion interactions in (\ref{Eq:EFTNf1}) should be interpreted as a source for a small perturbation about the instanton background, which shifts the eigenvalues of the Dirac-operator for the fermions enough to procure the 't Hooft operator. As prescribed, one first integrates over the fermionic Hilbert space and then takes care of the integrating over the gauge instanton background. And in principle, integration over the full instanton moduli space is required to extract the coupling strength of the instanton-induced fermion-couplings. In practice, however, one can infer that the integral over the instanton-size~$\rho$ is dominated by the instantons with sizes set by the strong coupling scale $\rho \sim \Lambda_s^{-1}$, such that the one-instanton pre-factor $Z_1$ scales as:
\begin{eqnarray}\label{Eq:InstPreScaling}   
Z_1 \varpropto \int d^4x  \left(\frac{8\pi^2}{g_2^2(\Lambda_s)}\right)^{2N_c} \Lambda_s^{4-3n_f} e^{-\frac{8\pi^2}{g_2^2(\Lambda_s)}}  \left[ \det(\ov\psi_L \psi_R) + h.c. \right]
\end{eqnarray}
To obtain the full contribution of the gauge instanton background to the path integral, we follow the dilute instanton gas approximation~\cite{tHooft:1986ooh} and consider the instantons as non-interacting, identical particles with characteristic size smaller than the average distance between two neighboring (spacelike separated) instantons. This allows us to sum all instantons and anti-instantons located at different positions $x=x_i$, while relying on the declustering assumption for the instantons $(Z_1)$ and anti-instantons $(\ov Z_1)$:
\begin{equation}\label{Eq:OneInstOneGeneration}
\sum_{\nu_+=-\infty}^{+\infty} \frac{1}{\nu_+!} Z_1^{\nu_+}  \sum_{\nu_-=-\infty}^{+\infty} \frac{1}{\nu_-!}  
\ov Z_1^{\nu_-} = \exp \left( \frac{e^{-\frac{8\pi^2}{g_2^2}}}{g_2^{4N_c}} \left[ {\cal C}_{n_f} e^{i\theta} \det\left( \ov \psi_L  \psi_R \right) + h.c. \right] \right),
\end{equation} 
such that the fermionic interactions captured by 't Hooft determinant in the effective lagrangian~(\ref{Eq:DynMassTermFermions}) are reproduced for a one-generational model, upon inclusion of the $\theta$-parameter. For a one-generational model the 't Hooft operator reduces to a mass term for the  chiral fermions, which explicitly breaks the chiral $U(1)$ symmetry, in line with equation~(\ref{Eq:U(1)ConsEq}).  

In deriving the one-instanton vacuum-to-vacuum transition amplitude~(\ref{Eq:InstPreFactor}) we have pointed out the role of the four-fermion interactions as a source term preventing a vanishing functional determinant of the fermionic degrees of freedom.  In a next phase, one has to evaluate the four-fermion interactions in the infrared vacuum of our theory, defined by the non-perturbative properties of the non-Abelian gauge theory. To this end, we transform the four-fermion interactions to a Yukawa interaction through the bosonization method and introduce auxiliary fields $(\sigma, \pi)$:
\begin{equation}\label{Eq:BosonMethod}
{\cal L}_{4\psi} = -  2  \frac{q_L q_R}{M_{st}^2} \Lambda^2 (\sigma \ov\psi \psi + \pi \ov \psi i \gamma^5 \psi)  -  \frac{q_L q_R}{M_{st}^2} \Lambda^4 (\sigma^2 + \pi^2),
\end{equation}
such that the equations of motion relate these Lagrange multipliers to the scalar and pseudo-scalar fermionic bilinears:  
\begin{equation}
\sigma = - \frac{1}{\Lambda^2} \ov\psi \psi, \qquad \qquad \pi = - \frac{1}{\Lambda^2} \ov\psi i\gamma^5 \psi, 
\end{equation}
with the scale $\Lambda$ an arbitrary energy scale introduced for dimensional reasons. For a fully consistent model we also require that the scalar field $\sigma$ and $\eta$ transform under the chiral $U(1)$ symmetry, as an irreducible (real) representation under the $SO(2)$ group:
\begin{equation}
\left( \begin{array}{c} \sigma \\ \pi \end{array} \right) \rightarrow \left( \begin{array}{cc}  \cos 2\alpha q_- & \sin2\alpha q_- \\ -\sin 2\alpha q_- &  \cos2\alpha q_-    \end{array} \right)  \left( \begin{array}{c} \sigma \\ \pi \end{array} \right),
\end{equation}
in addition to the transformation~(\ref{Eq:ChiralU1Fermions}) of the chiral fermions. Furthermore, one can show that the auxiliary field $\sigma$ acquires a non-vanishing vacuum expectation value in the presence of gauge instantons. This property can be traced back to the fermionic zero-modes $\psi_0$~\cite{tHooft:1976rip,tHooft:1976snw}, which have a vanishing eigenvalue of the Dirac-operator in the one-instanton background:
\begin{equation}
\psi_0 (x,\rho)  = \left(\frac{2}{\pi^2}\right)^{1/2} \frac{\rho}{(x^2 + \rho^2)^{3/2}} u,
\end{equation}
where the left-handed spinor $u$ is normalized in the sense $u^\dagger u = 1$.
When computing the fermion two-point function $\langle \theta | \ov\psi  \psi |\theta \rangle \equiv \langle \ov\psi  \psi \rangle_\theta$ in the $\theta$-vacuum, one has to evaluate the fermion propagator in the instanton background~\cite{Callan:1977gz,Shifman:1979uw,Vafa:1983tf} to arrive at the expression: 
\begin{equation}
\langle \ov\psi  \psi \rangle_\theta  = - \int d\rho \int d^4x \frac{D(\rho)}{\rho^5} \sum_n \frac{1}{i\lambda_n   -  2  \frac{q_L q_R}{M_{st}^2} \Lambda^2  \sigma (\rho) } \psi_n^\dagger (x) \psi_n (x),
\end{equation}
with $D(\rho) \sim \left(\frac{8\pi^2}{g_2^2(\rho)}\right)^{2N_c} e^{-\frac{8\pi^2}{g_2^2(\rho)}} $ the instanton density and $\lambda_n$ the eigenvalues of the Dirac-operator in the instanton background corresponding to the eigenmodes $\psi_n$. Solving this integral equation to obtain an explicit solution for $\langle \ov\psi  \psi \rangle_\theta$ turns out to be quite complicated, but one can infer that the sum over the eigenvalues is dominated by the zero-mode solution for a massless fermion. Integrating the renormalizable zero-modes over the position coordinates leaves only an integral over the size $\rho$ of the instantons. If we furthermore consider a profile $\sigma(\rho) \sim \rho^{-1}$ and integrate over the instantons with size $\Lambda_s^{-1} \leq\rho \leq M_{\rm string}^{-1}$ (in line with the dilute gas approximation), we obtain a non-vanishing vacuum expectation value for the scalar bilinear, namely  $\langle \ov\psi  \psi \rangle_\theta \sim - \Lambda_s^3$. Hence, we find that evaluating the four-fermion interactions in the $\theta$-vacuum leads to an effective mass term for the fermions of the form:
\begin{equation}\label{Eq:1GenEffMass4Ferm}
{\cal L}_m =  - m_q  \ov \psi \psi , \qquad m_q  \equiv -2 \frac{q_L q_R}{2 M_{\rm St}^2}\langle \ov \psi \psi \rangle_\theta .
\end{equation}
This also implies that the $U(1)$ current conservation equation evaluated in the $\theta$-vacuum should be supplemented by this effective mass term, such that equation~(\ref{Eq:U(1)ConsEq}) is modified to:
\begin{equation}\label{Eq:U(1)ConsEq1Generation}
 \partial_\mu J^\mu_{U(1)}   = 2 q_-  m_q \ov\psi i\gamma^5 \psi   +  \frac{n_f q_- g^2}{16\pi^2} \Tr(G_{\mu \nu} \tilde G^{\mu \nu}) .
\end{equation} 
The effective mass term can be modelled as a one-loop generated mass term through a diagram analogous to the one in figure~\ref{Fig:4FermSelfInter}, for which the grey blob corresponds instead to the interaction with the gauge instanton background. It is important to stress that the gauge instantons are not able to confine the fermions by themselves, such that the two-point function $\langle \ov\psi  \psi \rangle_\theta$ should not be confused with the fermion condensate formed in the confining phase. As such, we can treat the mass $m_q$ as an explicit $U(1)$-breaking fermion mass, whose microscopic origin is tied to the perturbative four-fermion interactions and the gauge instanton background.

Until now our attention went entirely to the low energy effective interactions for the chiral fermions, but the action~(\ref{Eq:EFTNf1}) also contains a closed string axion. If it were not for its anomalous coupling to the topological density of the non-Abelian gauge theory, the action would also be invariant under the shift symmetry $\xi \rightarrow \xi + \beta$ with $\beta$ a constant. As such, we can associate a separate current $J^\mu_\xi$ to this shift symmetry, whose conservation is violated in the presence of gauge instantons: 
\begin{equation}
\partial_\mu J^\mu_\xi = \frac{1}{8\pi^2} \Tr(G_{\mu \nu} \tilde G^{\mu \nu}). 
\end{equation}
Also the closed string axion is then expected to acquire a mass term from the non-perturbative gauge instantons, when the action~(\ref{Eq:EFTNf1}) is evaluated in the $\theta$-vacuum. 

\subsubsection{From Fermions to Interacting Infladrons}\label{Sss:InterInfladrons}
Around and below the strong coupling scale $\Lambda_s$ we expect the effective description of (\ref{Eq:EFTFermionConfinement}) in terms of fermionic excitations propagating over the non-trivial $\theta$-vacuum to break down. Instead, a proper effective description should consist of interacting scalar bound states, similar to the chiral perturbation theory of pions for the strong interactions below the QCD-scale, see e.g. \cite{Ecker:1994gg,Scherer:2002tk}. To describe the low energy behavior of the bound states $\ov \psi_L \psi_R$, we introduce a complex scalar field $\Phi$:
\begin{equation}\label{Eq:PhiParaSigmaEta}
\ov \psi_L \psi_R \qquad \longrightarrow \qquad \Phi(x) = \sigma(x) e^{i \frac{\eta(x)}{f}}
\end{equation}
which corresponds to a fermionic bound state whose fluctuations consist of a CP-even scalar $\sigma$ and a CP-odd scalar $\eta$ with decay constant~$f$. The field  also corresponds to a non-linear representation of the chiral $U(1)$ symmetry:
\begin{equation}
\Phi \quad  \rightarrow \quad e^{2i q_- \alpha} \Phi.
\end{equation}
As argued in section~\ref{Ss:ChiralSymBreaking}, the condensation of the chiral fermions into fermion-antifermion bound states breaks the $U(1)$ chiral symmetry (spontaneously)~\cite{Casher:1979vw}. This spontaneous symmetry breaking translates to a non-vanishing vacuum expectation value $\langle\Phi\rangle \neq  0$, setting the value of the decay constant, namely $\langle\Phi\rangle = \langle \sigma\rangle = f \sim \Lambda_s$. Unfortunately, it is not known how to derive the action for the scalar field upon integrating out the fermions directly from the UV theory (\ref{Eq:EFTNf1}), not even by a clever use of Lagrange-multipliers. Instead, we will use the toolbox of effective field theory and write down the effective action for $\Phi$ compatible with the symmetries of the UV-theory lagrangian and add the $U(1)$ symmetry-breaking effects in line with (\ref{Eq:DynMassTermFermions}). Based on these considerations we can write down the most generic (renormalisable) action for $\Phi$ and axion~$\xi$:
\begin{eqnarray}\label{Eq:EFTNL1Gen}
{\cal L}_{\rm EFT}^{\rm cl} &=& \frac{1}{2} (\partial_\mu \Phi)^\dagger \partial^\mu \Phi + \frac{f_\xi^2}{2} \partial_\mu \xi \partial^\mu \xi  + \mu^2 \Phi^\dagger \Phi - \frac{\lambda}{2} ( \Phi^\dagger \Phi)^2 - \Lambda_{cc}  \\
&& + \Lambda^2_s \kappa\, e^{i (\theta +  \xi )}   \det (\Phi) + \Lambda^2_s  \kappa\, e^{-i (\theta +  \xi)} \det (\Phi^\dagger) + \Lambda^2_s M \Phi + \Lambda^2_s   M\Phi^\dagger . \notag
\end{eqnarray}
The upper line contains (canonical) kinetic terms for $\Phi$ and axion $\xi$, a scalar potential responsible for breaking the $U(1)$ symmetry spontaneously and a four-dimensional cosmological constant $\Lambda_{cc} \geq 0$ accounting for the total energy of the vacuum.\footnote{At this stage, the four-dimensional cosmological constant $\Lambda_{cc}$ is added by hand to ensure the field theory setting is situated in a Minkowski or de Sitter spacetime, suited for inflationary purposes. In a consistent and global string theory vacuum solution, this cosmological constant ought to be computed as the vacuum expectation value of the full scalar potential for the geometric moduli, after stabilisation of the geometric moduli. } The real parameters $\mu^2 > 0$ and $\lambda > 0$ cannot be computed directly from the fermionic UV-action~(\ref{Eq:EFTNf1}), yet we assume $\lambda < 1$ such that a perturbative approach remains applicable. The lower line captures the terms in $\Phi$ that break the $U(1)$ symmetry explicitly and which arise as effective interactions upon evaluating action (\ref{Eq:EFTNf1}) in the $\theta$-vacuum, following the considerations in the previous section. We can employ the $U(1)_{\rm chiral} \times U(1)_\xi$ symmetries of the UV-description (\ref{Eq:EFTNf1}) to constrain the form of these terms in the IR, while treating the $\theta$-parameter and scalar condensate {\it vevs} as spurion fields. More precisely, the transformation properties of the $\theta$-parameter and the scalar field $\Phi$ under the chiral $U(1)$ symmetry indicate that they can only appear in the combination $e^{i\theta} \det(\Phi)$ and its complex conjugate. Under the axion shift symmetry $\xi \rightarrow \xi +\beta$ the $\theta$-parameter transforms as $\theta \rightarrow \theta - \beta$, such that the parameter $\theta$ and the axion $\xi$ can only appear in the combination $\theta + \xi$ in the IR effective action. These reflections support the first two terms in the second line of equation (\ref{Eq:EFTNL1Gen}) and the $\theta$-dependence signals that these terms result from the interactions with the gauge instanton background. To justify the last two terms we return to the effective mass~(\ref{Eq:1GenEffMass4Ferm}) and remind the reader that it arises from the Yukawa-coupling to the auxiliary field $\sigma$ which acquires a non-vanishing expectation value in the $\theta$-vacuum due to the gauge instanton background. To arrive at the last two terms of (\ref{Eq:EFTNL1Gen}) we have to combine the auxiliary fields $(\sigma, \pi)$ into a complex spurion field $M$ which transforms non-linearly under $U(1)_{\rm chiral}$:
\begin{equation}
M \rightarrow e^{-2i q_- \alpha } M. 
\end{equation} 
Hence, invariance under $U(1)$ symmetry allows to add a term of the form $M \Phi$ with its complex conjugate in the IR theory, reflecting the effective mass term found in~(\ref{Eq:1GenEffMass4Ferm}). The precise values for the parameters $\kappa$ and $M$ are difficult to determine from first principles, yet by assuming equivalence between the hamiltonian of the UV-theory~(\ref{Eq:DynMassTermFermions}) and the one constructed for the IR-theory~(\ref{Eq:EFTNL1Gen}), we can deduce the parametric dependence for $\kappa$ on the UV-parameters $\Lambda_s$ and $g_2$:
\begin{equation}\label{Eq:MassRelationGaugeCond}
\kappa  \sim A \,  \left(\frac{8\pi^2}{g^2_2(\Lambda_s)}\right)^{2N_c} e^{- \frac{8\pi^2}{g_2^2(\Lambda_s) } }  \Lambda_s^{4-3n_F},
\end{equation} 
with $n_F=1$ and where the factor $ A = \frac{ C_1}{(N_c-1)! (N_c-2)!} e^{C_2}$ collects all $\Lambda_s$- and $g_2$-independent contributions from (\ref{Eq:InstPreFactor}) into one single constant. The gauge coupling $g_2^2(\Lambda_s)$ is evaluated at the strong coupling scale $\Lambda_s$. The $e^{- 8\pi^2/g_2^2}$ dependence expresses the non-perturbative origin of this interaction due to gauge instantons. We can also relate the parameter $M$ to the effective mass in (\ref{Eq:1GenEffMass4Ferm}):  
\begin{equation}\label{Eq:MassRelationMFermCond}
M  = - \frac{q_L q_R}{M_{\rm St}^2} \langle \ov \psi \psi \rangle_\theta.
\end{equation}
If we consider a string theory compactification characterized by a string scale $M_{\rm string}$, St\"uckelberg scale $M_{\rm St}$ and strong coupling scale $\Lambda_s$:
\begin{equation} 
M_{\rm string} \sim {\cal O} (10^{17} -10^{18} \text{GeV}), \quad M_{\rm St} \sim {\cal O} (10^{16} -10^{17} \text{GeV}), \quad  \Lambda_s \sim {\cal O} (10^{15} -10^{16} \text{GeV}),  
\end{equation}
the parameters $M$ and $\kappa$ in (\ref{Eq:EFTNL1Gen}) would fit within the parameter window:
\begin{equation}\label{Eq:ParameterkappaM}
M \sim {\cal O} (10^{13} -10^{16} \text{GeV}), \qquad \kappa \sim  {\cal O} (10^{15} -10^{16} \text{GeV}).
\end{equation} 
It is important to point out that the strong coupling scale $\Lambda_s$ can be located near the GUT-scale $10^{16}$~GeV, provided that the $SU(N_c)$ gauge coupling $g_2^2$ evaluated at the string mass scale is not too small, e.g.~for a $SU(3)$ gauge group with one generation of chiral fermions ($n_F=1$) we need to require $g_2^2  \gtrsim 0.62$ at the string mass scale $M_{\rm string} \sim 10^{18}$~GeV.

Next, we consider the vacuum configuration for the effective action in equation~(\ref{Eq:EFTNL1Gen}). To this end, we expand the fields about their supposed minima with only $\sigma$ acquiring a non-zero {\it vev}, such that $ \sigma (x)  = f + s(x)$. We rescale the closed string axion $\xi$ by its axion decay constant, so that we find the following effective lagrangian:
\begin{eqnarray}
{\cal L}_{\rm EFT}^{\rm non-lin}&=& \frac{1}{2} (\partial_\mu s)^2 + \frac{1}{2} \left( \frac{f +s}{f} \right)^2 (\partial_\mu \eta)^2  + \frac{1}{2} (\partial_\mu \xi)^2   - \Lambda_{cc} +  \mu^2 (f+s)^2 - \frac{\lambda}{2} (f+s)^4 \notag \\
&& + 2 M \Lambda^2_s (f+ s) \cos \left(   \frac{\eta}{f} \right)  + 2 \kappa \Lambda^2_s (f+ \sigma)  \cos \left(\frac{\eta}{f} + \frac{\xi}{f_\xi} + \theta \right),  \label{Eq:EFTNL1GenSigEtaAx}
\end{eqnarray}
and determine the minima for all fields appearing in the full scalar potential:
\begin{equation}\label{Eq:VacuumEFTRelations}
\langle \xi \rangle = - f_\xi \theta, \qquad \langle \eta \rangle = 0, \qquad \langle \sigma^2 \rangle\lambda = \mu^2  + \frac{M + \kappa}{\langle \sigma \rangle} \Lambda_s^2.
\end{equation}
In this vacuum configuration the CP-even scalar decouples from the other two scalars and acquires a heavier mass:
\begin{equation}
m_{\sigma}^2= 4 \langle \sigma^2 \rangle \lambda + 2 \frac{M+\kappa}{\langle \sigma \rangle} \Lambda^2_s.
\end{equation}
If we further assume that the $\theta$-parameter takes the parity-conserving value $\theta = 0$~\cite{Vafa:1984xg} (or absorb the $\theta$-parameter into the dynamical axion $\xi$), we find that the CP-odd scalars give rise to a symmetric two-by-two mass matrix: 
\begin{equation}
M_{\rm axions}^2 = 2 \frac{\Lambda^2_s}{f} \left( \begin{array}{cc} M+\kappa & \kappa   \frac{f}{f_{\xi}} \\  \kappa   \frac{f}{f_{\xi}}  & \kappa  \left(\frac{f }{f_{\xi}}\right)^2    \end{array}\right),
\end{equation}
which comes upon diagonalisation with the eigenvalues:
\begin{equation}\label{Eq:MassRelationsAxions1Gen}
\begin{array}{rcl}
m^2_+ &=& \Lambda^2_s \frac{f^2 \kappa + f_{\xi}^2 (M+ \kappa) + \sqrt{ ( f^2 \kappa +  f_{\xi}^2 (M+ \kappa) )^2 - 4 f^2  f_{\xi}^2 M \kappa  } }{f f_{\xi}^2}, \\
m^2_- &=& \Lambda^2_s  \frac{f^2 \kappa + f_{\xi}^2 (M+ \kappa) - \sqrt{ ( f^2 \kappa +  f_{\xi}^2 (M+ \kappa) )^2 - 4 f^2  f_{\xi}^2 M \kappa  } }{f f_{\xi}^2}.
\end{array}
\end{equation}
For a better physical appreciation of these mass eigenstates, we consider two limiting cases for the mass spectrum depending on the relative value between the decay constants $f$ and $f_\xi$:
\begin{itemize}
\item[(i)] \underline{$f = \langle \sigma \rangle \ll f_{\xi}$:} in the limit where the closed string axion $\xi$ has a significantly larger decay constant than the open string axion, the mass eigenvalues can be reduced to:
\begin{equation}\label{Eq:AxionMasssit1}
m_+^2 \sim 2 \Lambda^2_s \frac{M+\kappa}{f} \left( 1 + {\cal O}\left(\frac{f^2}{f_\xi^2}\right)   \right), \qquad m_-^2 \sim 2 \Lambda^2_s \frac{ f M \kappa}{f_{\xi}^2 (M+\kappa)} \left( 1 - {\cal O}\left(\frac{f^2}{f_\xi^2}\right)  \right), 
\end{equation}
for the corresponding eigenvectors:
\begin{equation}
a_+ = \eta + \frac{f \kappa}{f_{\xi}(M+\kappa)} \xi, \qquad   a_- = - \frac{f}{f_{\xi}} \eta + \xi .
\end{equation}
Hence, in this limit the open string axion $\eta$ aligns with the heaviest mass eigenvector and acquires a mass that is significantly larger than the one of the axion~$\xi$, whose mass arises through a conspiracy of both explicit $U(1)$-breaking effects represented by $M$ and $\kappa$ respectively, in a similar spirit as the Gell-Mann-Oakes-Renner relations~\cite{GellMann:1968rz}. Taking into account the CP-even scalar mass, we obtain the following spectrum:
\begin{equation}\label{Eq:MassHierarchysit1}
m_-^2 \ll m_+^2 <  m_{\sigma}^2 = 4 f^2 \lambda + m_+^2,
\end{equation}
with the closed string axion $\xi$ corresponding to the lowest lying mass state in this vacuum configuration. The masses of the other two scales are of the same order and significantly larger than $m_-$ due to the hierarchy between the axion decay constants. Considering the parameter space (\ref{Eq:ParameterkappaM}) for $\kappa$ and $M$, we find that $m_\sigma, m_+ \sim {\cal O}(10^{15}-10^{16})$~Gev, while the range for $m_-$ is much broader:
\begin{equation}
m_- \sim {\cal O} (10^{10} -10^{14} \text{GeV}).
\end{equation}
The closed string axion can perfectly serve as an inflaton candidate with a mass $m_- \sim 10^{13}$~GeV in a UV-complete set-up that offers a dynamical and non-perturbative explanation for the low mass of the inflaton. This case will be worked out in more detail in section~\ref{Ss:NatInfIR}.
\item[(ii)] \underline{$f_{\xi}\ll f = \langle \sigma \rangle$:} if there exists an opposite hierarchy between the decay constants w.r.t.~the first case, the mass eigenvalues reduce to different values:
\begin{equation}
m_+^2 \sim 2 \Lambda^2_s \frac{f \kappa}{f_{\xi}^2} + 2 \frac{\kappa }{f} \Lambda^2_s + {\cal O}\left(\frac{f^2_\xi}{f^2}\right), \qquad m_-^2 \sim \Lambda^2_s \frac{2 M}{f} \left( 1 - {\cal O}\left(\frac{f^2_\xi}{f^2}\right)  \right)
\end{equation}
with the corresponding eigenvectors:
\begin{equation}
a_+ = \frac{f_{\xi}}{f} \eta +\xi  , \qquad a_- = \eta -\frac{f_{\xi}}{f} \xi.
\end{equation}
In this limit, the closed string axion $\xi$ aligns with the heavier of the two mass eigenvectors and we observe a reverse see-saw mechanism, where $m_+$ turns out to be heavier than both composite masses: 
\begin{equation}
m_-^2 < m_{\sigma}^2 \ll m_+^2,
\end{equation}
due to the hierarchy in the axion decay constants. If we consider once again the parameter space (\ref{Eq:ParameterkappaM}) for $\kappa$ and $M$, we expect the composite masses $m_-$ and $m_{\sigma}$ to lie in the range ${\cal O}(10^{14}-10^{16} \text{GeV})$  and ${\cal O}(10^{15}-10^{16} \text{GeV})$ respectively, while the mass of the closed string axion is estimated to lie within ${\cal O}(10^{18}-10^{19} \text{GeV})$. In this scenario, the closed string axion decouples from the other fields, while the two scalars constitute an inflationary set-up that requires further study. The mass hierarchy suggests to take the lightest axion $a_-\simeq\eta$ as the inflaton candidate, though the sub-Planckian decay constant $f$ a priori excludes a realisation of natural inflation due to an obvious violation of the slow-roll conditions. Moreover, the mass hierarchy between the scalar $\sigma$ and the axion~$\eta$ might not be large enough to treat this case as a single field inflationary model, such that only a proper two-field analysis can determine the viability of this limit as an inflationary model. 
\end{itemize}

\noindent Irrespective of the limiting cases, we can conclude that all three scalars in the classical action (\ref{Eq:EFTNL1Gen}) acquire a non-zero mass due to the instanton and fermionic condensate background. The non-negligible effects of the gauge instantons and the fermion condensate are equally required in order for both CP-odd scalars to acquire a mass. Note that the closed string axion $\xi$ and open string axion $\eta$ have a different microscopic origin. The axion $\xi$ can be seen as a fundamental CP-odd scalar, while the scalar excitations~$\sigma$ and $\eta$ are inherently bound states. In analogy to the nomenclature in QCD, from now onwards we refer to them as {\it infladrons}: the hadrons in an inflationary setting.\footnote{Spanish or Italian native speakers will recognize the word ``ladr\'on" or ``ladrone" (thief) hiding within the word infladron, which happens to be a coincidental tongue-in-cheek as these bound states ``steal" mass from the closed string axion.} The open string axion $\eta$ should thus be seen as the pseudo-Nambu Goldstone boson associated to the breaking of the chiral $U(1)$ symmetry, similar to the $\eta(')$-meson for the strong interactions.

In setting up the classical action~(\ref{Eq:EFTNL1Gen}), we employed a non-linear parametrisation of the scalar field $\Phi$ in~(\ref{Eq:PhiParaSigmaEta}) by virtue of the bound states $\sigma$ and $\eta$, as well as symmetry arguments inherited from the UV-theory~(\ref{Eq:EFTFermionConfinement}). Furthermore, we also required renormalizability and equivalence between the Hamiltonians of the UV-theory and the IR-theory. In general, additional non-renormalizable corrections to the IR-theory are expected to occur as well, suppressed by the St\"uckelberg scale $M_{\rm St}$ or by the strong coupling scale $\Lambda_s$ acting as cut-off scales for the IR-theory. Given that the IR-theory with scalar fields $\sigma$, $\eta$ and $\xi$ cannot be directly computed from the UV-theory by integrating out the fermions and gauge bosons, we have to resort to the toolbox of effective field theory. Following Weinberg's theorem~\cite{Weinberg:1978kz} we can write down all renormalizable and non-renormalizable terms compatible with the symmetries in the UV (global $U(1)$ symmetry, charge conjugation $C$ and parity $P$), Lorentz-invariance, causality and locality. To this end, we continue treating the parameters $M$ and $e^{i \theta}$ as spurion fields under the $U(1)$ symmetry and use a momentum-expansion as a book keeping device to quantify the importance of the non-renormalizable terms when $p^2 \ll \Lambda_s^2, M_{\rm St}^2$.\footnote{In this book keeping device, a derivative counts for order one in momentum ${\cal O}(p)$, while the parameters $M$ and $\kappa$ count for order two in momentum ${\cal O}(p^2)$ due to their intimate relation to the fermion mass.} With these considerations in mind, we can infer which type of non-renormalizable interactions are expected to appear, suppressed by the UV-scale $M_{UV}$ lying in the energy range $M_{UV} \lesssim \Lambda_s , M_{\rm St}$:
\begin{itemize}
\item \underline{Higher order derivative terms:} our inability to dynamically solve the strong dynamics involving non-Abelian gauge bosons and fermions forces us to consider corrections to the kinetic terms for the scalar field $\Phi$ containing four derivatives, such as:
\begin{equation}
\frac{c_1}{M_{UV}^4} \left( \partial_\mu \Phi^\dagger \partial^\mu \Phi \right)^2, \qquad \qquad  \frac{c_2}{M_{UV}^4} \partial_\mu \Phi^\dagger \partial_\nu \Phi \,  \partial^\mu \Phi^\dagger \partial^\nu \Phi, 
\end{equation} 
among others. When computing scattering amplitudes involving $\Phi$, however, these terms are in practice mostly converted into non-renormalizable operators in the scalar potential by imposing the equations of motion for $\Phi$, following from the classical action (\ref{Eq:EFTNL1Gen}). Apart from the higher order derivative terms, also corrections of the form $c_3\left(\frac{\Phi^\dagger \Phi}{M_{UV}^2}\right)^n \partial_\mu \Phi^\dagger \partial^\mu \Phi$ are invariant under the symmetries of the classical action and turn the classical action (\ref{Eq:EFTNL1Gen}) into a full-fledged non-linear sigma-model, even for the real CP-even infladron $\sigma$. 
\item \underline{Higher order instanton corrections:} the classical action (\ref{Eq:EFTNL1Gen}) considers only the effective interactions between the bound states and the one-(anti-)instanton background, while contributions of instanton solutions with topological charge $n\geq 2$ are suppressed by the UV-cut-off scales:
\begin{equation}
\frac{1}{\Lambda_s^{n-3} M_{\rm St}^{n-1}} \kappa^n e^{i\,n (\theta + \xi)} \Phi^n + h.c., 
\end{equation}
Cross-terms between the instanton and anti-instanton solutions do not occur due to the declustering ansatz in the dilute gas approximation. With respect to the one-instanton gauge field configurations, the amplitudes of the higher order instanton solutions (with $n\geq 2$) are suppressed by a factor $e^{- \frac{8\pi^2}{g_2^2}(n-1)}$ and therefore only represent corrections of the order ${\cal O} (10^{-5})$ and smaller, which can be safely discarded. 
\item \underline{Higher order fermionic condensate corrections:} in case interactions between the fermion-antifermion bound state and the fermionic condensate background do not limit themselves to linear order in $\Phi$, quadratic (or higher order) terms of the form can be included:
\begin{equation}
c_4\frac{f^4}{M_{UV}^4}(M\Phi + M^\dagger \Phi^\dagger)^2,
\end{equation} 
which are invariant under the global $U(1)$ symmetry with the spurion $M$ transforming as $M\rightarrow M e^{-2i q_-\alpha}$. 
\item \underline{Mixed corrections:} upon integrating out heavy degrees of freedom, also terms that mix different types of interactions might arise. In line with the symmetry arguments, such mixing terms can include the kinetic terms, such as $c_5 \frac{f^2}{M_{UV}^2} |\partial_\mu \Phi|^2 (M\Phi + M^\dagger \Phi^\dagger)$ and $c_6 \frac{f^2}{M_{UV}^2} |\partial_\mu \Phi|^2 (\kappa  e^{i (\theta +  \xi )} \Phi + \kappa^\dagger  e^{-i (\theta +  \xi )} \Phi^\dagger)$, or just involve the interactions for the bound state, such as $ c_7  \frac{f^4}{M_{UV}^4} (M\Phi + M^\dagger \Phi^\dagger)(\kappa  e^{i (\theta +  \xi )} \Phi + \kappa  e^{-i (\theta +  \xi )} \Phi^\dagger)$.
\end{itemize}   
The aforementioned corrections represent the set of expected non-renormalisable interactions of order ${\cal O}(p^4)$ in the momentum-expansion, which in practice offer small corrections to the correlator amplitudes computed from the classical theory (\ref{Eq:EFTNL1Gen}). From the theory side, the constants $c_{i \in \{1, \ldots, 7 \}}$ remain undetermined, such that their value ought to be settled experimentally by matching the data with the amplitudes computed from the effective field theory. With respect to the inflationary paradigm, one can investigate how the non-renormalizable terms alter the cosmological predictions of the model, such as the spectral index and the tensor-to-scalar ratio due to changes in the slow-roll parameters. For our purposes, however, it suffices to estimate the order of the correction to figure out whether a correction is numerically relevant, once the accuracy level~\cite{Burgess:2007pt} of the effective field theory is chosen. We have already seen that the higher order instanton corrections are suppressed by a factor ${\cal O}(10^{-5})$ with respect to the one-instanton background in the classical action. The other terms are suppressed by powers of the UV-cut-off scale or by factors $f^{2n}/M_{UV}^{2n}$, which would imply a suppression by a factor ${\cal O}(10^{-2})$ or smaller. If we specify the corrections in terms of the $(\sigma, a_+, a_-)$ basis, we observe additional suppression in powers of the axion decay constants for the axions $(a_+,a_-)$. More precisely, in the regime where $f\ll f_\xi$, the non-renormalisable corrections involving the $a_-$ direction (mostly aligned along $\xi$) are suppressed by an additional factor $f^2/f_\xi^2$, such that it suffices to consider only the classical action (\ref{Eq:EFTNL1Gen}). In the opposite regime $f_\xi \ll f$, the axion $a_-$ aligns primarily along the open string axion $\eta$ and non-renormalisable corrections such as the higher derivative terms could present non-negligible effects along the inflationary trajectory for $a_-$, along the lines of~\cite{Weinberg:2008hq,Shiu:2002kg,Pedro:2017qcx}.

\subsubsection{Infladron Quantum Corrections}\label{Sss:InfQM}
In the previous section we considered the classical lagrangian (\ref{Eq:EFTNL1Gen}) for two composite scalars $\sigma$ and $\eta$ and one fundamental axion $\xi$ as the infra-red description of a chiral gauge theory at strong coupling with one generation of fermions and anomalously coupled to a closed string axion $\xi$. Loyal to the true spirit of effective field theory, we also indicated which kinds of (non-renormalisable) corrections are expected in this toy-model, given our inability to produce the IR theory directly from action (\ref{Eq:EFTNf1}) upon integrating our the fermions and gauge bosons. Yet, these terms do not yet capture the quantum corrections associated to the (perturbative) self-interactions of the scalar field $\Phi$, which are indispensable to understand~\cite{Coleman:1973jx} the (quantum) vacuum of the IR theory.

\begin{figure}[h]
\begin{center}
\begin{tabular}{c@{\hspace{0.4in}}c}
\includegraphics[scale=0.7]{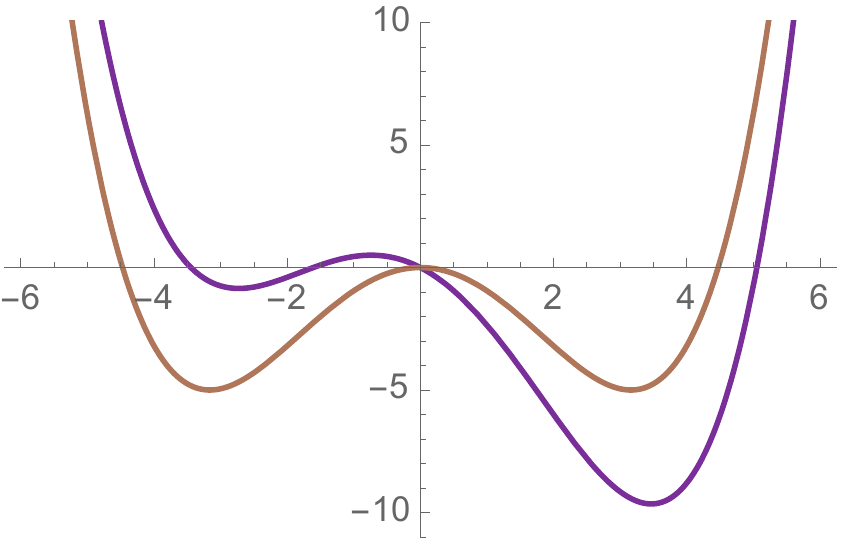} \begin{picture}(0,0)  \put(-145,85){\color{myaubergine}$V_{\rm cl}$} \put(0,50){$\sigma$}    \put(-45,80){\color{mybrown}$V_{\rm MH}$} \end{picture}  & \includegraphics[scale=0.75]{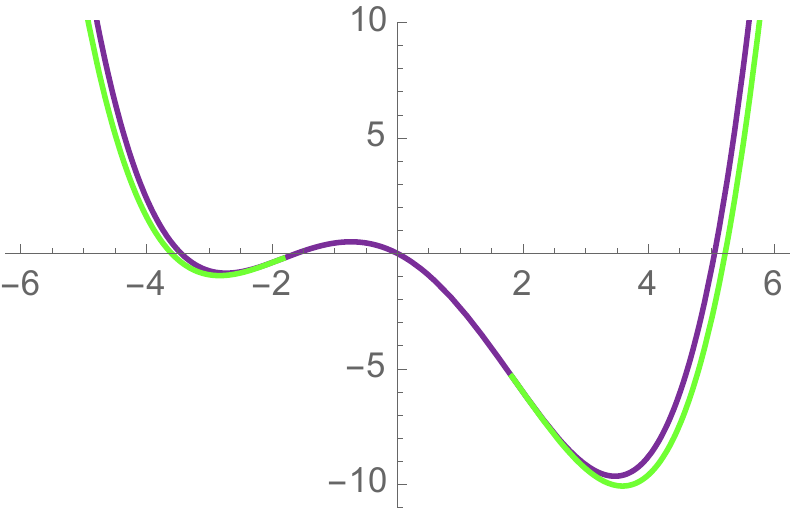}  \begin{picture}(0,0)  \put(-145,85){\color{myaubergine}$V_{\rm cl}$} \put(0,50){$\sigma$}  \put(-10,80){\color{CycleGreen}$V_{\rm qu}$} \end{picture} 
\end{tabular}
\caption{Effects of the interaction between composite scalar $\Phi$ and the non-trivial vacuum (due to non-perturbative physics) on a mexican hat potential $V_{\rm MH} (\sigma) = - \mu^2 \sigma^2 + \frac{\lambda}{2} \sigma^4$ along the infladron direction $\sigma$ (left), expressed in  units of $\Lambda_s$. Schematic representation of the one-loop corrections in the effective potential in comparison to the classical potential $V_{\rm cl}$ from (\ref{Eq:EFTNL1GenSigEtaAx}) along the infladron direction $\sigma$ (right). The one-loop effective potential (green curve) is only depicted in the range where it takes real values. In the field range where $-2 \mu^2 + 6\lambda \Phi^\dagger \Phi <0$ the effective potential becomes imaginary and non-convex, encoding information~\cite{Weinberg:1987vp} about unstable fluctuations due the tachyonic (local) maximum of the scalar potential.  \label{Fig:PotClEffQu}}
\end{center}
\end{figure}
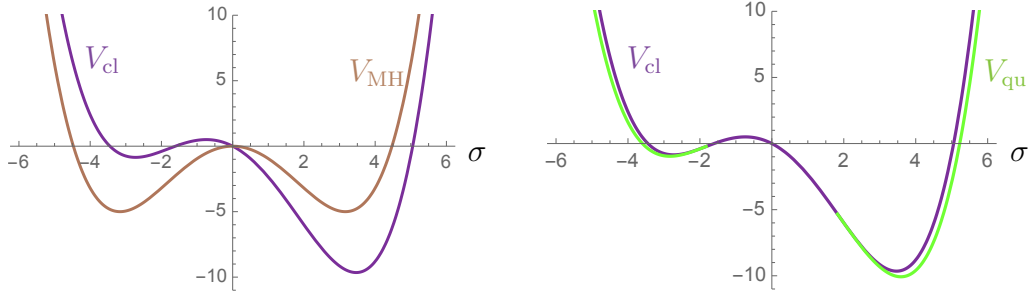

In first instance, we consider the one-loop effective potential associated to the perturbative interactions involving the scalar field $\Phi$:  
\begin{eqnarray}
V_{\rm eff}^{\rm 1-loop} &=&-\mu^2 \Phi^\dagger \Phi + \frac{\lambda}{2} ( \Phi^\dagger \Phi)^2 + \frac{1}{4(4\pi)^2}( -2 \mu^2 + 6 \lambda \Phi^\dagger \Phi)^2 \left[  \ln \left( \frac{-2 \mu^2 + 6 \lambda \Phi^\dagger \Phi}{\Lambda_r^2} \right) - \frac{3}{2}\right]  \notag \\
&&  + \Lambda^2_s M \Phi + \Lambda^2_s   M\Phi^\dagger + \Lambda^2_s \kappa\, e^{i (\theta +  \xi/f_{\xi} )}   \det (\Phi) + \Lambda^2_s  \kappa\, e^{-i (\theta +  \xi/f_{\xi})} \det (\Phi^\dagger), \label{Eq:1LoopEffPot}
\end{eqnarray}
where the Minimal Subtraction (MS) scheme was used to regulate the one-loop corrections to the potential and $\Lambda_r$ is an arbitrary mass scale at which we impose the renormalisation conditions. One could naively assert that the one-loop corrections to the potential destabilise the classical vacuum in favour of a distinct quantum vacuum, as depicted in figure~\ref{Fig:PotClEffQu}, yet this interpretation would require the logarithmic correction $\lambda \ln\left[(-2\mu^2 + 6 \lambda \Phi^\dagger \Phi)/\Lambda_r^2 \right]$ to take values of order one. This would in turn imply that the $n$-loop corrections $\left(\lambda \ln\left[(-2\mu^2 + 6 \lambda \Phi^\dagger \Phi)/\Lambda_r^2 \right] \right)^n$ cannot be ignored either. The resolution to this conundrum lies in a proper resummation of the loop corrections, which can be successfully performed by solving the Callan-Symanzik equation for the scalar potential. More precisely, one has to rewrite the Callan-Symanzik equation (or renormalization group equation for the effective action) into a renormalization group equation (RGE) for the effective potential and the wave-function renormalization $Z(\Phi)$ and solve both in terms of the running scale $\Lambda_r$. This technique was originally proposed for a massless scalar field with quartic couplings in~\cite{Coleman:1973jx}, but also works successfully for the massive case~\cite{Kastening:1991gv,Bando:1992np,Ford:1992mv}.   
The bottom line of the computation\footnote{A thorough and pedagogical review involving the effective potential and its resummation can be found in chapter 7 of~\cite{Miransky:1994vk}.} consists in evaluating the effective potential at a particular UV energy scale, for instance the energy scale at which the logarithmic term vanishes, and extrapolate the effective potential at lower energy scales by virtue of the running parameters in the theory. If we follow this logic, the effective potential takes the form:
\begin{eqnarray}
V_{\rm eff} &=& -\ov\mu^2 \Phi^\dagger \Phi + \frac{\ov\lambda}{2} ( \Phi^\dagger \Phi)^2 - \frac{3}{128 \pi^2} ( -2 \ov\mu^2 + 6\ov \lambda \Phi^\dagger \Phi)^2 \label{Eq:ResumEffPot}\\
&&+ \Lambda^2_s M \Phi + \Lambda^2_s   M\Phi^\dagger + \Lambda^2_s \kappa\, e^{i (\theta +  \xi/f_{\xi} )}   \det (\Phi) + \Lambda^2_s  \kappa\, e^{-i (\theta +  \xi/f_{\xi})} \det (\Phi^\dagger), \notag 
\end{eqnarray}
where the parameters $\mu$ and $\lambda$ have to be distinguished from the functions $\ov\mu$ and $\ov\lambda$:
\begin{eqnarray}\label{Eq:FunctSolutionParameters}
\ov \mu^2 &=& \frac{\mu^2(\Lambda_r)}{\left( 1 - \frac{9 \lambda(\Lambda_r)}{8 \pi^2} \ln \frac{ -2 \mu^2 + 6 \lambda \Phi^\dagger \Phi}{\Lambda_r^2}  \right)^{\frac{1}{3}}},\\
\ov \lambda &=& \frac{\lambda(\Lambda_r)}{ 1 - \frac{9 \lambda(\Lambda_r)}{8\pi^2} \ln \frac{-2 \mu^2 + 6 \lambda \Phi^\dagger \Phi}{\Lambda_r^2}}.
\end{eqnarray}
These functions are field-dependent solutions to the renormalization group equations, with the coupling constants evaluated at an arbitrary IR energy scale $\Lambda_r$. At one-loop the relevant renormalisation group equations for the parameters read in terms of the running scale $\Lambda$:
\begin{equation}\label{Eq:RGEParameters}
\Lambda \frac{d \ln \mu}{d \Lambda} = \frac{3}{8\pi^2}  \lambda, \qquad  \Lambda \frac{d \lambda}{d \Lambda} = \frac{9}{4\pi^2} \lambda^2,
\end{equation}
which can be solved sequentially starting from the solution of the RGE for $\lambda$. The explicit solution for the running of the quartic coupling $\lambda$ in the IR is given by:
\begin{equation}
\lambda (\Lambda_r) = \frac{\lambda (\Lambda_0)}{1 +  \frac{9}{4\pi^2}  \lambda(\Lambda_0) \ln \frac{\Lambda_0}{\Lambda_r}} 
\end{equation}
with $\Lambda_r < \Lambda_0$ naturally. This implies that the quartic coupling runs to smaller values in the IR and undergoes a screening effects at lower energies. Plugging this solution into the RGE for the mass-parameter $\mu^2$ allows to extract an expression for the running mass parameter, which translates to the functional relation for $\ov \mu^2$ in~(\ref{Eq:FunctSolutionParameters}). Note that the scalar field $\Phi$ itself does not acquire a wavefunction renormalization at one-loop, such that an expansion of the field-dependent parameters $\ov \mu^2$ and $\ov \lambda$ in the resummed effective potential~(\ref{Eq:ResumEffPot}) for small field ranges $ \left|-2 \mu^2 + 6 \lambda \Phi^\dagger \Phi\right| \sim \Lambda_r^2$ suffices to reproduce the one-loop corrected effective potential. Moreover, as the resummed effective potential~(\ref{Eq:ResumEffPot}) has the same structure as the perturbative part of the classical superpotential, the relations for the classical vacuum configuration in~(\ref{Eq:VacuumEFTRelations}) will remain valid, though the parameters $\mu^2$ and $\lambda$ need to be replaced by their effective counterpart.\footnote{For a massive scalar field the RGE improved scalar potential also requires a careful treatment of the running cosmological constant. In this paper, we have added the cosmological constant by hand and refrain ourselves from discussing its RGE flow until we have obtained a fully consistent picture of the cosmological constant by successfully embedding the entire set-up in a string compactification with stabilised moduli.}

\begin{figure}[h]
\begin{center}
\begin{tabular}{c@{\hspace{0.6in}}c}
\includegraphics[scale=0.4]{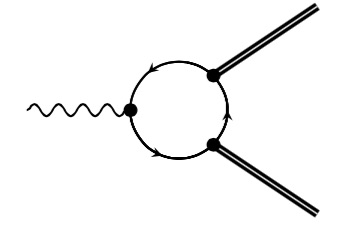} \begin{picture}(0,0) \put(-10,0){$\Phi^\dagger$}  \put(-10,90){$\Phi$} \put(-120,58){$g_{\mu \nu}$} \end{picture} & \includegraphics[scale=0.4]{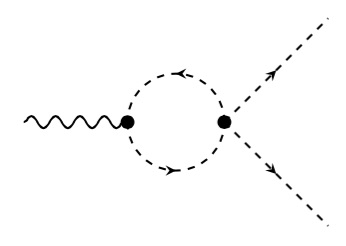} \begin{picture}(0,0) \put(-10,0){$\sigma$}  \put(-10,90){$\sigma$} \put(-120,58){$g_{\mu \nu}$} \end{picture}
\end{tabular}
\caption{Non-minimal couplings to gravity for the composite field $\Phi$ (double solid line) induced from internal fermion loops (single solid line) on the lefthand side. The running of the non-minimal coupling parameter $\varpi$ at one-loop is due to the presence of the 1 PI diagram on the righthand side involving the (radial) infladron $\sigma$.   \label{Fig:NonMinimalCoupling}}
\end{center}
\end{figure}
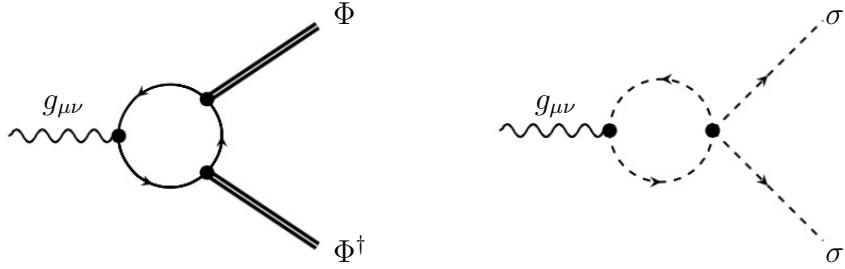

The effective potential and the running of the couplings are not the only loop-effects that have to be considered in the full quantum theory. In the UV theory (\ref{Eq:EFTNf1}), the chiral fermions couple to gravity and are simultaneously subject to the four-fermion interactions in (\ref{Eq:NJLInter1Gen}). Loop-corrections involving the fermions induce a non-minimal coupling~\cite{Hill:1991jc} of the composite field $\Phi$ to gravity with coupling $\varpi$:
\begin{equation}\label{Eq:NonMinimCoupling}
{\cal S}_{\rm EFT} \ni  \int d^4x \sqrt{g}\,  \left(  - \frac{1}{2} \varpi \Phi^\dagger \Phi R \right),
\end{equation} 
as depicted on the lefthand side of figure \ref{Fig:NonMinimalCoupling}. In the EFT theory expressed in terms of the scalar $\Phi$, the coupling $\varpi$ is by itself also subject to running due to loop diagrams resulting from the non-minimal coupling to gravity and the quartic couplings for $\Phi$, see righthand side of figure \ref{Fig:NonMinimalCoupling}. 
A one-loop computation reveals the running of $\varpi$ to be determined by the RGE:
\begin{equation}
\Lambda \frac{d\ln \varpi}{d\Lambda} = \frac{3 \lambda}{16 \pi^2},
\end{equation}
where $\varpi$ and $\lambda$ represent the bare couplings and $\Lambda$ the energy scale with which the couplings run. Given that we know the solution to the RGE (\ref{Eq:RGEParameters}) for the quartic coupling, we can solve the RGE for $\varpi$ as well, in terms of the renormalized coupling~$\varpi$:
\begin{equation}
\varpi (\Lambda_r) = \varpi(\Lambda_0) e^{ -\int_{\lambda(\Lambda_0)}^{\lambda(\Lambda_r)} \frac{3 \tilde\lambda}{16\pi^2} \frac{d\tilde\lambda}{\beta (\tilde\lambda)}  }  .
\end{equation}
We know from the RGE (\ref{Eq:RGEParameters}) that the beta-function $\beta (\lambda) > 0$ is positive, which implies that the function under the integral is a strictly positive function. Hence, the non-minimal coupling parameter $\varpi$ runs to smaller values at lower energies (or equivalently taking $\Lambda_0 \rightarrow \infty$) and has an IR fixed point at $\varpi = 0$, in line with the findings of~\cite{Voloshin:1982eb}. This result has to be contrasted with the non-minimal coupling for Nambu-Jona-Lasinio models (or equivalently gauged Yukawa-Higgs models) for which the RGE for $\varpi$ results from one-loop corrections involving only internal fermions, such that the non-minimal coupling parameter has an IR fixed point $\varpi=\frac{1}{6}$~\cite{Hill:1991jc}. Given these considerations, we consider the non-minimal coupling in (\ref{Eq:NonMinimCoupling}) to be characterised by a parameter $\frac{1}{6}\leq \varpi < 1$ in the UV, such that its value runs to decreased values in the IR following the renormalization group flow. The presence of a non-minimal coupling of $\Phi$ to gravity implies that the effective action is written in the Jordan frame and has to be reverted to the Einstein frame by virtue of a conformal transformation on the spacetime metric in order to make contact with the cosmological observables, see for instance~\cite{Fujii:2003pa} for technical details. In our set-up the conformal factor $\Omega$ only involves the radial infladron $\sigma$:
\begin{equation}
\Omega^2 = 1 + \frac{\varpi }{M_{Pl}^2} \Phi^\dagger \Phi = 1 + \frac{\varpi }{M_{Pl}^2} \sigma^2, \qquad \varpi < 1,
\end{equation}
and changes the kinetic term for $\sigma$ into a non-canonical one upon conformal transformation. To bring the kinetic term for infladron $\sigma$ back to its canonical form, we have to make a field redefinition:
\begin{equation}
\left(\frac{d\varphi}{d \sigma}\right)^2 = \frac{1}{\Omega^2} \left( 1 + \frac{6 \varpi^2 \sigma^2}{M_{Pl}^2 \left( 1 + \frac{\varpi \sigma^2}{M_{Pl}^2} \right)} \right) \sim \frac{1}{\Omega^2},
\end{equation}
where we approximated the field redefinition factor by assuming $\varpi^2 \ll1$ and that the infladron $\sigma$ is not subject to trans-Planckian displacements. Under these two assumptions, we can solve for $\sigma$ analytically as a function of $\varphi$:
\begin{equation}
\sigma=  \frac{M_{Pl}}{\sqrt{\varpi}} \sinh \frac{\sqrt{\varpi} \varphi}{M_{Pl}}.
\end{equation} 
Hence, the field redefinition becomes trivial, i.e.~the scalar fields $\sigma$ and $\varphi$ coincide, as long as the two assumptions are satisfied. The smallness of the parameter $\varpi$ has been argued above using the renormalization group flow. To prove the second assumption one has to show that the displacement of the infladron $\sigma$ due to the motion of the other scalar fields along their potentials remains small such that its {\it vev} does not undertake field excursions of the order of the Planck mass scale. We will not make any attempt to prove this assumption in all generality here, yet the next section contains a proof in case the hierarchy $f\ll f_\xi$ is realized.

Finally, a last set of perturbative corrections that can potentially destabilize the classical vacuum configuration of the effective action~(\ref{Eq:EFTNL1GenSigEtaAx}) are perturbative loop-corrections involving gravitons. It was shown, however, that these one-loop corrections can be grouped~\cite{Smolin:1979ca} into a one-loop effective potential, which depends on the (classical) perturbative potential through powers of $V_{\rm per}/M_{Pl}^4$, $V'_{\rm per}/M_{Pl}^3$ and  $V''_{\rm per}/M_{Pl}^2$. And even though this one-loop effective potential corrects the perturbative potential in (\ref{Eq:EFTNL1GenSigEtaAx}), one can argue that the resulting one-loop effective potential does not destabilize the vacuum configuration of the infladron $\sigma$, similarly to the infladron loop-corrections discussed above. Hence, perturbative quantum corrections involving gravitons are also not expected to deform the scalar potential in such a way that inflationary scenarios are potentially ruled out a priori.

\subsection{Natural-like Inflation in the Infra-red}\label{Ss:NatInfIR}
In the previous section, we established an effective field theory approach to describe the low energy limit of the confining chiral gauge theory in (\ref{Eq:EFTNf1}), coupled to a closed string axion and one generation of chiral fermions. Working out the one-loop and non-renormalizable corrections to the EFT in (\ref{Eq:EFTNL1Gen}) renders some initial intuition into the relevance of the corrections and their potential physical effects in an inflationary set-up. The non-renormalisable corrections  for instance are negligible in the region of the parameter space where $f\ll f_\xi$ and where the closed string axion $\xi$ acquires the smallest mass. Given the vacuum structure and the structure of the scalar potential in (\ref{Eq:EFTNL1GenSigEtaAx}), natural inflation is the most obvious inflationary model to extract in this region of the parameter space with $\xi$ as the inflaton candidate.\footnote{The derivation of the gauge instanton contribution to the scalar potential depends heavily on the dilute instanton gas approximation. Abandoning this approximation and adapting the monodromic structure that appears to be a better description for the IR-vacuum structure of pure Yang-Mills theories does not alter our picture qualitatively. That is to say, the vacuum structure for the closed string axion with a trans-Planckian decay constant will still  be periodic, as the monodromy structure will expose itself in the vacuum for the infladron $\eta$.} To this end, the heavier infladrons need to be consistently integrated out to end up with the effective potential for $\xi$, realising natural inflation in the IR. A prerequisite for integrating out the heavy infladrons consists in showing that the back-reaction of the infladrons on the inflationary trajectory of axion $\xi$ does not destroy the flat direction along which the classical inflaton field slowly rolls during inflation. This constraint turns out to be intimately tied to a {\it conditio sine qua non} for the spontaneous symmetry-breaking process to be unaffected by a large non-minimal coupling (\ref{Eq:NonMinimCoupling}) of the radial infladron to gravity, as argued at the end of section~\ref{Sss:InfQM}.  

Let us first neglect any back-reaction effects and integrate out the heavy infladrons by virtue of their vacuum configuration. Then, we can distinguish two distinct situations, depending on the relative magnitude between the parameters $M$ and $\kappa$, in line with the parameter range in (\ref{Eq:ParameterkappaM}). The first situation occurs when $M\ll \kappa$, in which case the vacuum configuration for $\eta$ will be determined by the non-perturbative vacuum effect in (\ref{Eq:EFTNL1GenSigEtaAx}) with the largest amplitude:
\begin{equation}\label{Eq:VacuumEtaXiMsK}
\langle \eta\rangle = - \frac{f}{f_\xi} \xi,
\end{equation}
with the $\theta$-angle swallowed up into $\xi$, while the vacuum configuration for the infladron~$\sigma$ is then given by:
\begin{equation}\label{Eq:VacuumSigmaXiMsK}
\langle \sigma^2 \rangle\lambda = \mu^2  + \frac{M \cos \frac{\xi}{f_\xi} + \kappa}{\langle \sigma \rangle} \Lambda_s^2.
\end{equation}
In this vacuum the periodic potential for the closed string axion arises predominantly from the interaction with the fermion condensate vacuum and the mass relations for the respective scalars match the tree-level masses (\ref{Eq:AxionMasssit1}) and (\ref{Eq:MassHierarchysit1}). The second situation, for which ${\cal O}(M)\sim {\cal O}(\kappa)$, is elaborated in section~\ref{Sss:InterInfladrons} yielding the vacuum configuration (\ref{Eq:VacuumEFTRelations}), in which the tree-level mass of the closed string axion $\xi$ acquires equal contributions from both non-perturbative vacuum effects, as exhibited in (\ref{Eq:AxionMasssit1}). 

Despite their simplicity, the vacuum relations (\ref{Eq:VacuumEtaXiMsK}) and (\ref{Eq:VacuumSigmaXiMsK}) fully expose the profound implications of back-reaction effects in this set-up.\footnote{Even though the vacuum equations (\ref{Eq:AxionMasssit1}) and (\ref{Eq:MassHierarchysit1}) were derived for a specific hierarchy between the parameters $M$ and $\kappa$, these vacuum relations turn out to be generically valid for all values of the respective parameters.} Displacement along the inflationary direction $\xi$ in field space cause the radial infladron $\sigma$ to move away from its vacuum configuration dictated by equation (\ref{Eq:VacuumSigmaXiMsK}), which in turn  also induces displacements of the angular infladron $\eta$ through the $\langle \sigma \rangle$-dependent decay constant $f$ in (\ref{Eq:VacuumEtaXiMsK}). When these $\xi$-dependent displacements are taken into account in the scalar potential, the back-reaction of $\delta \sigma (\xi)$ and $\delta \eta (\xi)$ might distort the scalar potential for $\xi$ sufficiently as to obstruct flat directions required for the classical inflationary motion. Computing the field displacements $\delta \sigma$ as a function of $\xi$ directly from (\ref{Eq:VacuumSigmaXiMsK}) is cumbersome due to the cubic equation. As we want to discuss small displacements from the vacuum configuration for the infladrons $\sigma$ and $\eta$, we will limit ourselves to linear perturbations $\delta \sigma$ and $\delta \eta$ about the vacuum: 
\begin{equation}\label{Eq:LinearBackreaction}
\delta \sigma (\xi) = \frac{2 M \Lambda_s^2}{m_\sigma^2}  \left( \cos \frac{\xi}{f_\xi}  - 1 \right), \qquad \delta \eta (\xi) = - \frac{\delta \sigma (\xi)}{f_\xi} \xi,
\end{equation}
which matches the quadratic expansion of the scalar potential in the heavy field perturbations according to the method in~\cite{Buchmuller:2015oma,Landete:2016cix}. The linear approximation of the back-reaction to the inflationary potential is only valid for small perturbations, meaning $\delta \sigma \ll \sigma_0$ and $\delta \eta \ll \sigma_0$, which imposes a hierarchy between the inflaton mass $m_-$ and the infladron mass $m_{\sigma}$. By rewriting the linear shift $\delta \sigma$ of the infladron {\it vev} for small field excursions of the inflaton $\xi$:
\begin{equation}\label{Eq:RenormDeltaSigma}
\frac{\delta  \ov \sigma}{ \ov\sigma_0} \sim - \frac{3 H^2}{m_{\sigma}^2} \frac{M+\kappa}{\kappa} \frac{1}{\ov \sigma_0^2},
\end{equation}  
with dimensionless parameter $\ov \sigma \equiv \sigma/M_{\rm Pl}$, we can infer a hierarchy between the Hubble scale $H$ during inflation and  the infladron mass $m_{\sigma}$  given the sub-Planckian scale of the infladron {\it vev} ($\ov \sigma_0 \ll 1$). The hierarchy $H\ll m_{\sigma}$ is very reminiscent of the hierarchy to which closed string moduli masses or stabiliser field masses are subjected to in a four-dimensional supergravity set-up \cite{Dudas:2015lga}. For a Hubble-scale of the order $H \sim 10^{14}$ ${\rm GeV}$ the infladron mass has to be at least $m_{\sigma} \sim 10^{16}$ ${\rm GeV}$, which puts the strong coupling scale $\Lambda_s \sim 10^{16}$ ${\rm GeV}$ around that scale as well. Note that there is only a solution to $\delta \ov \sigma < \ov \sigma_0$ in (\ref{Eq:RenormDeltaSigma}) in case $M\leq \kappa$. The opposite region in the $(M, \kappa)$-parameter space is characterised by large displacements of the infladron vev $\sigma_0$, such that the large infladron back-reaction risks to lift the flat direction of the inflationary potential along the $\xi$-direction.
The structure of the linear displacements in (\ref{Eq:LinearBackreaction}) indicates that it is sufficient to constrain the shift $\delta \sigma$ for the heaviest infladron {\it vev} to ensure that the linear shift $\delta \eta$ in the {\it vev} of the other infladron remains under control. Furthermore, we can also deduce that the displacements of the infladron {\it vev}s along the inflationary trajectory will always be of the order ${\cal O}(10^{16} {\rm \, GeV})$, so that we can safely discard the non-minimal coupling to gravity discussed in the previous section. For large field excursions of the inflaton, i.e.~$\langle \xi \rangle \sim f_\xi$, the displacements (\ref{Eq:LinearBackreaction}) along the infladron {\it vev}s are of order ${\cal O}(M)$ and their back-reaction on the inflationary potential are destined to lift the inflationary trajectory substantially, unless we impose $M \leq {\cal O}(10^{16} {\rm \, GeV})$. The microscopic origin~(\ref{Eq:MassRelationMFermCond}) of $M$ as the order parameter for interactions between the chiral fermions and the fermion condensate background then implies that the St\"uckelberg mass scale $M_{\rm St}$ is of the sub-Planckian order ${\cal O}(10^{16} {\rm \, GeV})$ or higher. 

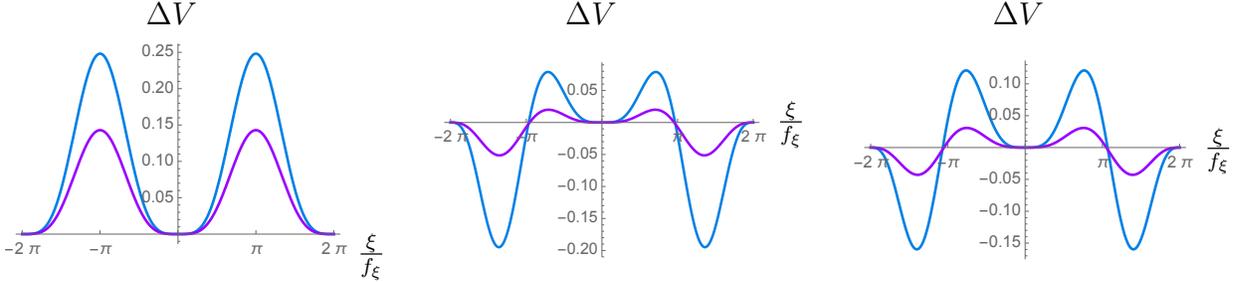
\begin{figure}[h]
\vspace*{0.2in}
\begin{center}
\hspace*{-0.2in} 
\begin{tabular}{c@{\hspace{0.4in}}c@{\hspace{0.4in}}c}
\includegraphics[scale=0.62]{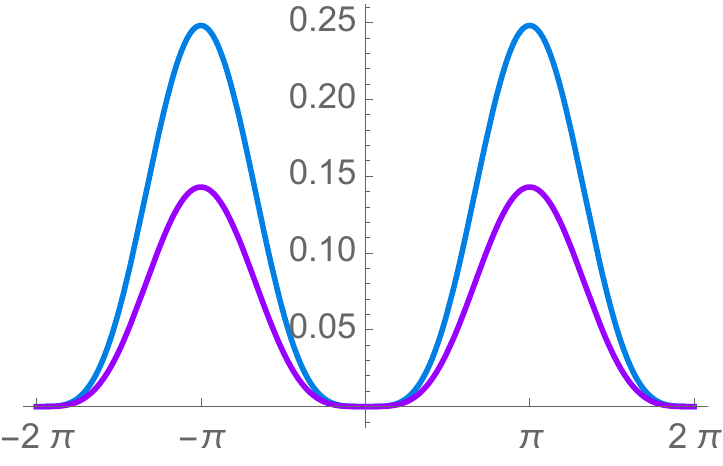} \begin{picture}(0,0) \put(0,0){$\frac{\xi}{f_\xi}$} \put(-80,90){$\Delta V$} \end{picture}
&\includegraphics[scale=0.6]{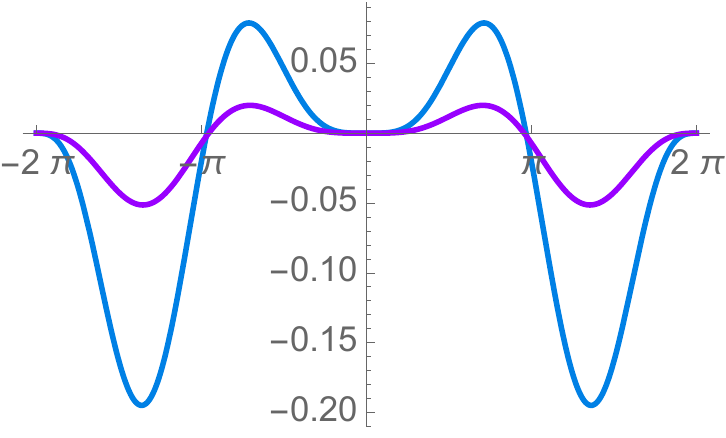} \begin{picture}(0,0) \put(0,50){$\frac{\xi}{f_\xi}$} \put(-80,90){$\Delta V$} \end{picture}
 & \includegraphics[scale=0.62]{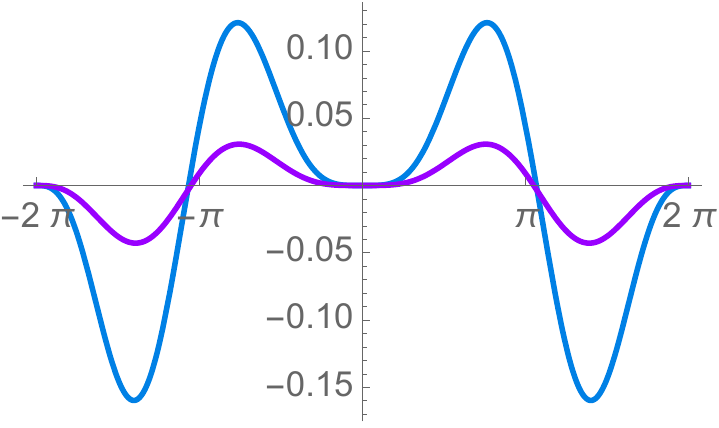} \begin{picture}(0,0) \put(0,40){$\frac{\xi}{f_\xi}$} \put(-80,90){$\Delta V$} \end{picture}
\end{tabular}
\caption{Comparison $\Delta V$ between the non-back-reacted potential and the back-reacted potential as a function of the inflaton $\xi$: the back-reaction due to infladron $\sigma$ (left), back-reaction due to infladron $\eta$ (middle) and back-reaction due to both infladrons $\sigma$ and $\eta$ (right). The potentials are drawn for parameter choice $\mu = 2 \kappa = 10 \lambda = 10^{16}$ GeV and $M = 10^{15}$ GeV (purple) or $M = 2 \times 10^{15}$ GeV (blue).  \label{Fig:DiffPotential}}
\end{center}
\end{figure}

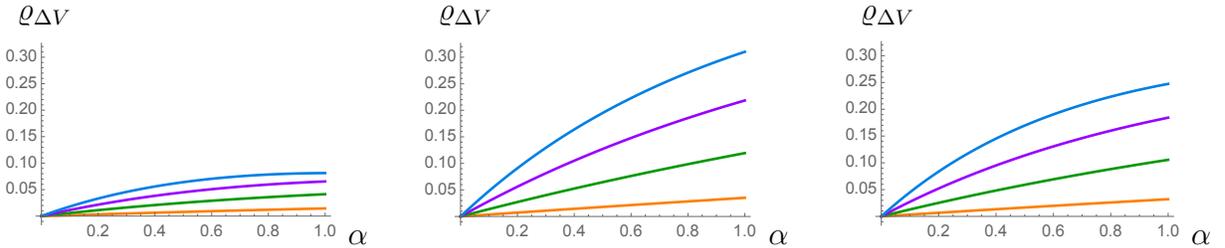
\begin{figure}[h]
\vspace*{0.2in}
\begin{center}
\hspace*{-0.2in}
\begin{tabular}{c@{\hspace{0.4in}}c@{\hspace{0.4in}}c}
\vspace*{0.2in} \includegraphics[scale=0.62]{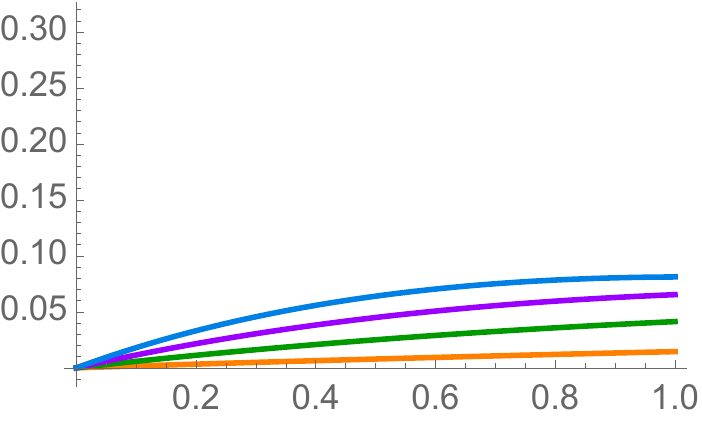} \begin{picture}(0,0) \put(0,0){$\alpha$} \put(-125,85){$\varrho_{\Delta V}$} \end{picture}
&\includegraphics[scale=0.62]{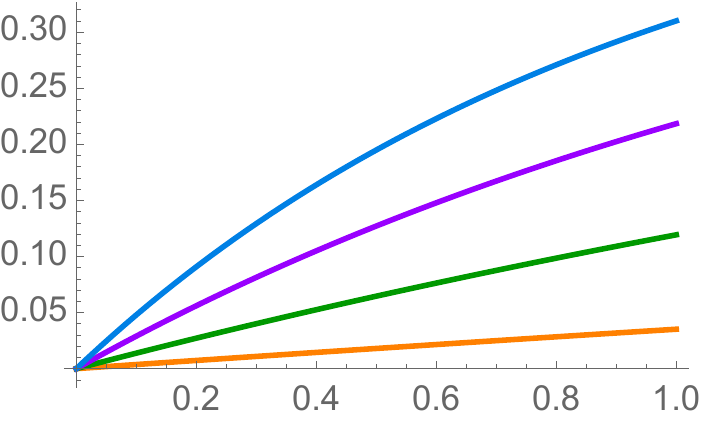}  \begin{picture}(0,0) \put(0,0){$\alpha$} \put(-125,85){$\varrho_{\Delta V}$} \end{picture}
& \includegraphics[scale=0.62]{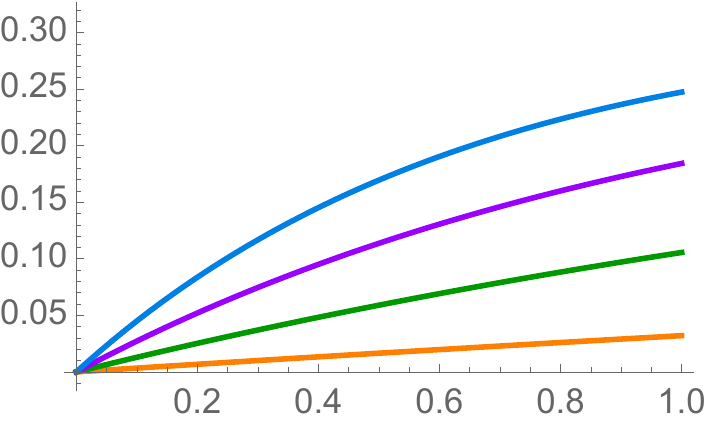}  \begin{picture}(0,0) \put(0,0){$\alpha$} \put(-125,85){$\varrho_{\Delta V}$} \end{picture}
\end{tabular}
\caption{Normalised back-reaction effects on the scalar potential expressed in terms of the ratio $\varrho_{\Delta V}$ defined in (\ref{Eq:RatioRDV}): $\varrho_{\Delta V}$ for the back-reaction from the infladron $\sigma$ (left), from the infladron $\eta$ (middle) and from both infladrons (right) for the parameter choice $\mu  = 10 \lambda = 10^{16}$ GeV and $\kappa = 0.25 \times 10^{16}$ GeV (orange), $0.5  \times 10^{16}$ GeV (green), $0.75  \times 10^{16}$ GeV (purple), and $10^{16}$ GeV (blue). The parameter $M$ is parameterized through $M= \alpha \, \kappa$ such that functional behaviour of $\varrho_{\Delta V}$ on $M$ can be evaluated over the range $0 \leq M\leq \kappa$.  \label{Fig:DiffRatio}}
\end{center}
\end{figure}

The back-reaction of the heavy infladrons on the inflationary potential can be systematically studied by inserting the expressions (\ref{Eq:LinearBackreaction}) for their shifted {\it vev}s back into the scalar potential in (\ref{Eq:EFTNL1GenSigEtaAx}). To visualize the back-reaction on the inflationary potential we consider the difference between the non-back-reacted scalar potential $V^{\text{no br}}$ and the back-reacted scalar potential $V^{\text{br}}$:  
\begin{equation}
\Delta V (\sigma, \eta, \xi) = V^{\text{no br}} (\sigma, \eta, \xi) - V^{\text{br}} (\sigma, \eta, \xi),
\end{equation}  
and choose a point in the $(M, \kappa)$-parameter space in line with the considerations above, such as in figure \ref{Fig:DiffPotential}.
By splitting up the back-reaction effects for each infladron direction we can investigate their effects separately. On the lefthand side of figure \ref{Fig:DiffPotential} we observe that the back-reaction effects due to the radial infladron $\sigma$ are largest when the inflaton~$\xi$ is farthest from a local minimum, namely $\langle \xi \rangle = \pi f_\xi k$ with $k\in \Z$. The back-reaction due to the infladron $\eta$ is depicted in the middle panel and shows a slightly different picture: the largest back-reaction effects occur for $ \pi f_{\xi} < \langle \left| \xi \right| \rangle < 2 \pi f_{\xi}$. The combined back-reaction of both infladrons then leads to the right part of figure \ref{Fig:DiffPotential}, with maximal back-reaction effects in the region $ \pi f_{\xi} < \langle \left| \xi \right| \rangle < 2 \pi f_{\xi}$. Figure \ref{Fig:DiffPotential} gives an overall picture of the back-reaction effects along the inflationary trajectory, and in order to appreciate the magnitude of the back-reaction we introduce the ratio:
\begin{equation}\label{Eq:RatioRDV}
\varrho_{\Delta V} \equiv \frac{\left|\Delta V\right|}{V^{\text{no br}}} \Big|_{\xi = \xi_{\rm max}}  
\end{equation} 
evaluated at the point $\xi_{\rm max}$ along the inflationary trajectory where $\left|\Delta V\right|$ reaches a local maximum. The result is presented in figure \ref{Fig:DiffRatio}, with the panels representing the back-reaction effects from the infladron $\sigma$ (left), from the infladron $\eta$ (middle) and from both infladrons combined (right). By parameterizing the parameter $M$ in terms of the parameter $\kappa$ through the equation $M = \alpha\, \kappa$ with $\alpha \in [0,1]$, we can evaluate how the relative magnitude of the back-reaction grows in terms of the parameters $M$ and $\kappa$. 
The plots in figure~\ref{Fig:DiffRatio} solidify the deductions from figure~\ref{Fig:DiffPotential}: the back-reaction due to the displacement of infladron $\sigma$ exhibits a different pattern along the inflationary trajectory from the back-reaction associated to the displacement of infladron $\eta$. Moreover, the back-reaction effects of both infladrons combined are not mutually constructive such that the ratio $\varrho_{\Delta V}$ for the full back-reacted solution takes values between the ratios for the back-reacted solution along one infladron direction alone. From the plots in figure~\ref{Fig:DiffRatio} we can also deduce that the total back-reaction can at most flatten the scalar potential by about 25\% with respect to the original scalar potential. Given the aforementioned constraints on the parameters $(M, \kappa)$ the fraction of the back-reaction will be much lower and can even be kept below or about the percent-level. To explore potential flat directions in the back-reacted scalar potential one can consider contour plots of the latter as a function of the various scalar fields, such as in figure~\ref{Fig:ContourPlotsBackRePot}. In these plots, the red curve identifies sufficiently flat directions, which can support  trans-Planckian field excursions of the inflaton $\xi$. Note that it suffices  for the inflaton $\xi$ to run over a distance $\Delta \xi = \pi f_\xi$ starting from a local maximum to a local minimum along the indicated red curve, in order for inflation to take place with a proper amount of e-folds.
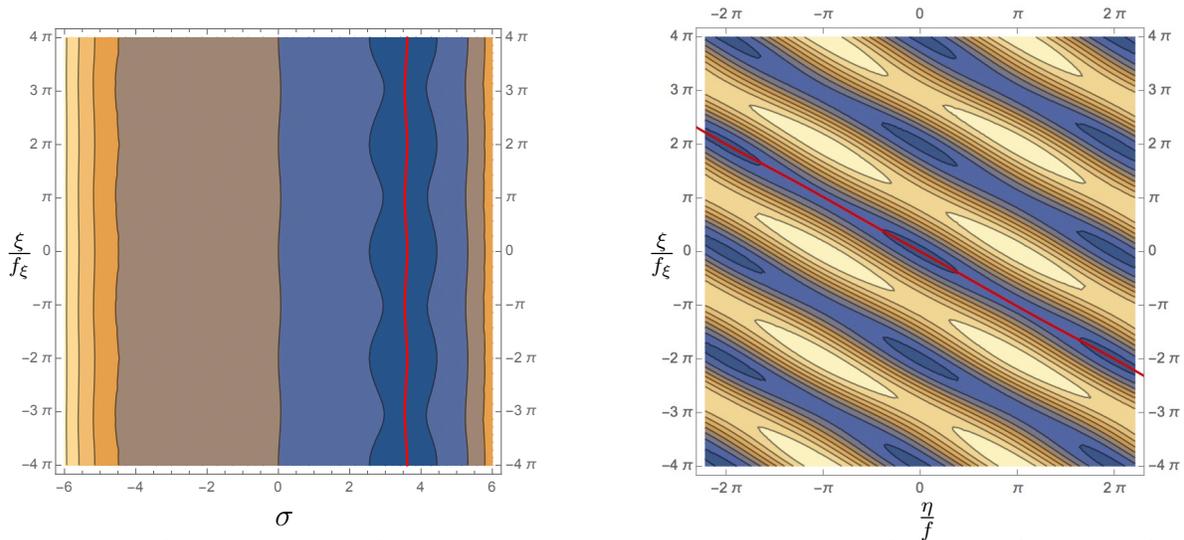
\begin{figure}[h]
\begin{center}
\begin{tabular}{c@{\hspace{0.6in}}c}
\includegraphics[scale=0.54]{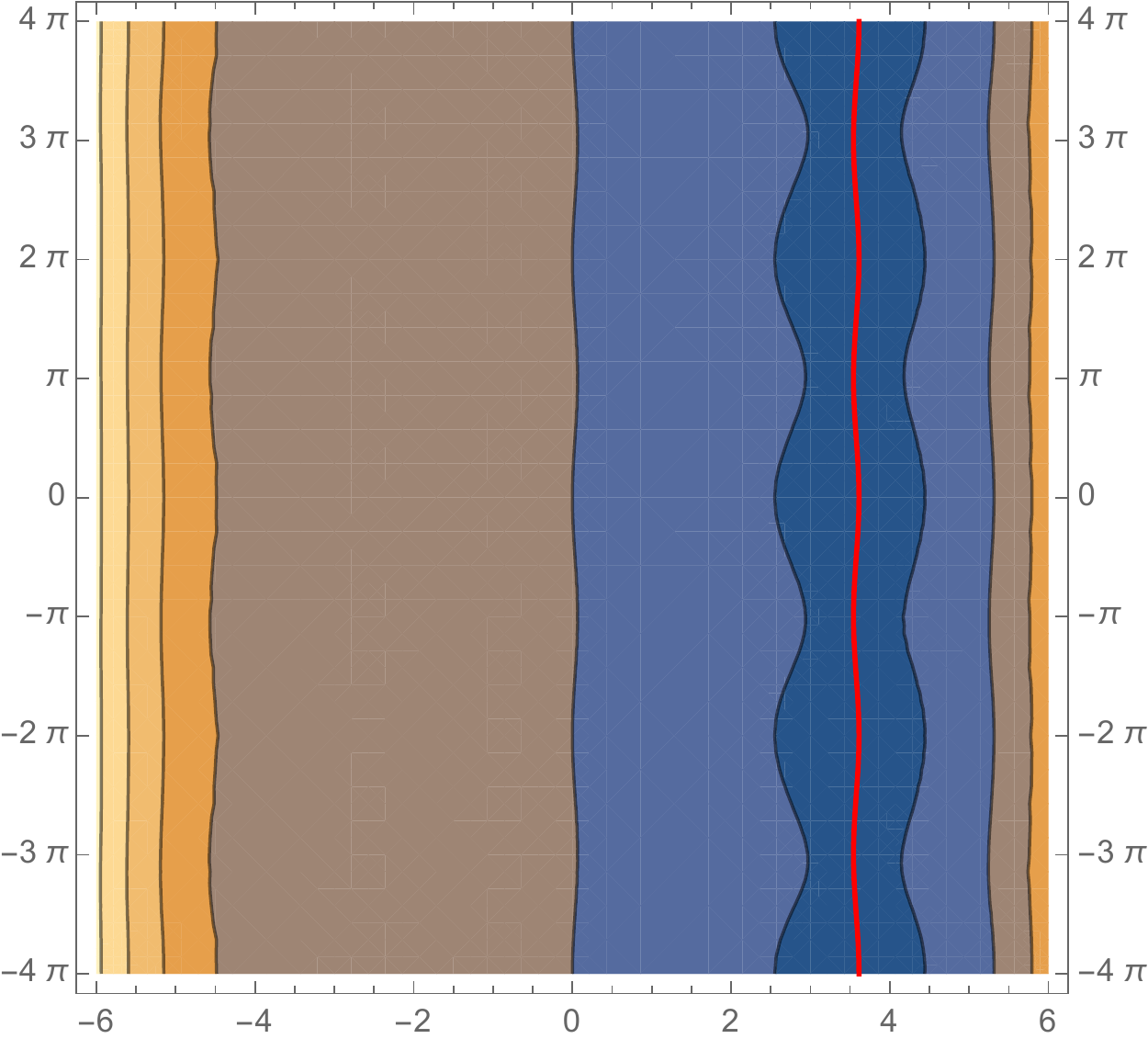} \begin{picture}(0,0) \put(-205,90){$\frac{\xi}{f_\xi}$} \put(-103,-10){$\sigma$} \end{picture}
&\includegraphics[scale=0.26]{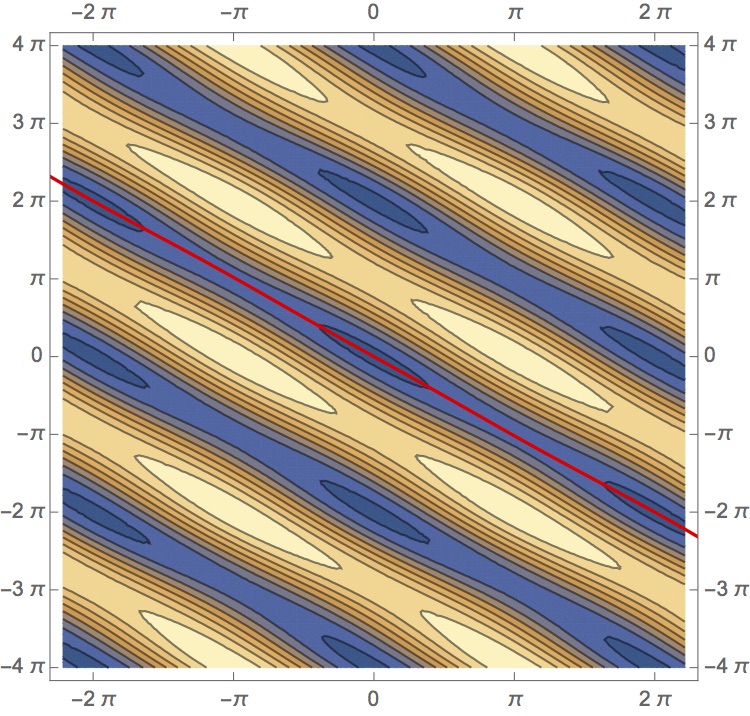}   \begin{picture}(0,0) \put(-205,90){$\frac{\xi}{f_\xi}$} \put(-103,-10){$\frac{\eta}{f}$} \end{picture}
\end{tabular}
\caption{Countour plots of the back-reacted scalar potential as a function of $(\sigma, \xi)$ (left) and $(\eta, \xi)$ (right). The valleys of the scalar potential are represented by the dark-blue regions, while the brown and orange regions correspond to steeper faces of the scalar potential. The red curve indicates the flat direction along which a potential inflationary trajectory can partly run. The back-reacted scalar potential is plotted for the parameter choice $\mu = 10 \lambda =\kappa = 10 M = 10^{16}$ GeV. \label{Fig:ContourPlotsBackRePot}}
\end{center}
\end{figure}
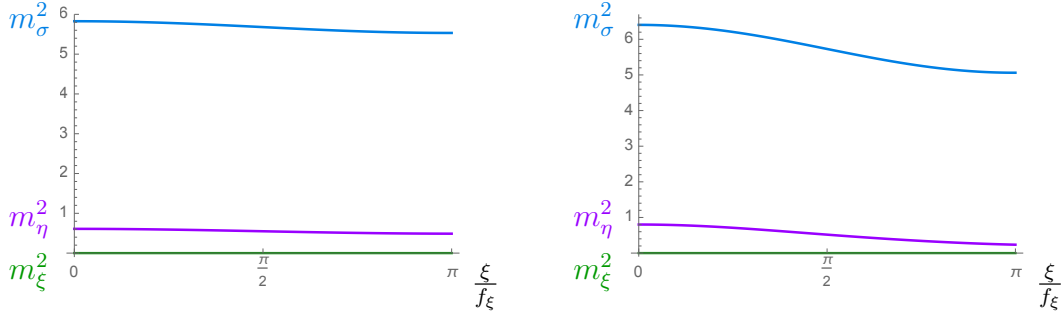
\begin{figure}[h]
\begin{center}
\begin{tabular}{c@{\hspace{0.8in}}c}
\includegraphics[scale=0.6]{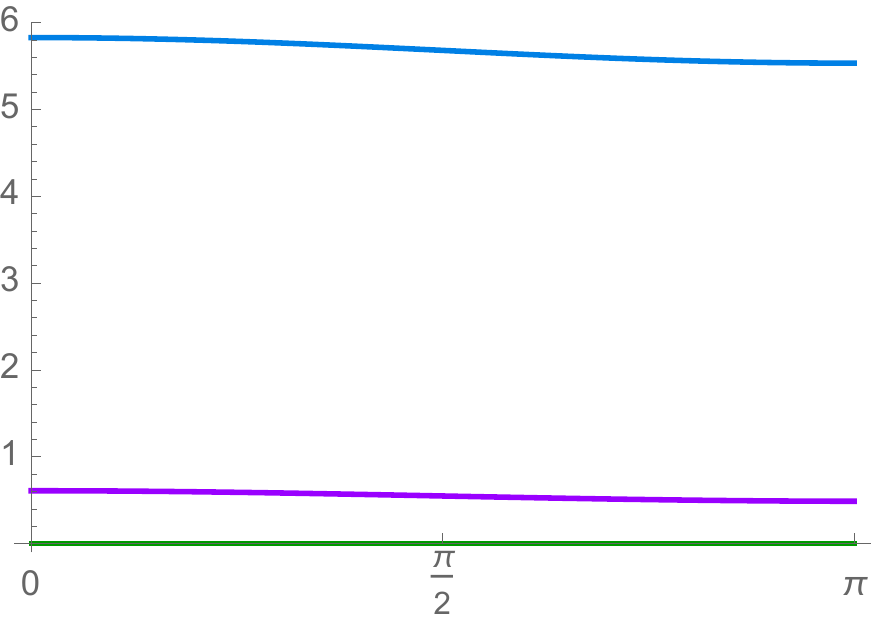} \begin{picture}(0,0) \put(0,0){$\frac{\xi}{f_\xi}$} \put(-175,100){\color{myblue}$m_{\sigma}^2$} \put(-175,25){\color{mypurple}$m_{\eta}^2$} \put(-175,5){\color{mygr}$m_{\xi}^2$} \end{picture}& \includegraphics[scale=0.6]{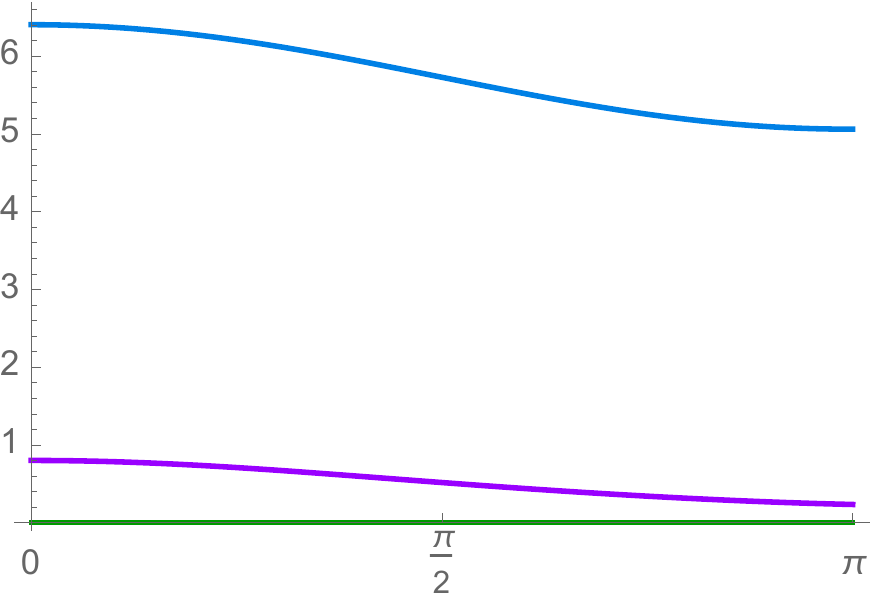} \begin{picture}(0,0) \put(0,0){$\frac{\xi}{f_\xi}$} \put(-175,100){\color{myblue}$m_{\sigma}^2$} \put(-175,25){\color{mypurple}$m_{\eta}^2$} \put(-175,5){\color{mygr}$m_{\xi}^2$} \end{picture} 
\end{tabular}
\caption{Mass-hierarchy between the infladrons $(\sigma, \eta)$ and inflaton $\xi$ (measured in units $10^{32}$ GeV$^2$) as a function of the inflaton {\it vev} for the parameter choice $\mu = 10 \lambda =\kappa = 10 M = 10^{16}$ GeV (left) and $\mu = 10 \lambda =\kappa = 2 M = 10^{16}$ GeV (right).  \label{Fig:MassSpecInTSRatio}}
\end{center}
\end{figure}

In order to be able to integrate out the heavy infladrons, it does not suffice to have control over their back-reaction effects on the scalar potential. One should in addition verify that the mass hierarchy between the inflaton $\xi$ and the infladrons persists along the full inflationary trajectory outside of the vacuum. To this end, we consider the back-reacted scalar potential and determine the eigenvalues of the associated mass matrix, which are evaluated along the inflationary trajectory. The result of this computation is given in figure~\ref{Fig:MassSpecInTSRatio}. The plots show a displacement of the infladron masses near the local maximum for $\xi$, where the inflaton mass becomes tachyonic. The functional dependence of the mass shifts on the parameter $M$ can be traced back to its presence in the linear perturbations for the infladron {\it vev}s in (\ref{Eq:LinearBackreaction}), such that larger values of~$M$ signify a larger mass shift near the local maximum for $\xi$. The infladron mass $m_{\sigma}$ remains parametrically larger than the inflaton mass along the inflationary trajectory for arbitrary points in the allowed parameter space $(M, \kappa)$, while the mass shifts in the infladron mass $m_\eta$ only remain parametrically small provided $M<\kappa$. This constraint can be deduced more easily by determining the mass spectrum from the classical, non-back-reacted scalar potential in the point $\xi = \pi f_\xi$:
\begin{equation}
m_-^2 = - 2 \Lambda^2_s \frac{ f M \kappa}{f_{\xi}^2 (\kappa-M)}, \qquad  m_+^2 = 2 \Lambda^2_s \frac{\kappa-M}{f}, \qquad  m_{\sigma}^2 = 4 f^2 \lambda + m_+^2. 
\end{equation}
These mass-relations indeed expose the tachyonic nature of the inflaton, yet they also force the infladron interactions with the gauge instanton background to be dominant over the infladron interactions with the fermionic condensate background, or equivalently $\kappa > M$, in order for the infladron mass $m_\eta$ not to become zero or tachyonic. Note that the hierarchy $f\ll f_\xi$ between the axion decay constants is paramount to arrive at this result and when the displacements (\ref{Eq:LinearBackreaction}) in the $\sigma$-direction (and thus also deviations in the decay constant $f$) remain at most of the order ${\cal O}(10^{16})$ GeV, the hierarchy $f\ll f_\xi$  perseveres along the inflationary trajectory. Hence, we can conclude that the mass hierarchy between the heavy infladrons and the light inflaton survives for all values of the classical inflaton {\it vev} due to the decay constant hierarchy. Subsequently,  the mass hierarchy allows us to integrate out the heavy infladrons through their equations of motion and determine the corrections to the inflationary potential arising from this process as an expansion in the infladron masses. The corrections of order ${\cal O}(m_\sigma^{-2})$ take the form:
\begin{equation}\label{Eq:NonRenIntSEm2} 
\Delta{\cal L}_{\frac{1}{m_\sigma^{2}}}  = \frac{m_+^2}{2 m_\sigma^2} m_-^2 f_\xi^2 \cos \frac{\xi}{f_\xi}, 
\end{equation}
and thus have the same functional dependence as the inflationary potential. This correction merely corresponds to a small correction $(m^2_+ < m_{\sigma}^2)$ to the amplitude of the cosine-potential. The (numerically relevant) corrections of order ${\cal O}(m_\sigma^{-4})$, ${\cal O}(m_+^{-4})$ and ${\cal O}(m_\sigma^{-2} m_+^{-2})$ take the respective forms:      
\begin{eqnarray}
\Delta{\cal L}_{\frac{1}{m_\sigma^{4}}} &\sim& m_-^2 f_\xi^2  \left(\sin\frac{\xi}{f_\xi}\right)^2 \left( \frac{M \kappa}{f_\xi^2} + \frac{\kappa^2}{f_\xi^2} \cos \frac{\xi}{f_\xi} \right),  \label{Eq:NonRenIntSEm41}  \\
\Delta{\cal L}_{\frac{1}{m_\sigma^{2} m_+^{2}}}  & \sim& 2  \frac{m_+^2}{m_\sigma^2}  m_-^2 f_\xi^2 \left(\sin\frac{\xi}{f_\xi}\right)^2 + 2 \frac{m_+^2}{m_\sigma^2} m_{+}^2 f^2    \left(\sin\frac{\xi}{f_\xi}\right)^2   \cos \frac{\xi}{f_\xi}  \label{Eq:NonRenIntSEm42} \\
\Delta{\cal L}_{\frac{1}{m_+^{4}}} &\sim& 3 \frac{m_+^2}{m_\sigma^2} m_{\sigma}^2 f^2    \left(\sin\frac{\xi}{f_\xi}\right)^2  \left(1-  \cos \frac{\xi}{f_\xi}\right),  \label{Eq:NonRenIntSEm43}
\end{eqnarray}
The order ${\cal O}(m_\sigma^{-4})$ is clearly negligible due to the large suppression factors $M\kappa/f_\xi^2  \sim {\cal O}(10^{-6})$ and $\kappa^2/f_\xi^2 \sim {\cal O}(10^{-6})$, while the order ${\cal O}(m_+^{-4})$ and ${\cal O}(m_\sigma^{-2} m_+^{-2})$ corrections receive a much softer suppression of the order $m_+^2/m_{\sigma}^2 \lesssim {\cal O}(10^{-1})$. However, none of the terms above present significant corrections to the scalar potential that will obstruct or destroy the inflationary trajectory. Similar considerations can be made for one-loop corrections associated to the heavy infladrons, which are equally suppressed by factors $\frac{\Lambda^4_s}{f_\xi^4} \sim {\cal O}(10^{-12})$, supporting the standard lore that the shift symmetry of the inflaton-axion fully constrains the perturbative corrections to the inflationary potential.  

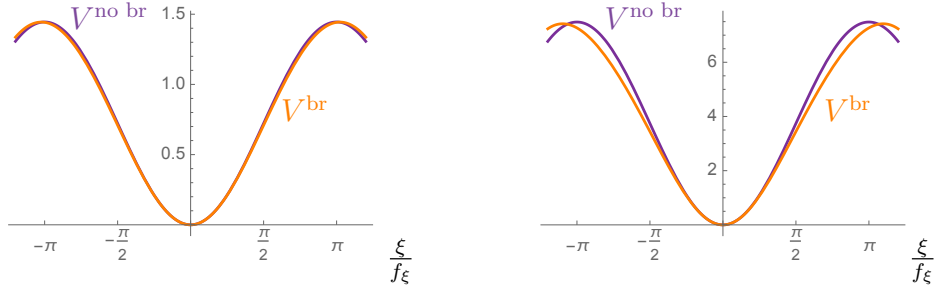
\begin{figure}[h]
\begin{center}
\begin{tabular}{c@{\hspace{0.8in}}c}
\includegraphics[scale=0.65]{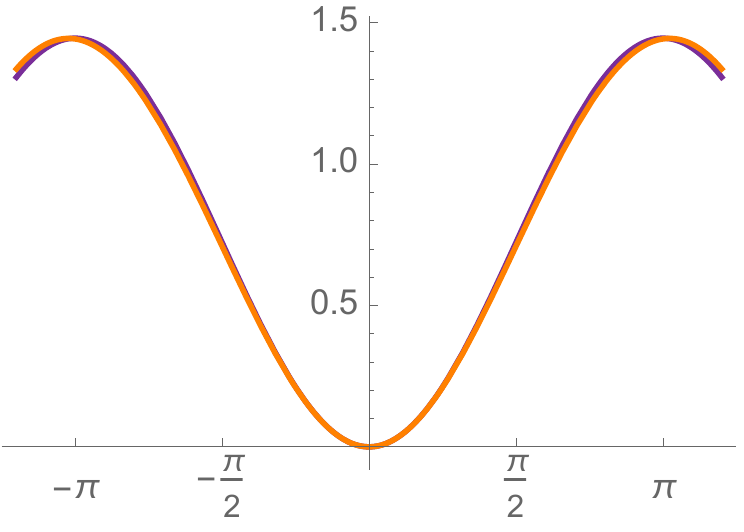} \begin{picture}(0,0) \put(0,0){$\frac{\xi}{f_\xi}$} \put(-120,90){\color{myaubergine}$V^{\text{no br}}$} \put(-40,55){\color{myorange}$V^{\text{br}}$} \end{picture}
&\includegraphics[scale=0.65]{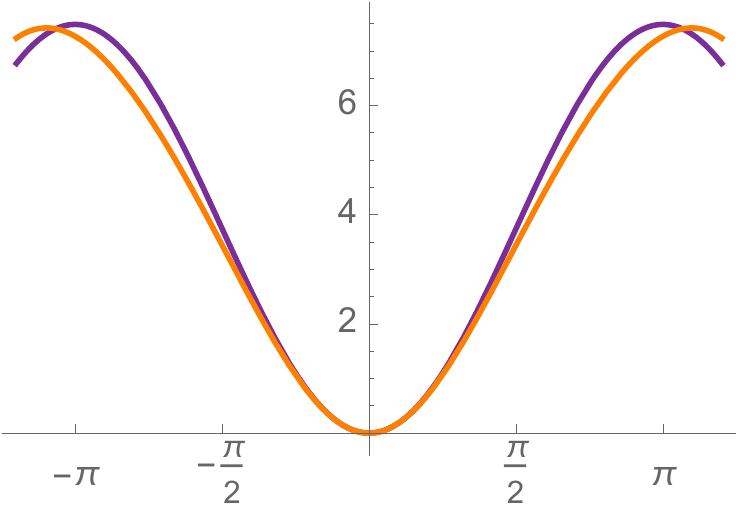}\begin{picture}(0,0) \put(0,0){$\frac{\xi}{f_\xi}$} \put(-115,90){\color{myaubergine}$V^{\text{no br}}$} \put(-32,55){\color{myorange}$V^{\text{br}}$} \end{picture}
\end{tabular}
\caption{Back-reacted scalar potential as a function of inflaton $\xi$ with parameter choice $\mu = 10 \lambda =\kappa = 10 M = 10^{16}$ GeV (left) and $\mu = 10 \lambda =\kappa = 2 M = 10^{16}$ GeV (right). The purple curve represents the non-back-reacted scalar potential $V^{\text{no br}}$, while the orange curve corresponds to the back-reacted scalar potential $V^{\text{br}}$.\label{Fig:BackReactPot}}
\end{center}
\end{figure}

\begin{figure}[h]
\begin{center}
\begin{tabular}{c@{\hspace{0.8in}}c}
\includegraphics[scale=0.6]{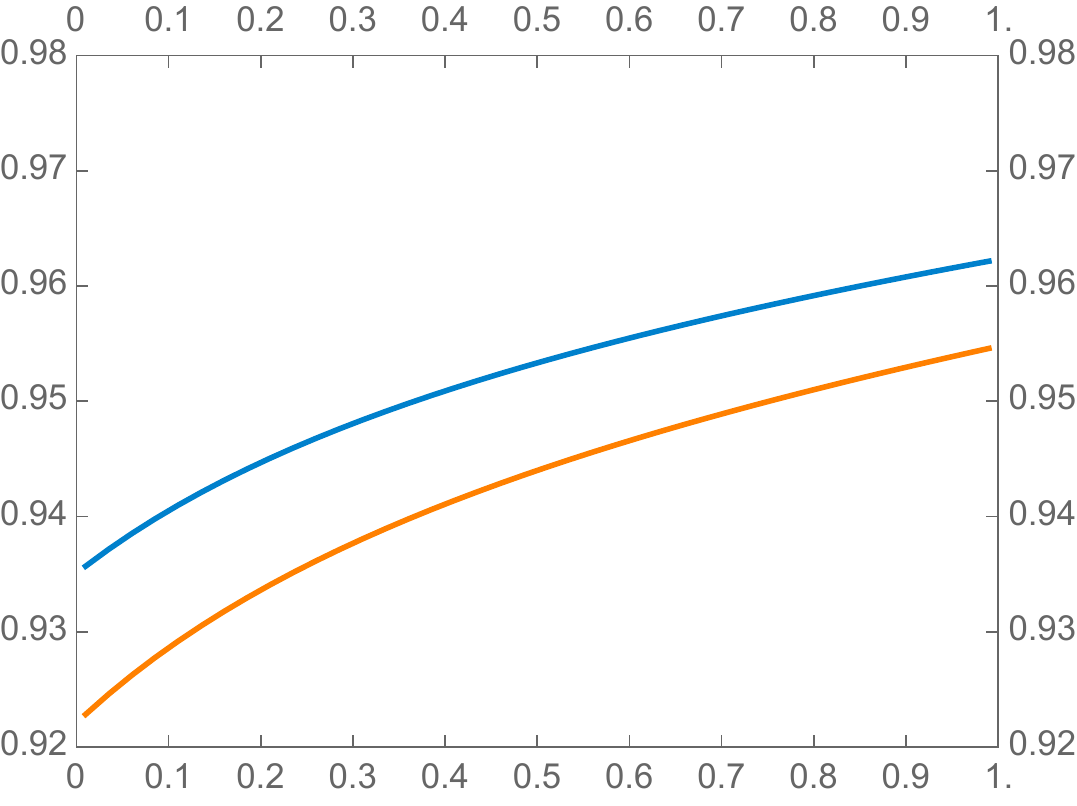} \begin{picture}(0,0) \put(-210,70){$n_s$}  \put(-105,-12){$M$} \put(-90,50){\scriptsize \color{myorange}$N_* = 50$} \put(-90,90){\scriptsize\color{myblue}$N_* = 60$}  \end{picture} 
& \includegraphics[scale=0.6]{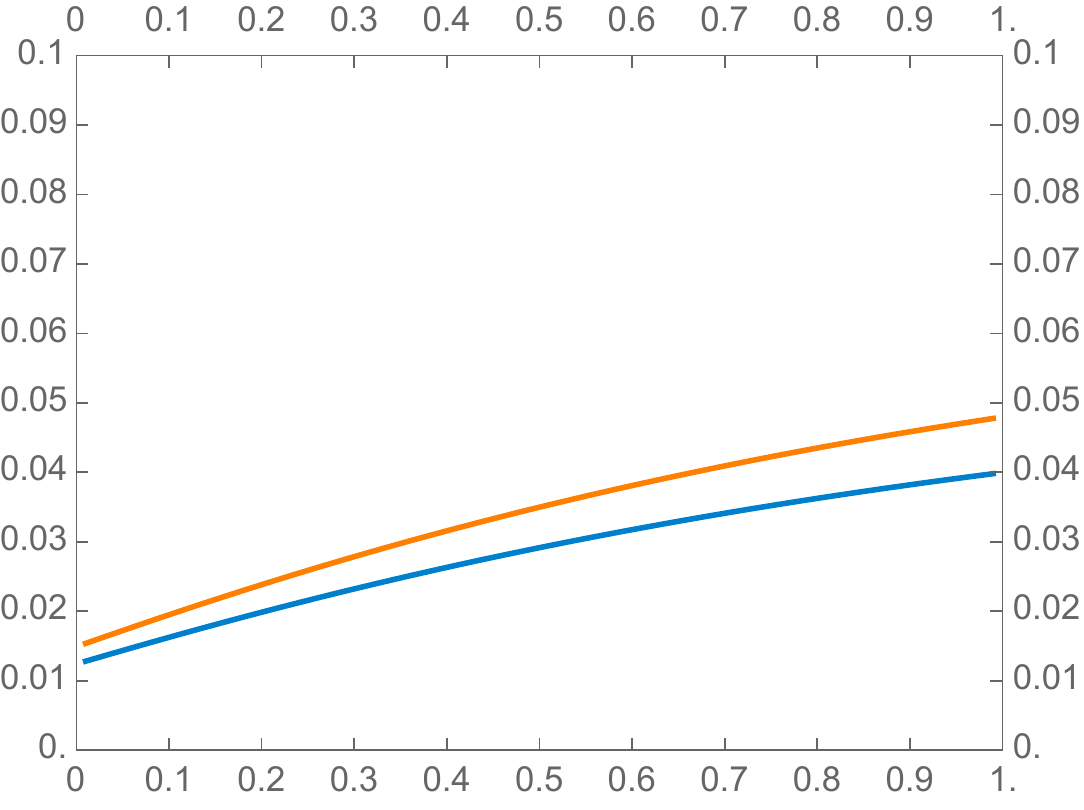} \begin{picture}(0,0) \put(-205,70){$r$} \put(-105,-12){$M$} \put(-90,65){\scriptsize \color{myorange}$N_* = 50$} \put(-90,40){\scriptsize\color{myblue}$N_* = 60$}     \end{picture} 
\end{tabular}
\caption{Spectral index $n_s$ as a function of $M$ in units of $10^{16}$ GeV (left) and tensor-to-scalar ratio $r$ as a function of $M$ in units of $10^{16}$ GeV (right) with a decay constant $f_\xi \sim 4 M_{Pl}$. The blue (orange) curve follows the evolution of the cosmological observables for $N_*=60$ ($N_*=50$) e-folds.\label{Fig:NsRvaryingM}}
\end{center}
\end{figure}
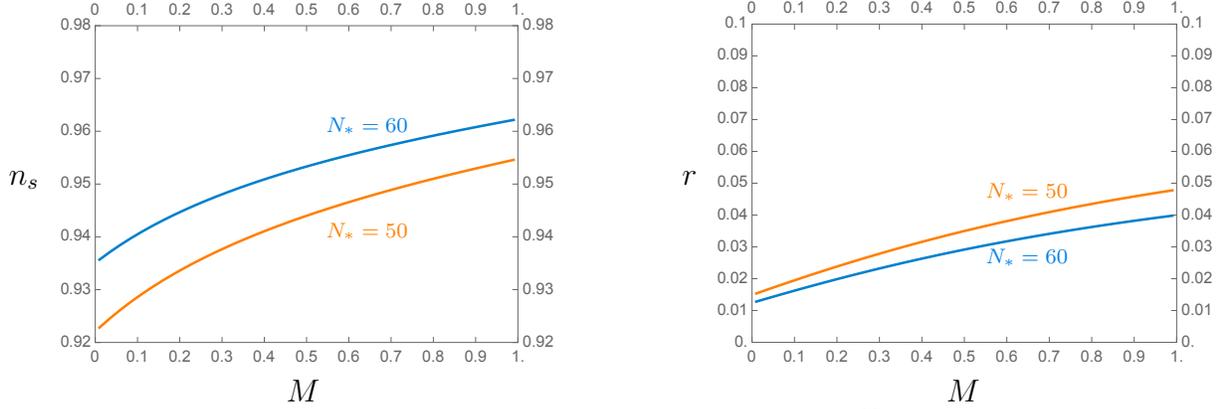

\begin{figure}[h]
\begin{center}
\includegraphics[scale=0.55]{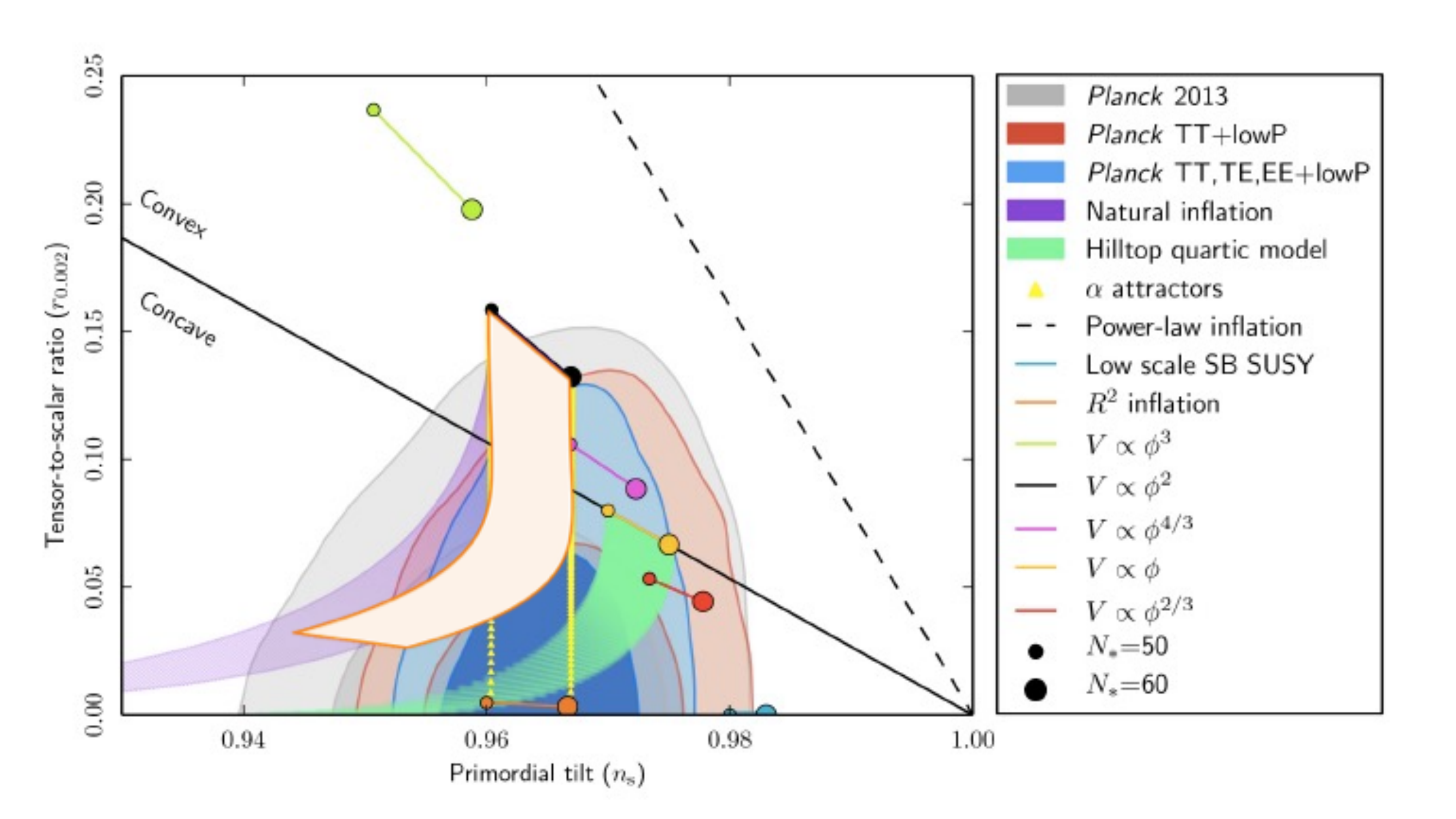}
\caption{$(n_s,r)$-plane for various inflationary models (taken from~\cite{Ade:2015lrj}) with the predictions for the back-reacted natural inflation model superimposed through the orange strip, for parameter choice $\Lambda_s=\mu = 10 \lambda =\kappa = 2 M = 10^{16}$ GeV and $\log_{10} (f_\xi/M_{Pl})$ within the prior $[0.57,2.55]$. \label{Fig:nsrplanebackreaction}}
\end{center}
\end{figure}

As the non-renormalizable corrections in (\ref{Eq:NonRenIntSEm2})--(\ref{Eq:NonRenIntSEm43}) are fully under control and form negligible corrections to the scalar potential, it suffices to consider the back-reacted scalar potential in order to discuss cosmological predictions of the single field inflationary model. To this end, we first consider the effect of the infladron back-reaction on the inflationary potential, as depicted in figure~\ref{Fig:BackReactPot}.  
The left plot signals a negligible back-reaction effect due to the smallness of the mass parameter, namely $M= 10^{15}$ GeV; but if we crank up the parameter $M$, the flattening of the scalar potential due to the infladron back-reaction becomes apparent at the maximum of the potential. The flattening in the scalar potential will also have clear repercussions for the cosmological observables, which in the case of a single field inflationary model (with standard kinetic terms) can be obtained from the canonical slow-roll parameters:
\begin{eqnarray}
\epsilon_V&\equiv&  \frac{M_{Pl}^2}{2} \left(\frac{V'(\xi)}{V(\xi)}\right)^2, \\
\eta_V&\equiv&  \frac{M_{Pl}^2}{2} \frac{V''(\xi)}{V(\xi)}, 
\end{eqnarray}      
where the potential is taken to be the back-reacted potential $V^{\text{br}}$, depending only on the inflaton $\xi$. The flattening of the scalar potential scales primarily with the parameter~$M$ and by varying its value we can follow the flattening effect on the spectral index $n_s$ and tensor-to-scalar ratio $r$, defined as:
\begin{eqnarray}
n_s &=& 1 + 2 \eta_V^* - 6 \epsilon_V^*, \\
r &=& 16 \eta_V^*,
\end{eqnarray}  
with the slow-roll parameters evaluated at the entry point of inflation $\xi=\xi_*$ along the inflationary trajectory. Figure~\ref{Fig:NsRvaryingM} shows the spectral index $n_s$ and tensor-to-scalar ratio~$r$ as a function of the mass parameter $M$, taken in the region $0.01 \times 10^{16} \text{ GeV } <M<0.99 \times 10^{16} \text{ GeV }$. We observe that an increasing back-reaction on the scalar potential, as a function of $M$, raises both the spectral index and the tensor-to-scalar ratio, bringing the model into viable patches of the $(n_s,r)$-plane, which are not in tension with current experimental data.
This observation is confirmed by looking at the $(n_s,r)$-plane in figure~\ref{Fig:nsrplanebackreaction}, with the results for the back-reacted scalar potential incorporated through the orange strip. 
To arrive at this result, we fixed the parameter $M=0.5 \times 10^{15}$ GeV, as well as all other parameters in the scalar potential, namely $\Lambda_s= \mu = 10 \lambda =\kappa = 10^{16}$ GeV, while letting the decay constant $f_\xi$ vary for a range of trans-Planckian values in the window $3.5 M_{Pl} \lesssim f_\xi \lesssim 350 M_{Pl}$.
Taking into account the back-reaction from the infladrons in case of a sufficiently large parameter $M$ brings the natural inflation strip within the 95\% confidence region of the combined observational data. Note that the infladron back-reaction effects only influence the spectral index and the tensor-to-scalar ratio, while keeping the speed of sound untouched as the kinetic terms for the inflaton $\xi$ remain canonical. In this sense, the proposed natural inflation scenario with infladrons in (\ref{Eq:EFTNL1GenSigEtaAx}) can be distinguished from the angular natural inflation with a very massive radial scalar field as discussed in~\cite{Achucarro:2015rfa}. This concludes our analysis of the inflationary prospects for the set-up in (\ref{Eq:EFTNL1GenSigEtaAx}), with the closed string axion $\xi$ playing the role of the inflaton in a consistent natural-like inflation scenario with all (perturbative) corrections under control.

\section{Phases of Inflation}\label{S:PhaseInf}
By working out the infra-red vacuum configuration for the one-generational model~(\ref{Eq:EFTNf1}) we were able to identify a natural-like inflationary model with the closed string axion~$\xi$ as the inflaton. This scenario assumes a trans-Planckian effective decay constant $f_\xi$, whose microscopic origin is tied to the axion mixing processes in the UV theory discussed in section~\ref{Ss:StrAxII} and depicted pictorially in figure~\ref{Fig:AxionModuliSpace}. More precisely, the linear combination of closed string axions coupling to the non-Abelian gauge group should not coincide with nor be perpendicular to the axionic direction that is gauged under the $U(1)$ symmetry. This leaves the question which inflationary models can be extracted from this set-up for the two limiting cases. Let us first consider the case where the axionic direction coupling anomalously to the non-Abelian gauge groups aligns with the axionic direction perpendicular to the St\"uckelberg axion, or in terms of figure~\ref{Fig:AxionModuliSpace}, where the green line coincides with the blue line. In that case, for certain isotropic regions in the complex structure moduli space the St\"uckelberg axion will not have an anomalous coupling to the non-Abelian gauge theory, such that the set-up is quantum mechanically consistent without the presence of chiral fermions to cancel gauge anomalies. Then, the model further simplifies to the first line of action~(\ref{Eq:EFTNf1}), where the remaining closed string axion retains its anomalous coupling to the gauge theory with the sub-Planckian decay constant discussed in~(\ref{Eq:SpCaseFeffPerp}). In the absence of chiral fermions, the infra-red vacuum structure originates purely from the non-Abelian gauge sector, in which case the dilute instanton gas approximation does no longer offer a proper description of the infra-red vacuum. Instead, the true vacuum for a pure Yang-Mills theory in the large $N_c$ limit is characterized by an energy with a multibranched structure~\cite{Witten:1979vv,Witten:1997ep}, such that the potential for the closed string axion is expected to take the form: 
\begin{equation}\label{Eq:YMEnergyMultiBranch}
V_{YM} (\xi) = N_c^2 \stackrel[k]{}{\rm min} {\cal H}\left( \frac{\xi + 2 \pi k f_\xi}{ f_\xi N_c} \right).
\end{equation}   
Some clarification is in order here: the vacuum of the pure Yang-Mills theory consists of many candidate vacua labeled by an integer number $k$ and whose energy profile is encoded in the multibranched function ${\cal H}$. Upon a shift of the axion $\xi \rightarrow \xi + 2\pi f_\xi$, the periodicity of the energy functional is preserved by ``jumping" from vacuum branch $k$ to $k+1$. In order to obtain the stable vacuum for each $\xi$ the energy functional ${\cal H}$ has to be minimized with respect to the vacuum integer $k$. For a CP-invariant theory, the absolute minimum is located at $\xi = 0$ and for a unique vacuum the minimum is located in the branch $k=0$. In leading order of $N_c$ one can argue that the energy functional ${\cal H}$ is quadratic, such that the axion potential emulates the basic  of chaotic inflation:
\begin{equation}    
V_{YM} (\xi)  =\; \stackrel[k]{}{\rm min} \left(  \frac{\xi + 2 \pi k f_\xi}{ f_\xi }\right)^2.
\end{equation}
As such, this configuration presents an explicit UV-realization of {\bf axion monodromy inflation} as proposed in~\cite{Kaloper:2008fb,Kaloper:2011jz}. As remarked in these papers, in order for inflation to be able to take place along one branch of the vacuum, transitions between different vacuum branches have to be suppressed, which roughly requires that the transition lifetime is much longer than the time scale of inflation. In this set-up the transition between two vacuum branches would be instigated by the nucleation of a gauge instanton bubble bounded by a domain wall and with the new vacuum inside. The tension $T_g$ of such a domain wall is expected to scale with $N_c$, while the energy difference $\Delta E$ between the two vacua is of order ${\cal O}(N_c^0)$~\cite{Shifman:1998if,Gabadadze:2002ff}. This implies that the expected decay width\footnote{Here, we are ignoring the dynamics of the field $\xi$ during the transition. Taking into account the dynamics of a rolling field can introduce further subtleties~\cite{Brown:2017wpl}.} for such a transition (evaluated in Minkowski spacetime) is exponentially suppressed with~$N_c$:
\begin{equation}
{\Gamma}_{\rm bubble} \propto  e^{- \frac{T_g^4}{(\Delta E)^3}} \sim e^{- N^4_c},
\end{equation}   
and that transitions betweens two adjacent vacuum branches are suppressed for sufficiently large $N_c$.

In the other limiting case, where the St\"uckelberg axion is the only axion that couples anomalously to the non-Abelian gauge group (or when the green line and orange line align in figure~\ref{Fig:AxionModuliSpace}), chiral fermions are still required to ensure anomaly cancelation. The action~(\ref{Eq:EFTNf1}) for a one-generational model remains valid with the anomalous coupling between the axion $\xi$ and the non-Abelian gauge group eliminated, implying that the closed string axion remains massless at low energies. In case the infra-red vacuum is still dominated by the gluodynamics of the non-Abelian gauge sector, the chiral fermions form bound states whose masses are generated dynamically. One generically assumes that the mass of the scalar infladron $\sigma$ is much heavier than the pseudo-scalar infladron~$\eta$, such that the latter infladron serves as an inflaton candidate now. Its mass and inflationary potential result primarily from the breaking of the chiral $U(1)$ symmetry by gauge instanton effects. This particular set-up is very reminiscent of the axial $U(1)_A$ symmetry breaking in QCD and we therefore expect to see similar structures in the infra-red vacuum configuration~\cite{Witten:1979vv}. This results in a vacuum with a multi-branched structure as in equation (\ref{Eq:YMEnergyMultiBranch}) upon replacing $\xi/f_\xi$ by $\eta/f$. One would be inclined to immediately adopt the considerations made above, but this would completely neglect the presence of the scalar infladron $\sigma$. We expect this more massive state to play a non-negligible role through flattening effects to the scalar potential for infladron $\eta$, leading to an {\bf axion monodromy-like} inflationary scenario. We leave the details of this inflationary scenario for future work.       

\begin{figure}[h]
\begin{center}
\begin{tabular}{ccc}
\includegraphics[width=2.4cm]{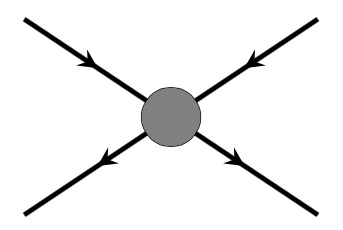} \begin{picture}(0,0) \put(-24,20){$\Gamma$}   \put(-56,20){$\Gamma$} \put(-4,20){=} \end{picture}  &  \includegraphics[width=2.4cm]{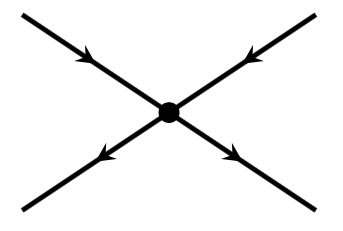}  \begin{picture}(0,0)  \put(-26,20){$\Gamma$}   \put(-56,20){$\Gamma$} \put(-4,20){+} \end{picture}   & \includegraphics[width=3.2cm]{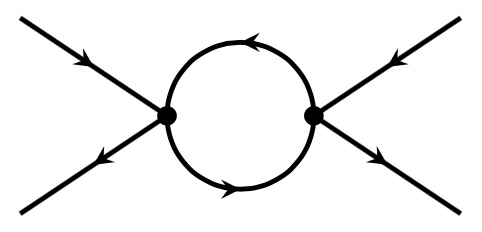}  \begin{picture}(0,0)  \put(-24,20){$\Gamma$}   \put(-78,20){$\Gamma$} \put(0,20){$+\ldots$} \put(-60,-5){$J_*(q^2)$} \end{picture}   
\end{tabular}
\caption{Perturbative expansion of the T-matrix elements $T_{0^\pm} (q^2)$ regarding the current interaction $\ov \psi \Gamma \psi$ for the scalar channel $\Gamma = {\bf 1}\, (\sigma = 0^+)$ and the pseudo-scalar channel $\Gamma = i\gamma^5\, (\pi~=~0^-)$.\label{Fig:AmplitudesTMatrixModel}}
\end{center}
\end{figure}
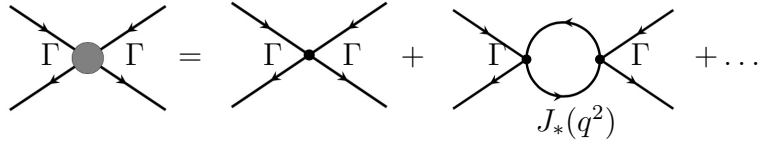

\begin{figure}[h]
\begin{center}
\vspace{0.2in}\includegraphics[scale=0.5]{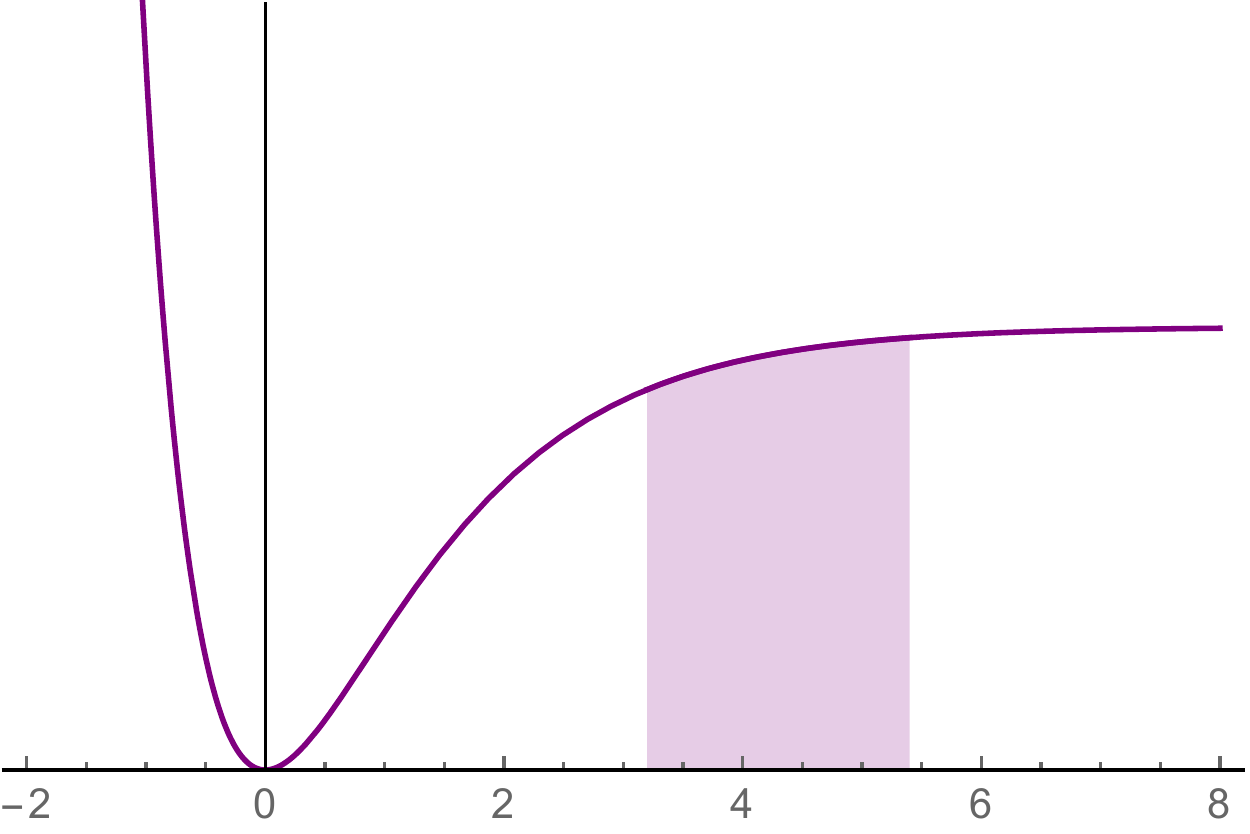} \begin{picture}(0,0) \put(0,0){$\frac{\varphi}{M_{Pl}}$} \put(-98,70){\tiny$\varphi_{\rm end}$} \put(-58,75){\tiny$\varphi_\star$} \put(-170,130){\color{myaubergine}$V_{\rm cl}$} \end{picture}
\caption{Pictorial representation of the Starobinsky-like inflationary model for $\varphi>M_{Pl}$ subject to the scalar potential in (\ref{Eq:RescaledScalarPotPhases}) with parameter choice $m = 7.20 \times 10^{11}$ GeV and $7.75 \times 10^{-11}\lesssim \lambda \lesssim 1.05  \times 10^{-10}$.  \label{Fig:NJLInflation}}
\end{center}
\end{figure}
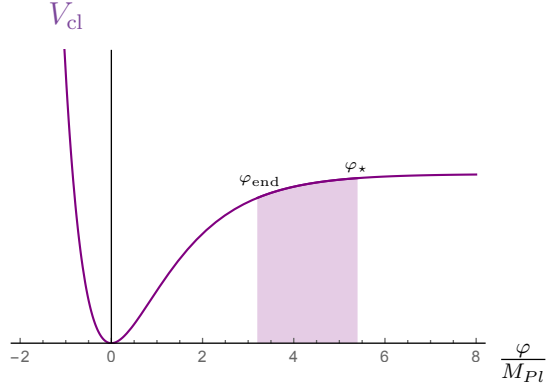

\begin{figure}[h]
\begin{center}
\includegraphics[scale=0.55]{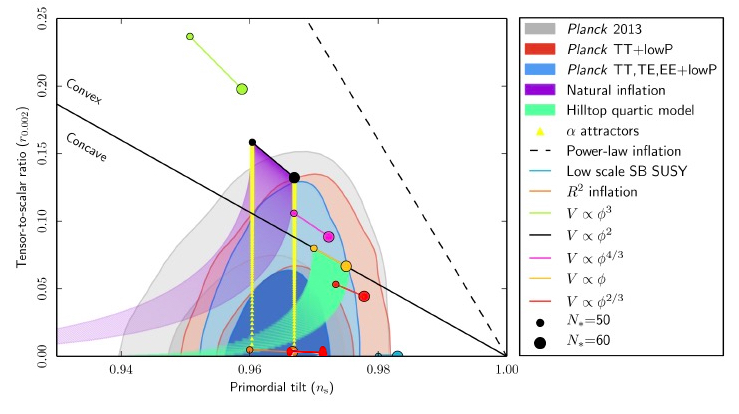}
\caption{$(n_s,r)$-plane for various inflationary models (taken from~\cite{Ade:2015lrj}) with the predictions for the Starobinsky-like inflationary model superimposed through the (lower) red line, for parameter choice $m = 7.20 \times 10^{11}$ GeV and $7.75 \times 10^{-11}\lesssim \lambda \lesssim 1.05  \times 10^{-10}$.
\label{Fig:nsrplaneStarobinsky}}
\end{center}
\end{figure}

The inflationary scenarios discussed thus far all have one element in common: a confining phase for the non-Abelian gauge such that masses for the various fundamental and composite fields are generated by strong $SU(N_c)$ dynamics in the infra-red. However, one could imagine that the non-Abelian gauge theory becomes strongly coupled at energy scales well below the Hubble scale $H_{\rm \inf}$ during inflation, due to a very weakly gauge coupling for the $SU(N_c)$ gauge theory at the string scale $M_{\rm string}$. In that case, the chiral $U(1)$ symmetry has to be broken spontaneously by the Nambu-Jona-Lasinio mechanism given the presence of four-fermion interactions, as anticipated on page~\pageref{P:NJLMech}. Translating these considerations to the one-generational model (\ref{Eq:EFTNf1}), it is obvious that the Wigner phase with $\langle \ov \psi \psi \rangle = 0$ leads to a vacuum configuration in which only the non-Abelian gauge instanton effects break the chiral $U(1)$ symmetry (explicitly). In the Nambu-Goldstone phase on the other hand, the groundstate of the theory corresponds to a non-vanishing fermion condensate $\langle \ov \psi \psi \rangle \neq 0$ that breaks the $U(1)$ symmetry spontaneously and induces a dynamically generated fermion mass 
\begin{equation}\label{Eq:NJLMassFermion}
m_\psi = - \frac{q_L q_R}{M_{st}^{2}  } \langle \ov\psi \psi \rangle .   
\end{equation}
In order for the Nambu Goldstone-phase to set in, a non-trivial solution to the mass gap equation \cite{Nambu:1961tp} needs to exist, 
\begin{equation}\label{Eq:MassGapU1}
m_\psi= |q_L q_R| N_c \frac{m_\psi}{4 \pi^2} \left[ 1 - \frac{m^2_\psi}{M_{St}^2}  \ln \left( 1 +  \frac{M_{St}^2}{m^2_\psi} \right) \right].
\end{equation}
This relation is a self-consistency relation for the dynamical mass $m_\psi\neq 0$ in the Nambu-Goldstone phase, which can easily be reproduced from (\ref{Eq:NJLMassFermion}) by evaluating the fermion condensate $ \langle \ov\psi \psi \rangle$ for a fermion with non-vanishing mass $m_\psi$. A non-trivial solution to~(\ref{Eq:MassGapU1}) exists for a dynamical mass $m_\psi < M_{\rm St}$ below the St\"uckelberg mass scale (as the natural cut-off scale), provided that the gauge coupling of the $U(1)$ gauge theory exceeds a critical value,    
\begin{equation}\label{Eq:U1NJLCon}
\exists\, \Lambda_{\rm comp} \lesssim M_{St}: \qquad \alpha_{U(1)} (\Lambda_{\rm comp}) \gtrsim \frac{\pi}{N_c},
\end{equation}
which is equivalent to the condition in (\ref{Eq:NJLU1GaugeCond}) with $n_F=1$. One might be worried that the condition following from (\ref{Eq:MassGapU1}) is only valid at tree-level and that a more rigorous condition has to be imposed in order for the non-trivial vacuum to persist in the fully quantum theory. However, by minimizing the one-loop effective potential for the N-JL model in the large $N_c$-limit~\cite{Gross:1974jv,Dudas:1993cj,Miransky:1994vk}, it can be shown that the non-trivial vacuum $\langle \ov \psi \psi \rangle \neq 0$ corresponds to the true minimum of the theory, as long as the $U(1)$ gauge coupling condition (\ref{Eq:U1NJLCon}) is satisfied. In the Nambu-Goldstone phase, the particle excitations of the vacuum $\langle \ov \psi \psi \rangle \neq 0$ are no longer described in terms of fermionic modes, but rather in terms of collective bosonic modes. From the action (\ref{Eq:EFTNf1}) one straightforwardly deduces the emergence of a scalar or $0^+$ mode $\sigma$ ($\sim \ov\psi \psi$) and a pseudo-scalar or $0^-$ mode $\pi$ ($\sim \ov\psi i\gamma^5 \psi$). 
The standard way to determine the masses for these two bound states consists in solving the Bethe-Salpeter equation~\cite{Salpeter:1951sz} with respect to the current interactions $\ov \psi \Gamma \psi$ with $\Gamma = \{{\bf 1}, i \gamma^5\}$ for the scalar and pseudo-scalar mode respectively. Alternatively, a more modern approach consists in looking for poles $q^2 = m_{0^\pm}^2$ in the four-fermion correlation functions $\langle \ov \psi \Gamma \psi (q)  \ov \psi \Gamma \psi (-q)  \rangle_{NG}$ evaluated in the NG-phase, which are interpreted as effective exchanges of bosonic bound states between external fermionic legs. Thus, it suffices to reconstruct the T-matrix element for the respective current interaction up to one-loop order:
\begin{equation}
T_{0^\pm} (q^2) = K_{0^\pm} + K_{0^\pm} J_{0^\pm} (q^2)  K_{0^\pm} + \ldots,
 \end{equation}
 where $K_{0^\pm} = |q_L q_R|/M_{st}^2$ represents the tree-level interaction (Born term) and $J_{0^\pm}(q^2)$ the one-loop contribution. A pictorial representation of the T-matrix up to one-loop is given in figure \ref{Fig:AmplitudesTMatrixModel}.
A closer look at the T-matrix $T_{0^\pm}(q^2)$ reveals that the poles appear in regions of momentum space where the following relations are satisfied:
\begin{equation}\label{Eq:ConditionPoles}
K_{0^\pm} J_{0^{\pm}} (q^2 = m^2_{0^{\pm}}) = 1.
\end{equation}
An explicit computation of the one-loop factors $J_{0^\pm}(q^2)$ yields the following result:
\begin{equation}\label{Eq:OneLoopFactor}
J_{0^\pm} (q^2) = - \frac{\langle \ov \psi \psi \rangle}{m_\psi}  - \frac{1}{2} \left\{ \begin{array}{c} \vspace{0.1in}  q^2 - 4 m^2_\psi  \\  q^2  \end{array} \right\} I(q^2), 
\end{equation}
where the momentum-dependent factor $I(q^2)$ is given by:
\begin{equation}\label{Eq:FIPoleStructure}
I(q^2) = 4 N_c i \int \frac{d^4 p}{(2\pi)^4} \frac{1}{\left[(p+q)^2 - m^2_\psi + i\varepsilon\right] \left[ p^2 - m^2_\psi + i\varepsilon\right]}.
\end{equation}
Imposing the conditions (\ref{Eq:ConditionPoles}) then reveals the masses $m_{\sigma}^2 = 4 m^2_\psi$ and $m_{\pi}^2 = 0$ for the bosonic bound states $\sigma$ and $\pi$ respectively. Hence, in case the four-fermion interactions in our model~(\ref{Eq:EFTNf1}) allow for a Nambu-Goldstone phase, its low-energy effective field theory is described in terms of a massive composite scalar $\sigma$, a massless composite pseudo-scalar $\pi$ and a massless closed string axion $\xi$. At that point, the infladron $\sigma$ represents the most suitable inflaton candidate, whose inflationary potential is extracted from the effective action upon integrating out the fermionic degrees of freedom from the path integral. To arrive at the effective potential, one first rewrites the four-fermion couplings in terms of Yukawa interactions involving the bound states $(\sigma,\pi)$, analogously to the discussion surrounding action (\ref{Eq:BosonMethod}). As a next step, one integrates over the fermionic degrees of freedom in the path integral, which produces a one-loop effective potential, whose divergent behaviour can be regularized by imposing that the ground state of the scalar potential reproduces the gap-equation~(\ref{Eq:MassGapU1}). Following these steps meticulously, one finds a linear sigma-model in terms of the bound states,
\begin{eqnarray}\label{Eq:NJLHYEquiv}
{\cal L}_{\rm lin} &=&  \frac{1}{2} g^{\mu \nu} \big( Z_\sigma \partial_\mu  \sigma \partial_\nu \sigma + Z_\pi  \partial_\mu  \pi \partial_\nu \pi \big)  + \frac{1}{2} m^2 \big(\sigma^2 + \pi^2\big)- \frac{\lambda}{4} \big(\sigma^2 + \pi^2\big)^2 \\
&&  - \frac{1}{64 \pi^2} \left( -m^2 + 3 \lambda (\sigma^2 + \pi^2)  \right)^2 \left[ \log\left(\frac{-m^2 + 3 \lambda (\sigma^2+\pi^2)}{\Lambda^2}\right)  -\frac{3}{2}\right]  - \frac{1}{2} \varpi R \big(\sigma^2 + \pi^2\big), \notag
\end{eqnarray}
whose wave function $(Z_\sigma, Z_\pi)$ and parameters ($m^2$, $\lambda$) are subject to the standard RGE's. The non-minimal coupling of the infladrons to gravity is equally a result of radiative corrections due to coupling of the fermions to gravity as depicted in the left diagram of figure~\ref{Fig:NonMinimalCoupling} and needs to be taken into account when considering this model as an inflationary set-up. Furthermore, by virtue of a one-loop RGE analysis one shows~\cite{Hill:1991jc} that the model flows to an attractive IR fixed point $\varpi = \frac{1}{6}$. Given that only the infladron $\sigma$ acquires a mass in the Nambu-Goldstone phase, it suffices in first order to consider the classical motion along the $\sigma$-direction and forget the pseudo-scalar field by ignoring potential isocurvature perturbations in the CMB, which a more complete analysis of the cosmological scenario certainly has to take into account. Due to the non-minimal coupling, a conformal transformation on the metric has to be performed to bring the set-up back to the Einstein frame, with conformal factor $\Omega^2 = 1 + \frac{\varpi\, \sigma^2}{M_{Pl}^2}$. As a result also the infladron $\sigma$ will undergo a field redefinition $\sigma \rightarrow \varphi (\sigma)$ dictated by the differential equation,
\begin{equation}\label{Eq:TransSigPhiNonMinCoup}
\left(\frac{d\varphi}{d\sigma}\right)^2 = \frac{1}{\Omega^2} \left( 1 + \frac{6 \varpi^2 \sigma^2}{M_{Pl}^2 \left( 1 + \frac{\varpi \sigma^2}{M_{Pl}^2} \right)} \right). 
\end{equation} 
Under the conformal transformation of the metric, the scalar potential will also undergo a rescaling, such that the effective potential for the infladron $\sigma$ takes the form: 
\begin{equation}\label{Eq:RescaledScalarPotPhases}
V_E (\sigma) = \frac{1}{\Omega^4(\sigma)} \left( V_{tree} (\sigma) + V_{one-loop} (\sigma) \right) , \qquad  V_{tree} (\sigma) = \frac{\left( m^2 - \lambda \sigma^2 \right)^2}{4 \lambda}.
\end{equation}
In order to proceed and identify a satisfactory inflationary scenario, we consider the regime where the classical inflationary motion of the infladron $\sigma$ occurs at {\it trans-Planckian} field displacements away from the minimum. In that case the right part of the expression between brackets in~(\ref{Eq:TransSigPhiNonMinCoup}) is dominant and the field redefinition is determined by, 
\begin{equation}
\frac{d\varphi}{d\sigma} = \frac{1}{\Omega^2} \frac{\sqrt{6} \varpi\, \sigma }{ M_{Pl} } \quad \rightsquigarrow \quad  \sigma(\varphi) =\langle \sigma \rangle +  \frac{M_{Pl}}{\sqrt{\varpi}}  \sqrt{e^{\sqrt{\frac{2}{3}} \frac{\varphi}{M_{Pl}}} - 1},
\end{equation}
where we included the infladron {\it vev} in the field redefinition to describe the classical inflationary motion in terms of the field $\varphi$ about the minimum $\langle\varphi \rangle = 0$. Inserting the expression for infladron $\sigma (\varphi)$ back into the classical potential (\ref{Eq:RescaledScalarPotPhases}) yields an inflationary potential for $\varphi$ that matches precisely the scalar field description~\cite{Baumann:2014nda} of {\bf Starobinsky-like inflation}:
\begin{equation}
V(\varphi) = \frac{M_{Pl}^4 \lambda}{4 \varpi^2} \left( 1 - e^{-\sqrt{\frac{2}{3}} \frac{\varphi}{M_{Pl}}} \right)^2.
\end{equation}
In this regime, inflation occurs along a plateau-like part of the scalar potential where the scalar field $\varphi$ takes on field values larger than the Planck-scale $M_{Pl}$ during the last 50-60 e-foldings of inflation, as depicted schematically in figure~\ref{Fig:NJLInflation}.
At the IR fixed point $\varpi = \frac{1}{6}$ the quartic coupling $\lambda$ has to lie in the window $ 7.75 \times 10^{-11} \lesssim \lambda \lesssim 1.05 \times 10^{-10}$ in order for the scalar perturbations amplitude to be of the order $\Delta_{{\cal R}}^2 \sim 2.196 \times 10^{-9}$, with the lower bound set for $N_\star=60$ e-folds and the upper bound for $N_\star=50$ e-folds. The predictions for the cosmological observables $(n_s,r)$ have been superimposed through a red line onto the $(n_s, r)$ diagram in figure~\ref{Fig:nsrplaneStarobinsky} taken from the Planck data. The corresponding field range for the inflationary trajectory is $3.2 M_{Pl} < \varphi < 5.58 M_{Pl}$ for $N_\star =60$ e-folds and $3.2 M_{Pl} < \varphi < 5.40 M_{Pl}$ for $N_\star =50$ e-folds. For these parameter choices the Hubble scale during inflation is of the order $H_\star \sim  1.30 \times 10^{13}$ GeV, while the energy scale of inflation is set at $V_{\rm inf}^{1/4} \sim 8.80 \times 10^{15}$ GeV.  

In our ``rudimentary" derivation of the inflationary potential and the cosmological observables, the non-Abelian gauge theory has been completely ignored in favour of N-JL type four-fermion interactions allowing for a Nambu-Goldstone phase.
Taking the non-Abelian gauge interactions into account, one can use the equivalence between the gauged N-JL model and a gauged Higgs-Yukawa model~\cite{Bardeen:1989ds,Harada:1994wy}, provided a compositeness condition holds in the gauged N-JL model that translates to the boundary conditions (imposed at the compositeness scale) on the parameters in the gauged Higgs-Yukawa model. Based on this equivalence, inflationary models involving the scalar excitation~$\sigma$ were studied in~\cite{Inagaki:2015eza,Inagaki:2016vkf,Inagaki:2017ymx} in the language of the gauged Higgs-Yukawa model and with a RGE-improved effective potential in the same spirit as the discussion in section~\ref{Sss:InfQM}. For a finite compositeness scale, the authors were able to extract inflationary models interpolating between chaotic inflation and Starobinsky or Higgs-inflation. The equivalence between the N-JL four-fermion interactions and the Higgs-Yukawa system only holds at weak coupling for the non-Abelian gauge theory. The weak coupling assumption does not automatically mean that the intricacies of the non-Abelian field configurations are completely harmless. More precisely, one should verify under which conditions the mass-generating effects of the non-trivial $\theta$-vacuum can be discarded, such that the infladron $\pi$ remains (sufficiently) massless. As reviewed in section~\ref{Ss:ChiralSymBreaking}, the $SU(N_c)$ vacuum structure is accompanied by instanton interactions that explicitly break the chiral $U(1)$ symmetry at one-loop and conspire with the fermion condensate to generate a mass for the pseudo-scalar boson:
\begin{equation}\label{Eq:pionmass}
m_{\pi}^2 f_{\pi}^2 = - \kappa  \langle \ov \psi \psi \rangle,
\end{equation} 
where $ f_\pi  \equiv \langle\sigma\rangle$ corresponds to the decay constant for the pseudo-scalar $\pi$ and $\kappa$ represents the $U(1)$-breaking parameter (\ref{Eq:MassRelationGaugeCond}) tied to the gauge instantons for a one-generational model in the dilute instanton gas approximation. The mass-relation for the infladron $\pi$ can be computed~\cite{Guralnik:1967zz,Cheng:1985bj,Weinberg:1996kr} by virtue of the $U(1)$ current properties in vacua with spontaneous symmetry-breaking. Consequently, also the mass for the scalar excitation is lifted to $m_{\sigma}^2 \simeq 4 M^2_\psi + m_{\pi}^2$ due to the presence of gauge instantons. The mass uplift can be checked explicitly for instance from the one-loop effective potential arising from integrating out the fermions in the Nambu-Goldstone phase. Note that the fermion mass $M_\psi$ consists of two separate contributions, i.e.~$M_\psi = m_\psi + \kappa$, when the effects of the $\theta$-vacuum are properly taken into consideration. The bosonic bound states continue to appear as poles in the four-fermion correlators, from which we can equally derive the mass relations for the composite scalars:
\begin{equation}
m_\pi^2 = - \frac{2 \kappa}{M_\psi I(m_\pi^2) K_{0^-}}, \qquad m_{\sigma}^2 = 4 M_\psi^2 - \frac{2\kappa}{M_\psi \RE(I(m_\sigma^2)) K_{0^+}}.
\end{equation}
When considering the momentum limit $q^2 \rightarrow m_\sigma^2 $ for the scalar channel in equation~(\ref{Eq:OneLoopFactor}) the momentum-dependent factor $I(q^2)$ in~(\ref{Eq:FIPoleStructure}) develops a complex component ${\IM}(I(m_\sigma^2)) \neq 0 $ as well, implying that the $\sigma$-excitation can no longer be considered stable. If we imagine the $\sigma$-bound state to be a narrowly peaked resonance, we can deduce its narrow decay width $\Gamma_\sigma$ from the location of the pole in (\ref{Eq:OneLoopFactor}) for the scalar channel:
\begin{equation}
\Gamma_\sigma = \frac{\kappa}{m_\sigma M_\psi} \left|\frac{{\IM}(I(m_\sigma^2))}{{\RE}(I(m_\sigma^2))}\right|.
\end{equation}   
By expanding the function $I(q^2)$ in (\ref{Eq:FIPoleStructure}) around the pole $q^2 = m_{\sigma}^2$, we can extract the real and imaginary part to fix the scaling (to first order) of the decay width with the scales in the model:
\begin{equation} 
\Gamma_\sigma \propto \frac{\pi}{2} \frac{\kappa^{3/2} M_{st}}{m_\sigma M_\psi^{5/2}}.
\end{equation}
The inverse of the decay width measures the mean lifetime of the infladron $\sigma$ and by demanding that it is longer than the time scale of inflation, an upper bound on the strong coupling scale $\Lambda_s \leq 6.21 \times 10^{29} \text{ GeV}$ in (\ref{Eq:MassRelationGaugeCond}) can be deduced for a St\"uckelberg mass scale $M_{st} \sim 10^{16}$ GeV. As this upper bound exceeds the Planck mass $M_{Pl} \sim 2.4 \times 10^{18}$ GeV, the cut-off scale for quantum gravity, there is no inherent stability problem for the infladron $\sigma$. As long as the infladron mass scale $M_\psi$ and the Hubble scale in this scenario are sufficiently high with respect to the strong coupling scale, they ensure a very narrow decay width $\Gamma_\sigma$, such that the infladron $\sigma$ can serve effortlessly as an inflaton candidate without the risk of decaying before the end of the inflationary epoch.  

\begin{figure}[h]
\begin{center}
\vspace*{0.4in}
\includegraphics[scale=0.7]{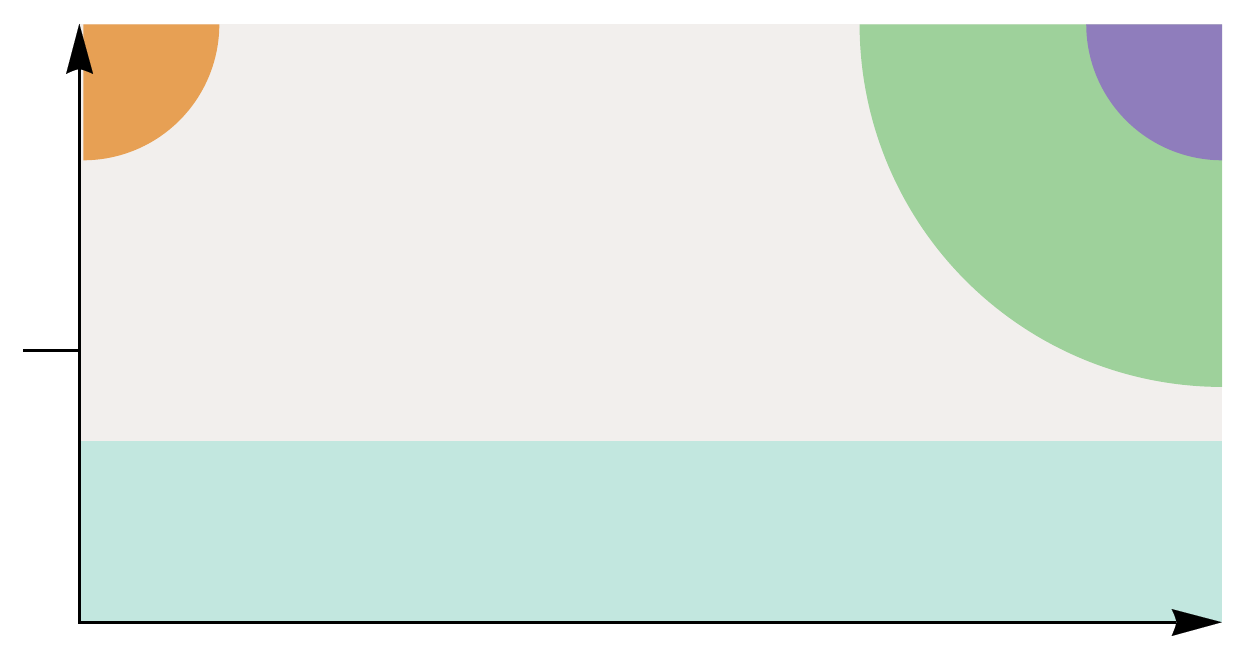} \begin{picture}(0,0) \put(0,0){$\log_{10}\frac{f_\xi}{f}$}  \put(-280,130){$\alpha_{SU(N_c)}$}  \put(-286,58){$\alpha = 1$} 
\put(-230,120){\begin{rotate}{40}$\eta$-axion\end{rotate}} \put(-230,100){\begin{rotate}{40}monodromy\end{rotate}}
\put(-30,140){\begin{rotate}{-40}$\xi$-axion\end{rotate}} \put(-30,120){\begin{rotate}{-40}monodromy\end{rotate}}
\put(-60,110){\begin{rotate}{-40}$\xi$-natural\end{rotate}} \put(-60,90){\begin{rotate}{-40} inflation\end{rotate}}
\put(-200,60){multi-field inflation} \put(-200,20){N-JL inflationary models for $\sigma$}
\end{picture}
\caption{Schematic phase diagram for inflationary models arising as infra-red theories from the string inspired action~(\ref{Eq:EFTNf1}).\label{Fig:PhaseDiagInf}}
\end{center}
\end{figure}
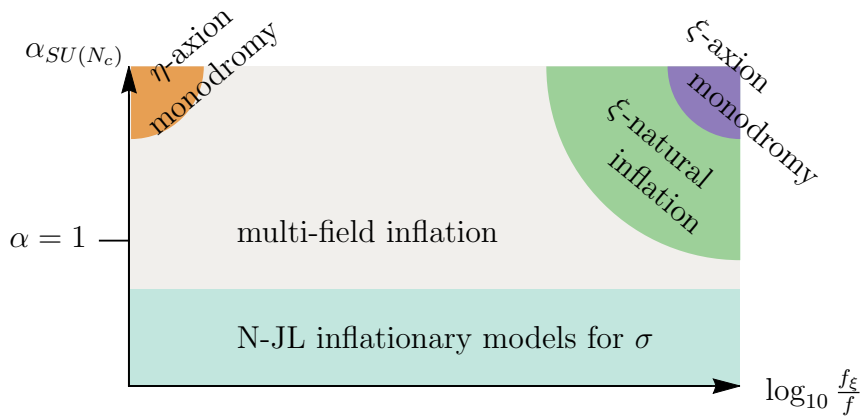

In summary, the four-dimensional effective field theory in (\ref{Eq:EFTNf1}) gives rise to different types of inflationary models, depending on the effective decay constants for the closed string axions and on the properties of the infra-red vacuum. The variety of inflationary models can be represented schematically in a phase diagram spanned by the ratio $f_\xi/f$ of the decay constants and the gauge coupling $\alpha_{SU(N_c)}$ of the non-Abelian gauge theory, as depicted in figure~\ref{Fig:PhaseDiagInf}. In this section, we have elaborated various corners of the phase diagrams, where explicit single-field inflationary models have been identified based on the phase of the non-Abelian gauge theory or the fermionic ground state in the infra-red. The largest area of the phase diagram is, however, taken by two- or three-field inflationary models, whose analysis is postponed for future research.

\section{Conclusion}\label{S:con}
In this paper, we work out the effective field theory description for chiral gauge theories coupled to closed string axions as they arise from the dimensional reduction of Type~II superstring theory with D-branes. The four-dimensional UV theory is characterized by a non-Abelian gauge group, an Abelian $U(1)$ gauge group, a set of chiral fermions transforming in the bifundamental representation under both gauge groups and a pair of closed string axions charged under the Abelian gauge group via St\"uckelberg couplings. Below the St\"uckelberg mass scale one of the closed string axions is turned into the longitudinal mode of the massive Abelian gauge boson, while the remaining closed string axion $\xi$ retains an anomalous coupling to the non-Abelian gauge group. Integrating out the massive $U(1)$ gauge boson induces four-fermion interactions among the chiral fermions, which are supplemented at strong non-Abelian coupling by the higher-order fermion interactions induced by the gauge instantons through the 't Hooft operator. Below the strong coupling scale, the fermionic and gauge degrees of freedom are no longer expected to appear as free asymptotic states, but instead be subject to the process of confinement. As the  chiral fermions confine in fermion-antifermion bound states, their excitations about the non-trivial vacuum can be captured by scalar bound states dubbed {\it infladrons}. In this IR-description, the masses for the infladrons and remaining closed string axion arise dynamically through the interactions with background gauge instantons and fermion condensate, which both lead to a breaking of the global remnant of the chiral $U(1)$ symmetry. 

Once the dynamical properties of the IR vacuum are known, the form of the low-energy effective action can be accordingly reconstructed by means of EFT tools to describe the fluctuations about the vacuum. Subsequently, the inflaton candidate is played by the lowest-lying massive state in the spectrum and is subject to an effective scalar potential compatible with the $U(1)$ breaking effects identified in the UV. The UV-richness of the string theory inspired set-up~(\ref{Eq:GeneralLagrangianN2}) implies an equally rich IR vacuum and thus a variety of various inflationary models. More precisely, for field theories where the IR vacuum structure is dominated by the non-Abelian gauge interactions, one can identify natural-like and monodromy-like inflationary scenarios with the inflaton being identified with one of the axions in the theory, be it a closed string axion or the phase of a chiral fermion bound state. In case the four-fermion interactions allow for a Nambu-Goldstone phase below the St\"uckelberg scale, the IR vacuum structure is rather the one of a Nambu-Jona-Lasinio model and the role of the inflaton is played by the scalar component of the chiral fermion bound state, which is subject to a Starobinsky-type potential. The variety of identifable inflationary models results in the phase diagram in figure~\ref{Fig:PhaseDiagInf}, where all the single field inflationary models have been located according to the values of the decay constants and the phase of the non-Abelian gauge theory. The regions of the phase diagram characterized by multi-field inflationary models is left for future research.

Apart from extracting the inflationary capabilities of the string inspired set-up~(\ref{Eq:GeneralLagrangianN2}), it is also crucial to have a full handle on all kinds of corrections to the IR theory that can distort the vacuum configuration. This is precisely the motivation behind section~\ref{S:EFTInfAx}, in which we work out the details for the natural-like inflationary scenario through the effective field theory approach in the IR and investigate how perturbative and non-perturbative corrections can compromise the inflationary scenario. True to the spirit of EFT one can estimate the order of magnitude for each non-renormalizable correction involving the infladrons and accordingly decide which corrections have to be taken into consideration based on the desired accuracy level for the computable observables. With the closed string axion as the inflaton and the infladrons as more massive states, many of the non-renormalizble corrections receive an additional suppression due to the parametrically large separation between the closed string axion decay constant and the cut-off scale of the EFT. Furthermore, higher-loop perturbative corrections involving the infladrons equally have to be taken computed to assure that they do not lift the vacuum configuration for the infladrons. Yet, an accurate resummation of the $n$-loop corrections follows from solving the renormalization group equation for the effective potential and reveals the same polynomial structure in the infladron fields as the classical potential. Hence, the classical vacuum structure persists also in the quantum theory provided that the radial infladron $\sigma$ does not undergo large displacements from the vacuum during inflation. The small displacement requirement also ensures a trivial field redefinition for the infladrons when considering the non-minimal coupling to gravity for these composite states. To investigate whether the radial infladron undergoes large excursions during the inflationary motion of the closed string axion, one can express the infladron displacements as a function of $\xi$ to linear order and conclude that they remain small provided that the Hubble scale during inflation is parametrically smaller than the infladron masses. The presence of two additional massive states and their displacements forces us to consider their back-reaction effects on the inflationary potential for $\xi$, which can entail significant flattening effects. Consequently, the cosmological observables for natural-like inflation can take value within the 95\% confidence region of the Planck data in the $(n_s, r)$-plane, as exhibited in figure~\ref{Fig:nsrplanebackreaction}.

In this paper, we have limited our IR analysis to models with one generation of chiral fermions, for which the low-energy theory is described in terms of three scalar excitations. A natural generalization would be to repeat the analysis for set-ups with multiple generations of chiral fermions. But adding more chiral fermions is bound to complicate the IR analysis significantly. In first instance, the increase in Goldstone bosons for the broken global $U(n_f)_L\times U(n_f)_R$ symmetry demands additional explicit symmetry-breaking effects to lift the masses of some Goldstone modes, if one wants to unequivocally identify an inflaton candidate and avoid measurable iso-curvature perturbations. Furthermore, for most scenarios we do not expect a clear mass separation between the pseudo-scalar states, such that single field inflationary models will form more the exception rather than the rule.

The most daring challenge to our variety of inflationary models will be the inevitable confrontation with quantum gravity. In first instance, one has to embed the various inflationary scenarios in a full-fledged Type II string compactification for which the geometric moduli have been stabilized and a precise hierarchy among the various scales in the theory can be established. It is also crucial to investigate how perturbative and non-perturbative quantum gravity effects can change the IR vacuum configuration of the EFT and thereby compromise the inflationary model at hand. These effects might not be enough to capture the full nature of quantum gravity, which appears to put additional constraints~\cite{ArkaniHamed:2006dz, Brown:2015iha,Brown:2015lia,Hebecker:2015rya, Rudelius:2015xta,Heidenreich:2015wga} on the UV-completions of EFT's. To probe the UV-consistency of our set-up with quantum gravity, it is indispensable to evaluate the criteria setting out the boundaries of the swampland~\cite{Vafa:2005ui,Ooguri:2006in} systematically. All these elements are currently under investigation and we plan to report on the implications from the swampland in the near future.

\acknowledgments
We would like to thank Kiwoon Choi, Dagoberto Escobar, Arthur Hebecker, Fernando Marchesano, Liam McAllister, Miguel Montero, Francisco G.~Pedro, Angel Uranga and Clemens Wieck for useful discussions and suggestions.  G.S. is supported in part by the DOE grant DE-SC0017647 and the Kellett Award of the University of Wisconsin.
W.S. is supported by the ERC Advanced Grant SPLE under contract ERC-2012-ADG-20120216-320421, by the grants FPA2015-65480-P (MINECO/FEDER EU) and IJCI-2015-24908 from the MINECO, and the grant SEV-2012-0249 of the ``Centro de Excelencia Severo Ochoa" Programme.





\addcontentsline{toc}{section}{References}
\bibliographystyle{ieeetr}
\bibliography{refs_InfladronsWGC}

\end{document}